\documentclass[11pt]{article}
\pdfoutput=1
\usepackage{jheppubnew}
\usepackage{amsmath}
\usepackage{amssymb}
\usepackage[usenames,dvipsnames,table]{xcolor}
\usepackage{slashed}
\usepackage{pstool}
\usepackage{mathtools}
\usepackage{color}
\usepackage{float}
\usepackage{array}
\usepackage{braket}
\usepackage{enumitem}
\usepackage{amssymb}
\usepackage{upgreek}
\usepackage{amsthm}
\usepackage{graphicx}
\usepackage{hyperref}
 \usepackage{url}
\usepackage{caption}
\usepackage[labelsep=quad]{subcaption}
\usepackage{epstopdf}
\usepackage{bm}
\usepackage{empheq}
\newcommand{\Sch}{\mathrm{Sch}}
	
\usepackage{epsfig}
\usepackage[percent]{overpic}
\usepackage[toc,page]{appendix}
\usepackage{tikz}
\usetikzlibrary{backgrounds}
\usetikzlibrary{decorations.markings}
\usetikzlibrary{decorations.pathmorphing, patterns,shapes}
\usetikzlibrary{shapes.misc}
\usepackage{tabularx}
\usepackage{tcolorbox}
\usepackage{empheq}
\usepackage{simpler-wick}
\definecolor{navy}{rgb}{0.05,0.1,0.75} 

\title{A universe field theory for JT gravity}
\author{Boris Post${}^{a}$, Jeremy van der Heijden${}^{a}$ and Erik Verlinde${}^{a}$}
\emailAdd{b.p.post@uva.nl, j.j.vanderheijden2@uva.nl, e.p.verlinde@uva.nl}
\affiliation{${}^{a}$Institute for Theoretical Physics, University of Amsterdam,
Science Park 904, Postbus 94485, 1090 GL Amsterdam, The Netherlands} 
\begin{document}
\newcommand{\tr}{\mathrm{tr}\,}
\newcommand{\bb}{\mathbb}
\newcommand{\HH}{\mathrm{HH}}
\newcommand{\beq}{\begin{equation}}
\newcommand{\eeq}{\end{equation}}
\newcommand{\fr}{\frac}
\newcommand{\mcal}{\mathcal}
\newcommand{\half}{\frac{1}{2}}
\newcommand{\del}{\partial}
\newcommand{\csch}{\mathrm{csch}}
\newcommand{\no}[1]{\!:\!{#1}\!:\!}
\newcommand{\res}[1]{\underset{#1}{\,\mathrm{Res}\,}}
\newcommand*\widefbox[1]{\fbox{\hspace{1em}#1\hspace{1em}}}
	
\abstract{We present a field theory description for the non-perturbative splitting and joining of baby universes in Euclidean Jackiw-Teitelboim (JT) gravity. We show how the gravitational path integral, defined as a sum over topologies, can be reproduced from the perturbative expansion of a Kodaira-Spencer (KS) field theory for the complex structure deformations of the spectral curve. We use that the Schwinger-Dyson equations for the KS theory can be mapped to the topological recursion relations. We refer to this dual description of JT gravity as a `universe field theory'. By introducing non-compact D-branes in the target space geometry, we can probe non-perturbative aspects of JT gravity. The relevant operators are obtained through a modification of the JT path integral with Neumann boundary conditions. The KS/JT identification suggests that the ensemble average for JT gravity can be understood in terms of a more standard open/closed duality in topological string theory.}

\maketitle

\newcommand{\RNum}[1]{\uppercase\expandafter{\romannumeral #1\relax}}

\section{Introduction}

Recently, the role of topology change in quantum gravity has found some renewed interest. In particular, questions about the definition of the gravitational path integral (GPI) (pioneered by Polyakov \cite{polyakovpathintegral}) and what it can tell us about the microscopic properties of gravity have resurfaced. Heuristically, the GPI is a recipe for any theory of quantum gravity that instructs us to sum over all fields of the theory, including the metric, weighted by the gravitational action. It has been a long-standing debate whether different topologies of the spacetime manifold should be included in this procedure or not
, but recent developments have shown that a great deal can be learned when we do. For example, it was shown \cite{replicawormholes, replicawormholes2} that one can obtain the Page curve for the entanglement entropy of Hawking radiation of an evaporating black hole by adding non-trivial topologies called `replica wormholes' to the gravitational path integral. 

How to interpret these non-trivial topologies from a microscopic point of view is still an open question, but developments of the past years have led to the following intuition: while semiclassical gravity is a low-energy effective description of some UV complete theory, the gravitational path integral still has access to some of the UV data, but only in an \emph{averaged} sense. The non-trivial topologies now probe certain statistical correlations within the model-dependent average. Although the general mechanism is not very well-understood, this idea has been concretely realized in some controlled settings. Let us highlight two viewpoints that have been influential:

\paragraph{\textbf{Matrix models.}} In 2-dimensional Euclidean Jackiw-Teitelboim (JT) gravity \cite{teitelboim,jackiw,almheiri2} the relevant averaging procedure has been identified by Saad, Shenker and Stanford \cite{saad2019jt} in terms of a double-scaled matrix integral. Instead of a single well-defined boundary quantum system described by a Hamiltonian $H$, it was argued that the bulk JT gravity theory is dual to an ensemble of boundary theories, whose Hamiltonians are random matrices drawn from some probability distribution. Each boundary theory is characterized by a partition function:
\beq \label{eq:MMpf}
Z(\beta_i) = \mathrm{Tr} \, e^{-\beta_i H}, \quad i = 1~,\dots, n~,
\eeq
where the inverse temperature $\beta_i$ corresponds to the (renormalized) length of the $i$-th boundary. This partition function becomes a random variable in an ensemble defined by a matrix integral $\braket{\cdots}_{\mathsf{MM}}$. The spacetime wormhole connecting $n$ boundaries now computes the $n$-th connected correlation function of the random boundary partition function
\beq
\mathcal{Z}_{\mathrm{wormhole}}(\beta_1, \dots, \beta_n) = \braket{\mathrm{Tr} \, e^{-\beta_1 H} \cdots \mathrm{Tr} \, e^{-\beta_n H} }_{\mathsf{MM}}^{\mathsf{c}}~,
\eeq
after taking some suitable double-scaling limit of the matrix model. See, for example, \cite{clocksandrods,dissecting,stanfordwittensuperjt,Saad:2019pqd,eigenbranes,cliffordjohnson1,cliffordjohnson2,mertensturiaci,maxfield,WittendeformationsJT,Blommaert:2021gha} for some related work on ensemble averaging in JT gravity, including the generalization to JT supergravity, non-perturbative effects and conical defect geometries. 

\paragraph{\textbf{Baby universes.}} An interesting interpretation of the ensemble average is given by Maxfield and Marolf \cite{marolfmaxfield}, building upon earlier ideas on spacetime wormholes \cite{coleman1,Giddings:1988cx,giddingsstrominger1}. Roughly speaking, a theory of dynamical gravity, where spacetime itself is allowed to change its topology, is most clearly formulated in a \emph{third-quantized} picture. This means that on top of the usual rules of quantum field theory we apply another quantization to account for the dynamics of the topology change. The quantum mechanical system consisting of states labeled by these topologically distinct universes is referred to as the Hilbert space of baby universes, since in Lorentzian signature such geometries can be viewed as modeling the emission and absorption of auxiliary baby universes \cite{Giddings:1987cg}. The ensemble now comes from a decomposition into \emph{$\alpha$-states}, which are defined as the eigenstates of certain boundary creation operators.

\paragraph{}

In this paper, we present a framework that naturally incorporates both viewpoints, in the case of JT gravity. Using intuition from string theory, where one can describe the topological expansion for the splitting and joining of closed strings in terms of a \emph{string field theory}, we introduce a quantum field theory for the non-perturbative splitting and joining of baby universes. This effective description lives on an auxiliary space $\mathcal{S}_\mathrm{JT}$ called the \emph{spectral curve}. The geometry of this space is determined by the leading order density of states and is given by
\beq\label{eq:speccurve}
\mathcal{S}_\mathrm{JT} : \quad y^2 - \fr{1}{(4\pi)^2}\sin^2(2\pi\sqrt{x}) = 0~,
\eeq 
where $x,y \in \mathbb{C}$. This curve can be uniformized by a single complex coordinate $z$ using
\beq\label{eq:uniformize}
x(z) = z^2~, \quad y(z) = \fr{1}{4\pi} \sin(2\pi z)~.
\eeq 
In the string field theory analogy, the spectral curve should be viewed as defining the target space geometry in which the JT universes (the equivalent of string world-sheets or `JT strings') propagate.  The quantum field theory living on the spectral curve now corresponds to the closed string field theory, in the sense that it describes the splitting and joining of JT universes by a cubic interaction vertex. Following \cite{anous}, where a useful analogy with world-line gravity is presented (see also \cite{Casali:2021ewu}), we use the term \emph{universe field theory} for this description. 

We will show that our universe field theory is the 2-dimensional Kodaira-Spencer (KS) theory of complex structure deformations of $\mathcal{S}_\mathrm{JT}$, originally found by Dijkgraaf and Vafa \cite{DVkodairaspencer}. It is obtained as a dimensional reduction of the topological B-model closed string field theory \cite{BCOV} to the spectral curve. This shows that JT gravity can be understood in terms of the well-established topological string theory framework (see \cite{vafamcnamara} where a similar statement was made). However, the interpretation from the gravity point of view is fundamentally different: the \emph{perturbative} expansion in the string coupling constant $\lambda$ corresponds to the \emph{non-perturbative} genus expansion in JT gravity via the identification
\beq
\lambda = e^{-S_0}~,
\eeq
where $S_0$ is proportional to $1/G_N$. Hence, higher loop corrections to the universe field theory amplitudes correspond to non-perturbative wormhole configurations on the gravity side. 

\begin{figure}[h]
\centering
\includegraphics[width=12cm]{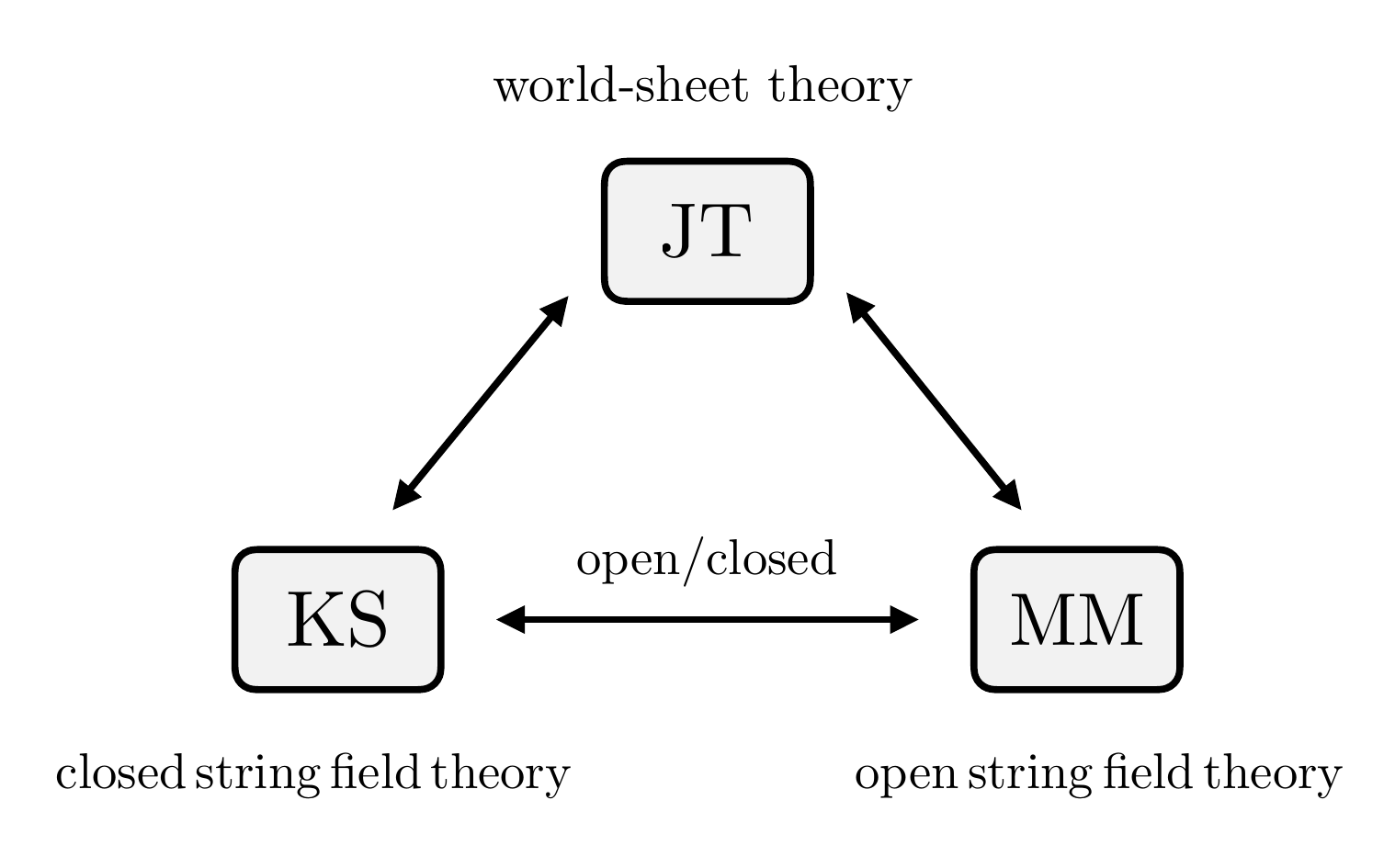}
\caption{A triangle of relations between JT, KS and MM, together with their interpretation in topological string theory.}
	\label{fig:1}
\end{figure}

The duality with KS theory connects the JT/matrix integral correspondence to earlier work on the relation between matrix integrals and topological string theory, e.g., \cite{vafadijkgraaf1,Dijkgraaf:2002fc}. It also nicely agrees with the viewpoint \cite{okuyamakdv,okuyama,Okuyama:2021eju,DijkgraafWitten} that JT gravity is equivalent to the world-sheet topological gravity \cite{wittentopogravity1,wittentopogravity2, kontsevich, dijkgraafverlinde2, dijkgraafverlinde1}. The formulation in terms of KS theory is in some ways more transparent than the matrix integral, as it is formulated directly in the double-scaling limit. Moreover, it makes the embedding in topological string theory manifest and thus provides many useful tools to study non-perturbative aspects of gravity. More importantly, it gives another explanation for why the random matrix ensemble of \cite{saad2019jt} arises in the study of JT gravity, namely as the dual open string field theory \cite{Witten:1992fb,Ooguri:1996ck} through a version of the open/closed duality \cite{openclosed1, openclosed2}. Therefore, we provide evidence for the claim that the JT/matrix integral correspondence is a special example of a more standard open/closed duality in topological string theory. This suggests the following triangle of relations between JT gravity, the matrix model (MM), and the KS theory as shown in Figure \ref{fig:1}.

We now outline the dictionary between the KS theory and JT gravity:
\paragraph{Observables.}
The basic field in KS theory is a $\bb{Z}_2$-twisted chiral boson $\mathcal{J}(z) = \del\Phi(z)$ which parametrizes the complex structure deformations of the spectral curve. The KS field theory contains a cubic interaction that is localized on a contour around the branch point $z=0$:
\beq\label{cubicint}
S_{\mathrm{int}} = \lambda \oint \fr{dz}{2\pi i} \fr{\Phi(z)}{\omega(z)} T(z)~,
\eeq
where $T(z) =\half (\mathcal{J}\mathcal{J})(z)$ is the holomorphic stress tensor, and $\omega = y(z)dx(z)$ is a holomorphic $(1,0)$-form that encodes the complex structure of $\mathcal{S}_\mathrm{JT}$. We will show that the $n$-point function of $\mathcal{J}(z)$ in the KS theory, after an inverse Laplace transform, computes the all-genus gravitational path integral for JT gravity with $n$ asymptotic boundaries:
\begin{equation} \label{eq:KSpathintegral}
	\mathcal{Z}_\mathsf{JT}(\beta_1,\ldots,\beta_n) = \int_{c-i\infty}^{c+i\infty} \prod_{i=1}^n \fr{dz_i}{2\pi i} e^{\beta_i z_i^2} \braket{\mathcal{J}(z_1)\cdots \mathcal{J}(z_n)}_{\mathsf{KS}}~.
\end{equation}
The renormalized boundary length $\beta_i$ of the $i$-th boundary has the interpretation of a fixed temperature in the boundary Schwarzian theory \cite{jensenchaos,Maldacena2016,Engelsoy:2016xyb,maldacenaSYK,schwarzianorigins}. Note that the left-hand side of \eqref{eq:KSpathintegral} only makes sense as a perturbative expansion in $\lambda^{2g-2+n}$, where $\lambda = e^{-S_0}$ and $g$ is the genus of the spacetime wormhole, while the right-hand side is a correlator in a well-defined Euclidean non-gravitational QFT. Expanding the interaction vertex \eqref{cubicint} and doing Wick contractions gives a matching expansion in $\lambda^{-\chi}$, where $\chi$ is the Euler number of the diagram. Thus, the right-hand side provides a non-perturbative completion of the topological expansion of the gravitational path integral in JT gravity. 

Coming back to the discussion of ensemble averaging, we see that \eqref{eq:KSpathintegral} expresses the gravitational path integral as an `average' $\braket{\cdots}_{\mathsf{KS}}$ of the following boundary operators:
\beq\label{boundarycrea}
Z(\beta) = \int_{c-i\infty}^{c+i\infty} \fr{dz}{2\pi i} e^{\beta z^2} \mathcal{J}(z)~.
\eeq

\paragraph{Recursion relations.} 
The argument for the identification \eqref{eq:KSpathintegral} is the universal recursive structure present in both descriptions. Computing the JT gravity path integral amounts to the computation of Weil-Petersson volumes $V_{g,n}(\bm{\ell})$ of the moduli space of bordered Riemann surfaces. These volumes can be found recursively, as was discovered by Maryam Mirzakhani \cite{Mirzakhani1, Mirzakhani2}, by iteratively `stripping off' 3-holed spheres in a modular invariant way. Mirzakhani's recursion is related via a Laplace transform to the topological recursion relations of Eynard and Orantin \cite{eynard1,eynard4,eynard5} for double-scaled matrix models. In this paper, we identify yet another recursion relation: we will show that the Schwinger-Dyson (SD) equations for $\Phi(z)$ in the KS theory imply the topological recursion relations. The SD equations can be expressed as a differential equation for the generating functional of connected correlation functions $W_\mathsf{KS}[\mu_\mathcal{J}]=-\log Z_\mathsf{KS}[\mu_\mathcal{J}]$, where $\mu_\mathcal{J}(z)$ is a source field for $\mathcal{J}(z)$. They take the following form: 
\begin{equation}  \label{eq:SDequation1}
\fr{\delta W_\mathsf{KS}}{\delta \mu_\mathcal{J}(z_0)} \Big\vert_{\chi<0}=\fr{\lambda}{4}\oint \fr{dz}{2\pi i}\fr{\braket{\mathcal{J}(z_0)\Phi(z)}_0}{\omega(z)}  \left[ \fr{\delta^2 W_\mathsf{KS}}{\delta \mu_\mathcal{J}(z)\delta\mu_\mathcal{J}(z)} + \fr{\delta W_\mathsf{KS}}{\delta \mu_\mathcal{J}(z)} \fr{\delta W_\mathsf{KS}}{\delta \mu_\mathcal{J}(z)}  \right]~.
\end{equation}
The details of this equation will be explained in section \ref{sec:schwingerdyson}. In particular, we show that expanding both sides in powers of $\lambda$ gives the topological recursion for the symplectic invariants $\omega_{g,n}(z_1,\dots,z_n)$, which are identified with connected correlation functions $\braket{\mathcal{J}(z_1)\cdots \mathcal{J}(z_n)}_\mathsf{KS,c}^{(g)}$. Using the map \eqref{eq:KSpathintegral} at fixed genus $g$, this gives a recursion relation between contributions from spacetime wormholes to the full GPI. In appendix \ref{ch:recursion}, we show that these recursion relations can be recast as a Virasoro constraint \cite{dijkgraafverlinde2} in the oscillator formalism of the KS theory.

 \paragraph{Non-perturbative effects.} The topological string perspective provides a natural setting to study non-perturbative effects due to D-branes. The spectral curve can be embedded in a 6-dimensional Calabi-Yau manifold, which defines the target space of the string:
\beq \label{eq:targetsp}
uv - y^2 + \fr{1}{(4\pi)^2}\sin^2(2\pi \sqrt{x})=0~.
\eeq
This geometry has non-compact subspaces $u=0$ and $v=0$, which can be wrapped by \emph{branes} and \emph{anti-branes} respectively \cite{integrablehierarchies}. In the KS theory, these branes can be described by a pair of complex fermions $\psi(E) = e^{\Phi(E)}$ and $\psi^\dagger(E) = e^{-\Phi(E)}$ in terms of the coordinate $E=-x$ on $\mathcal{S}_\mathrm{JT}$. We will identify the dual observables in JT gravity to be universes with fixed energy boundaries \cite{boundaryconditions}, ending on branes in the target space geometry \eqref{eq:targetsp}. Here, we do not fix the length of the boundary metric, but we fix the dilaton and its normal derivative, which corresponds to a fixed energy $E$ in the Schwarzian theory. The fixed energy boundary conditions are related to the asymptotically $\mathrm{AdS}_2$ boundary conditions by a Legendre transform. In fact, we will extend the dictionary \eqref{eq:KSpathintegral} for $n$ boundaries with energies $E_1,\dots,E_n$ to   
\begin{equation} \label{eq:KSpathintegral2}
	\mathcal{Z}_\mathsf{JT}(E_1,\ldots,E_n) = \int_{c-i\infty}^{c+i\infty} \prod_{i=1}^n \fr{d\beta_i}{\beta_i} e^{\beta_i E_i} \braket{Z(\beta_1) \cdots Z(\beta_n)}_{\mathsf{KS}}~.
\end{equation}
Using \eqref{boundarycrea}, the KS observables appearing on the right-hand side of \eqref{eq:KSpathintegral2} can be rewritten in terms of the discontinuity of the boson $\Phi$ across the branch cut of $z=\sqrt{-E}$: 
\beq
Z(E) \equiv \int_{c-i\infty}^{c+i\infty} \fr{d\beta}{\beta} e^{\beta E} Z(\beta) =\mathrm{disc}\,\Phi(E) = \int^E dE'\, \rho(E')~.
\eeq
At the last equality, the discontinuity of $\Phi$ is rewritten as an integrated density of states operator $\rho(E)$, to make clear that $Z(E)$ represents a microcanonical partition function, in the same way that $Z(\beta)$ is a canonical partition function \eqref{eq:MMpf} in the boundary theory. We will show that non-perturbative corrections to density correlators can be computed from insertions of the `energy brane' operators $e^{\pm \Omega(E)}$, where $\Omega(E)\equiv 2\pi i Z(E)$. In particular, we retrieve the universal sine-kernel for the connected part of the density-density amplitude:
\begin{align}
\braket{\rho_{\textrm{np}}(E_1)\rho_{\textrm{np}}(E_2)}^{\mathsf{c}}_\mathsf{KS} &\sim  \braket{\rho(E_1)\rho(E_2)}^{\mathsf{c}}_\mathsf{KS} +\frac{1}{4\pi^2}\left( \braket{e^{\Omega(E_1)}e^{-\Omega(E_2)}}_\mathsf{KS} +\braket{e^{-\Omega(E_1)}e^{\Omega(E_2)}}_\mathsf{KS}\right) \nonumber \\[0.8em]
 &\approx -\fr{1}{\pi^2(E_1-E_2)^2} \sin^2\left(\pi\,e^{S_0}\int_{E_2}^{E_1} \rho_0(E') dE'\right)~.
\end{align}
The remainder of this paper is organized as follows. 
\begin{itemize}
    \item In section \ref{universefield}, we introduce the KS field theory, and present a detailed derivation of the SD equations, which characterize the KS correlation functions up to all orders in perturbation theory.
    \item In section \ref{JTgravity}, we make the connection to JT gravity. In particular, we show that the SD equations of the universe field theory coincide with the topological recursion relations, with a choice of initial conditions given by the JT spectral curve. Therefore, the diagrams of the KS theory are in one-to-one correspondence with the JT universes in the asymptotic expansion of the GPI. 
    \item We generalize the setup in section \ref{boundaryconditions} to include fixed energy boundaries in JT. On the KS side of the duality we interpret these boundaries as attached to D-branes in the Calabi-Yau target space, which allows us to explore non-perturbative physics. In particular, we show how to derive the sine-kernel in the density-density correlator.
    \item We conclude with a discussion and a list of open questions in section \ref{discussion}. 
    \item Some KS calculations have been relegated to appendix \ref{freetwopoint}, and appendix \ref{toporecursion} involves a more detailed discussion of the topological recursion. In appendix \ref{ch:recursion} we have worked out the relation with topological gravity in more detail, which gives another perspective on the KS/JT duality in terms of the oscillator algebra of a twisted boson, the baby universe Hilbert space and the Virasoro constraints .  
\end{itemize}

\section{A universe field theory}\label{universefield}

In section \ref{sec:stringembedding} we present the universe field theory that describes dynamical topology change in JT gravity. We will argue that this is the 2-dimensional KS theory on the JT spectral curve. The precise identification follows from matching the topological recursion relations for JT gravity with the SD equations for the KS field theory, which will be derived in section \ref{sec:schwingerdyson}.

\subsection{Kodaira-Spencer theory on the spectral curve}\label{sec:stringembedding}
The KS theory on the spectral curve has the following path integral representation:
\beq\label{KSwithsources}
Z_{\mathsf{KS}}[\mu_\Phi, \mu_\mathcal{J}] = \int [d\mathcal{J}][d\Phi] \exp \left[-S_\mathsf{KS}[\Phi,\mathcal{J}] - \int_{\mathcal{S}_\mathrm{JT}}  \mu_\Phi  \Phi - \int_{\mathcal{S}_\mathrm{JT}}  \mu_\mathcal{J} \mathcal{J}\right]~,
\eeq
where $\mu_\Phi$ and $\mu_\mathcal{J}$ are external source fields, and the action is given by
\beq\label{KSaction}
S_\mathsf{KS} [\Phi,\mathcal{J}] = \int_{\mathcal{S}_\mathrm{JT}} \left[\half \,\del\Phi \wedge \overline{\del}\Phi - \mathcal{J}\wedge\overline{\del}\Phi \right] +  \oint_\gamma \left[\fr{\omega\, \Phi}{\lambda} + \fr{\lambda}{2}\fr{\Phi}{\omega} \mathcal{J}^2 \right]~. 
\eeq
Let us explain all the terms appearing in this action. First of all, the action consists of a `bulk' and a `boundary' contribution: the bulk integral is over the JT gravity spectral curve $\mathcal{S}_\mathrm{JT}$ given in \eqref{eq:speccurve}. We will use a uniformizing coordinate $z$ as in \eqref{eq:uniformize}. In particular, the relation $x= z^2$ shows that in terms of the variable $z$ the KS theory is defined on a branched double cover of the spectral $x$-plane, with a branch point at $z=0$. The boundary integral is over a closed curve $\gamma$ encircling the branch point, which does not enclose any other poles or zeroes of the holomorphic $(1,0)$-form:
\beq
\omega = \omega(z)dz = y(z)dx(z)~.
\eeq 
We have used complex differential notation in the sense of Dolbeault cohomology, so that for example $\del \Phi = \del\Phi(z) dz$, and
$
d = \del +\overline{\del}~.
$
We will always distinguish form fields and ordinary fields by writing fields with their argument and form fields without. For example,
$
\mathcal{J} = \mathcal{J}(z) dz
$
is a $(1,0)$-form field, while $\mathcal{J}(z)$ is a function of the local coordinate $z$ on the spectral curve. To further ease our notation, we define the integral $\int_{\mathcal{S}_{\mathrm{JT}}}$ to include a factor of $\fr{i}{2}$ to make the action real. This factor arises from the usual relation $d^2z = \fr{i}{2} dz \wedge d\overline{z}$. Similarly, we define the contour integral $\oint_\gamma$ to include a factor of $\fr{1}{2\pi i}$ to make the boundary action real. We will also often drop the wedge product when it is clear from the context. 

Having set the notation, we go on to analyze the field content of the theory. There are two dynamical bosonic fields $\Phi = \Phi(z)$ and $\mathcal{J} = \mathcal{J}(z) dz$. We do not explicitly write the anti-holomorphic dependence on $\overline{z}$, but on the level of the path integral $\Phi$ and $\mathcal{J}$ are not necessarily chiral. For now, this is just a notational convenience, but we will see that on-shell $\Phi$ and $\mathcal{J}$ will be chiral fields. The source fields $\mu_\Phi$ and $\mu_\mathcal{J}$ are $(1,1)$ and $(0,1)$-form fields, respectively. The holomorphic $(1,0)$-form $\omega$ appears in the boundary contribution to the action, and it serves to give the chiral boson $\mathcal{J}(z)$ a vacuum expectation value. The term proportional to $\fr{\Phi}{\omega} \mathcal{J}^2$ is the most interesting: this cubic interaction term encodes all the non-trivial dynamics of the splitting and joining of baby universes. 

The action \eqref{KSaction} was first written down by Dijkgraaf and Vafa \cite{DVkodairaspencer} in the context of topological string theory. There, it was obtained by reducing the 6-dimensional KS theory of the closed string B-model developed in \cite{BCOV} to a chiral boson on a Riemann surface. For this reason, we have labeled the action by $\mathsf{KS}$, for `Kodaira-Spencer'. This chiral boson perspective was used, for example, in the `re-modeling the B-model' program of \cite{remodelling}. See also \cite{vafadijkgraaf1, topologicalvertex, integrablehierarchies, lottehollands} for more work on the relation between topological strings, matrix models and integrable systems. In the next subsection, we will explain the origin of the universe field theory action \eqref{KSaction} in topological string theory.  

\subsubsection{Topological string theory origin of $S_\mathsf{KS}$}\label{calabiyau}
As stated in the introduction, we will embed the spectral curve into a non-compact Calabi-Yau manifold in the following way:
\beq\label{eq:calabiyau}
\mathsf{CY}: \quad uv = H(x,y)~, \quad u,v \in \mathbb{C}~,
\eeq
where $H(x,y)$ is given by:
\beq
H(x,y) \equiv y^2 - \fr{1}{(4\pi)^2} \sin^2(2\pi \sqrt{x})~. 
\eeq
The submanifolds where $u$ or $v$ vanish correspond to the spectral curve $\mathcal{S}_\mathrm{JT}: H(x,y) =0$, and $\mathsf{CY}$ can be viewed as a fiber bundle over the spectral curve. The defining relation \eqref{eq:calabiyau} shows that $\mathsf{CY}$ has three complex dimensions, and the complex structure of $\mathsf{CY}$ is encoded in the holomorphic $(3,0)$-form: 
\beq
\Omega_{\mathsf{CY}} = \fr{1}{u} \, du\wedge dx\wedge dy~.
\eeq
The KS field theory on $\mathsf{CY}$ describes deformations of the complex structure such that the cohomology class of $\Omega_{\mathsf{CY}}$ is unchanged. Upon reduction of the theory to the base Riemann surface $\mathcal{S}_\mathrm{JT}$, this translates to complex structure deformations of $\mathcal{S}_\mathrm{JT}$ such that the holomorphic $(1,0)$-form $\omega = y\,dx$ is preserved. To see this, consider a 3-cycle $\widetilde{C}$ in $\mathsf{CY}$. For a Calabi-Yau modeled on a Riemann surface, there is a one-to-one correspondence between 3-cycles in $\mathsf{CY}$ and 1-cycles on the Riemann surface \cite{lottehollands}. Explicitly, a 3-cycle $\widetilde{C}$ can be made by fibering an $S^1$ over a disk $D$, whose boundary $\del D$ is a non-trivial 1-cycle $C$ on the Riemann surface. Computing a period of $\Omega_{\mathsf{CY}}$ on $\widetilde{C}$ then reduces to a period integral of $\omega$ on $C$:
\beq\label{periodintegral}
\int_{\widetilde{C}} \Omega_{\mathsf{CY}} = \int_{\widetilde{C}} \fr{du \wedge dx \wedge dy}{u} = \fr{1}{2\pi i}\oint_{S^1} \fr{du}{u} \int_D dx \wedge dy = \int_C y\, dx~.
\eeq
At the last equality, we have evaluated the residue at $u=0$, followed by an application of Stokes' theorem. The complex structure deformations of $\mathcal{S}_\mathrm{JT}$ are captured by deforming the $\overline{\del}$ operator:
\beq
\overline{\del} \to \overline{\del} - \upmu \,\del~,
\eeq 
where $\upmu = \upmu^{z}_{\overline{z}} \, d\overline{z} \otimes \del_z$ is a so-called Beltrami differential. In the deformed complex structure, a function $f$ is holomorphic if and only if $(\overline{\del} - \upmu \,\del)f = 0$. As in the 6-dimensional KS theory \cite{BCOV}, the 2-dimensional KS theory is the quantization of fluctuations of the complex structure such that the cohomology class of $\omega$ is unchanged. That is, we demand that there is a vector field $\xi$ such that
\begin{equation}\label{combine1} \upmu = \overline{\partial} \xi \quad \mathrm{and} \quad \delta_\xi \omega = d\Phi~,
\end{equation}
where $\Phi$ is the basic field of the KS action \eqref{KSaction}. Explicitly, the variation of $\omega$ under a diffeomorphism $\xi$ is found by taking the Lie derivative in the direction of $\xi$:
\begin{equation}\label{combine2}
\delta_\xi \omega \equiv \mathcal{L}_\xi \omega = d(\iota_{\xi} \omega) - \iota_\xi d\omega~.
\end{equation}
Now we use that $\omega$ is a holomorphic $(1,0)$-form, so that \beq d\omega = (\overline{\partial} + \partial) \omega = \overline{\partial} \omega = 0~.\eeq
Here, we used that $\del\omega =0$: there are no $(2,0)$-forms on a Riemann surface.
Comparing \eqref{combine1} and \eqref{combine2} we conclude that $\iota_\xi \omega = \Phi$ up to a $d$-closed form, which we can conveniently write as:
\begin{equation}\label{vectorfield}
\xi = \frac{\Phi}{\omega}~.
\end{equation}
So we see that the Beltrami differential $\upmu$ depends on the field $\Phi$. Imposing that $\delta_\xi \omega$ is holomorphic in the deformed complex structure implies:
\beq
(\overline{\del} - \upmu \,\del) \delta_\xi \omega = (\overline{\del} - \upmu \,\del) \del\Phi = 0~.
\eeq
This should be implemented in the field theory as an equation of motion. So we see that the action should contain the term:
\beq
\half \int_{\mathcal{S}_{\mathrm{JT}}}\del\Phi \wedge (\overline{\del} - \upmu \,\del)\Phi~.
\eeq
This contains a kinetic term for $\Phi$, as well as the interaction:
\beq \label{eq:int1}
\int d^2 z\, \upmu^z_{\overline{z}} \,T(z)~,
\eeq
where we have written $T(z)$ for the stress tensor $T(z) \equiv \half \del\Phi(z)\del\Phi(z)$. In the quantum theory, $T(z)$ is normal ordered in the usual way using a point-splitting regularization, i.e., by subtracting the divergent part of the OPE. Plugging in the expression for $\upmu = \overline{\del} \xi$ explains the origin of the cubic interaction in the KS action \eqref{KSaction}. Before going into the details, let us pause and give some more intuition for why we have found the interaction \eqref{eq:int1}.

\begin{figure}[h]
\centering
\begin{overpic}[width=6cm]{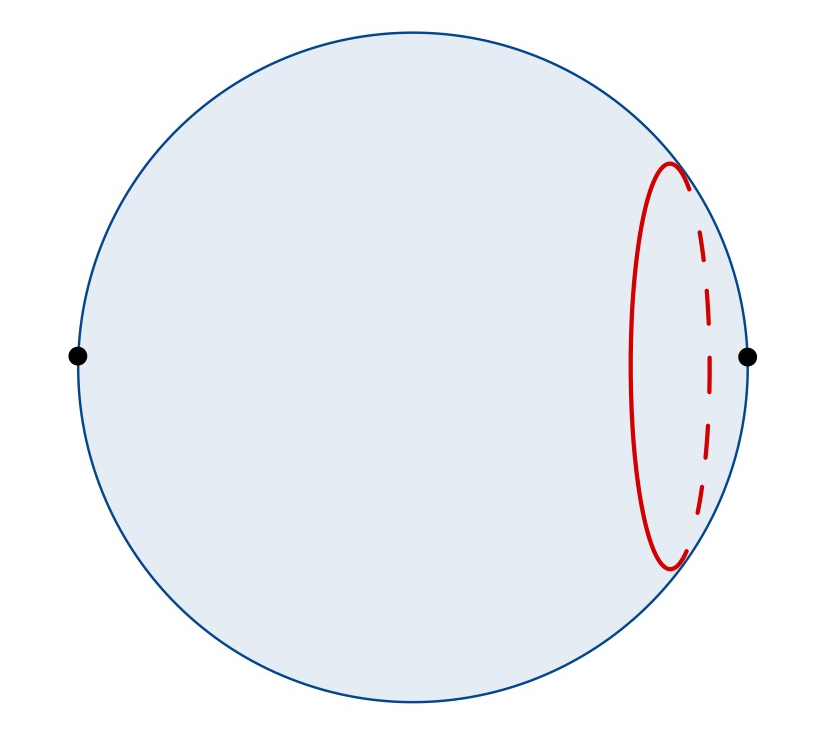}
 \put (1,44.5) {$\infty$}
  \put (95,44.3) {$0$}
  \put (72,58) {{\Large$\gamma$}}
  \put (13,44.5) {$\omega = \del\Phi_{cl}$}
  \put (61,30,5) {$\oint_\gamma \xi \,T$}
  \put (39,44.5) {$\xrightarrow{\hspace*{1.2cm}}$}
  \put (64,44.3) {$\del\Phi$}
  \put (5,80) {$\mathcal{S}_{\mathrm{JT}}$}
\end{overpic}
\caption{Fixing the behaviour of $\omega$ at $\infty$ determines the classical value $\del\Phi_{cl}$. As one moves away from infinity, quantum fluctuations of $\del\Phi$ can appear which deform the complex structure. At the contour $\gamma$ there is a coordinate change to the patch that covers $0$. }
\label{kscartoon}
\end{figure}
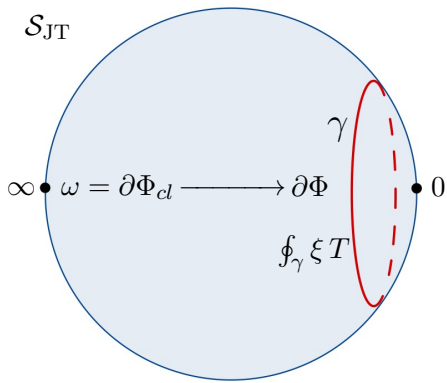

Consider the cartoon of our setup in Figure \ref{kscartoon}. 
We have drawn the (compactified) spectral curve $\mathcal{S}_{\mathrm{JT}}$ as a single Riemann sphere, by going to the covering space. The antiperiodicity is implemented by a twist field at $0$ and $\infty$. The Riemann sphere is covered by two coordinate patches, with a transition function that determines the complex structure. Basically, the complex structure defines what we mean by a holomorphic function in each patch. In the left patch covering $\infty$, the complex structure is such that $\omega = ydx$ is holomorphic. We want to think of $\omega$ as the classical vacuum expectation value of the basic field $\del\Phi$, set, for example, by some background gauge field. As we move away from infinity into the `bulk' of $\mathcal{S}_{\mathrm{JT}}$, the 1-form $\omega$ is allowed to fluctuate:
\beq
\omega + \delta_\xi \omega = \del\Phi_{cl}+ \del\Phi~.
\eeq
Classically, only negative frequencies are allowed in the mode expansion of the fluctuation $\del\Phi$ (corresponding to positive powers of $z$) with the same behaviour at infinity as $\omega$. But quantum mechanically, there can also be positive frequencies $\del\Phi_+$, which are expanded in powers of $z^{-1}$. These modes fall off to $0$ at $z\to \infty$, but they give non-zero contributions in the interior. Moving even further into the bulk, there is a coordinate change to the right patch, which is  implemented by the operator
\beq \label{eq:int2}
\oint_\gamma dz \,\xi(z) T(z)~.
\eeq
In the right coordinate patch, $\delta_\xi \omega$ is holomorphic in the deformed complex structure. 
Indeed, we will show in subsection \ref{inttheory} that after an integration by parts, the interaction \eqref{eq:int1} can be written as a contour integral \eqref{eq:int2}. This gives a nice interpretation of the KS interaction as a kind of $\Phi$-dependent coordinate transformation for the field $\del\Phi$.

\subsubsection{The free theory}\label{freetheory}
We will first analyze the free bosonic theory, i.e., without the interaction induced by the complex structure deformation. Consider the free action
 \beq\label{eq:freeaction}
 S_{\mathsf{KS}}^{(0)}[\Phi, \mathcal{J}] =  \int_{\mathcal{S}_\mathrm{JT}} \left[\half \,\del\Phi \wedge \overline{\del}\Phi - \mathcal{J}\wedge\overline{\del}\Phi \right]~.
 \eeq
 The equation of motion for $\mathcal{J}$ forces $\Phi$ to be chiral:
 \beq
 \overline{\del} \Phi = 0 \quad \textrm{on-shell}~.
 \eeq
This chirality constraint for $\Phi$ reflects the fact that the classical value of $\del\Phi$ is $\omega$, which is holomorphic. However, $\mathcal{J}$ is not merely a Lagrange multiplier: it is a dynamical field. In fact, the classical equation of motion for $\Phi$ shows that:
\beq
\mathcal{J} = \del\Phi \quad \textrm{on-shell}~.
\eeq
 This identification holds up to holomorphic forms. Now we see why we have used the notation $\mathcal{J}$: on-shell, it plays the role of the holomorphic current $\del\Phi$. The monodromy properties of $\del\Phi$ around the branch point follow from the fact that the 1-form $\omega$ is \emph{odd} in $z$. The variation under a diffeomorphism $\delta_\xi \omega$ should preserve the parity under the involution $z \to -z$ around the branch point, and so we conclude that $\del\Phi$ should also be odd. This shows that we are dealing with a $\bb{Z}_2$-twisted chiral boson on the spectral curve.  

Let us now compute the two-point functions of the free theory. Consider the free partition function including the sources:
 \beq\label{eq:freesources}
 Z^{(0)}_\mathsf{KS}[\mu_\Phi, \mu_{\mathcal{J}}] = \fr{1}{ Z^{(0)}_\mathsf{KS}[0]} \int [d\mathcal{J}][d\Phi] \exp \left[-S_\mathsf{KS}^{(0)}[\Phi,\mathcal{J}] - \int_{\mathcal{S}_\mathrm{JT}}   \mu_\Phi  \Phi - \int_{\mathcal{S}_\mathrm{JT}}  \mu_\mathcal{J} \mathcal{J}\right]~.
 \eeq
Since this is a Gaussian integral in $\Phi$ and $\mathcal{J}$, we can solve it using functional determinants. The determinants cancel against the normalization $Z^{(0)}_\mathsf{KS}[0]$. In appendix \ref{freetwopoint}, we show in detail how to compute the functional integral, which gives the result:
\beq\label{KSfreepartition}
 \log Z^{(0)}_\mathsf{KS}[\mu_\Phi, \mu_{\mathcal{J}}]  = \int d^2z \int d^2w \left[\half \mu_\mathcal{J}(z) \mathsf{B}(z,w)\mu_\mathcal{J}(w) + \mu_\mathcal{J}(z)\mathsf{G}(z,w) \mu_\Phi(w) \right]~,
\eeq
where we have defined \vspace{-1em}
\begin{align}\label{free2points}
\mathsf{B}(z,w) &= \fr{1}{(z-w)^2} + \fr{1}{(z+w)^2}~, \\
 \mathsf{G}(z,w) &= \fr{1}{z-w} - \fr{1}{z+w}~.
\end{align}
Defining connected correlation functions as functional derivatives of $\log Z^{(0)}_\mathsf{KS}[\mu_\Phi, \mu_{\mathcal{J}}]$, we find that the only non-zero two-point functions are:
\begin{align}
\braket{\mathcal{J}(z)\Phi(w)}^{\mathsf{c}}_0 &= \left.\frac{\delta^2 \log Z^{(0)}_\mathsf{KS}}{\delta \mu_\mathcal{J}(z) \delta \mu_\Phi(w)}\right |_{\mu=0} =  \mathsf{G}(z,w)~, \\
\braket{\mathcal{J}(z)\mathcal{J}(w)}^{\mathsf{c}}_0 &=\left. \frac{\delta^2 \log Z^{(0)}_\mathsf{KS}}{\delta \mu_\mathcal{J}(z) \delta \mu_\mathcal{J}(w)}\right |_{\mu=0} = \mathsf{B}(z,w)~.
\end{align}
In particular, we see that there are no contractions of $\Phi$ with itself. This will be an important fact, when we make the connection to the topological recursion in section \ref{sec:matchingtop}. It can be seen as a result of taking operator insertions inside correlation functions on-shell: since $\overline{\del}\Phi = 0$ on-shell, $\Phi$ only contains the negative frequencies (positive powers of $z$), and so it does not have a two-point function in the vacuum. 

However, we stress that we derived this fact purely in the functional formalism, without making reference to mode expansions of $\Phi$. Moreover, the functions $\mathsf{B}(z,w)$ and $\mathsf{G}(z,w)$ agree with the standard two-point functions of a free $\bb{Z}_2$-twisted chiral boson, as is explicitly verified in appendix \ref{chiralbosontwist}.

\subsubsection{The interacting theory}\label{inttheory}
Up till now, we have not given our bosonic fields a vacuum expectation value, although we argued that we want to think of $\omega$ as the classical value of $\del\Phi$. We can incorporate this shift in $\omega$ by noticing that: 
\beq\label{bdyintegral2}
\int_{\mathcal{S}_\mathrm{JT}}  \omega \wedge \overline{\del} \Phi = \oint_\gamma \omega\, \Phi~.
\eeq
So we can simply shift $\mathcal{J}$ by $\omega$ and integrate by parts, using that $\overline{\del}\omega = 0$:
 \beq
\int_{\mathcal{S}_\mathrm{JT}} \left[\half \,\del\Phi \wedge \overline{\del}\Phi - (\mathcal{J} - \omega)\wedge\overline{\del}\Phi \right] = \int_{\mathcal{S}_\mathrm{JT}} \left[\half \,\del\Phi \wedge \overline{\del}\Phi - \mathcal{J} \wedge\overline{\del}\Phi \right] + \oint_\gamma \omega \, \Phi~.
 \eeq
This does not change the e.o.m. \!for $\Phi$, since the identification $\del\Phi = \mathcal{J}$ only holds up to a $\overline{\del}$-closed form. The shift of $\mathcal{J}$ by $\omega$ also does not affect connected correlation functions, except for the classical one-point function:
\beq
\braket{\mathcal{J}(z)}_0 = \omega(z)~.
\eeq
The boundary integral in \eqref{bdyintegral2} is along a contour $\gamma$ that encircles the branch point at $z=0$. 
We can use a similar argument to show that the interaction term localizes to the branch point. Plugging in our expression for the Beltrami differential, and writing the stress tensor as $T = T(z)\, dz\otimes dz$, the interaction that implements the complex structure deformation is written as:
\beq
S_{\mathrm{int}} = 
\int_{\mathcal{S}_{\mathrm{JT}}} \upmu \cdot T = \int_{\mathcal{S}_{\mathrm{JT}}} \overline{\del}\xi \cdot T~. 
\eeq
The dot $\cdot$ is shorthand for contracting $ dw \partial_z = \delta_z^{w}$ in the second tensor factor and then integrating the (1,1)-form $\upmu_{\bar{z}}^z T(z) d\overline{z} \wedge dz$ coming from the first tensor factor. In perturbation theory, the stress tensor remains holomorphic:
\beq
\overline{\del} T = 0~.
\eeq 
We can thus integrate by parts in a region $V$ where $\xi(z)T(z)$ is holomorphic. 
Notice that the vector field $\xi = \fr{\Phi}{\omega}$ has poles at the branch point $z=0$ and at the other zeroes of $\omega(z) \sim z \sin(2 \pi z)$. The zeroes of $\omega(z)$ different from the branch point correspond to the `pinched cycles' of the Riemann surface $\mathcal{S}_{\mathrm{JT}}$ (see appendix \ref{app:spectralcurveJT}). We take $V$ such that its boundary is a collection of contours $\gamma_i$ surrounding the zeroes of $\omega$, including the branch point, and use Stokes' theorem:
\beq\label{Sint3}
S_{\mathrm{int}} = \int_V \overline{\del} (\xi \cdot T ) = \int_V d (\xi \cdot T ) = \sum_i \oint_{\gamma_i} \xi \cdot T~.
\eeq
Now we argue that only the contribution from the branch point gives a non-zero result inside correlation functions. To see this, recall that the spectral curve is a branched double cover of the spectral plane, with a twist field $\sigma(0)$ inserted at the branch point. As explained in appendix \ref{chiralbosontwist}, the twisted vacuum is related to the conformally invariant free boson vacuum by $\ket{\sigma} = \sigma(0) \ket{0}$. For any contour $\gamma_j$ which does \emph{not} surround the branch point, the operator $\oint_{\gamma_j} \xi\cdot T$ commutes with $\sigma(0)$ because their operator product is trivial ($\gamma_j$ never gets close to $0$). It then annihilates the untwisted vacuum:
\beq
\oint_{\gamma_j} \xi\cdot T \ket{\sigma} = \sigma(0)\oint_{\gamma_j} \xi\cdot T \ket{0} =0~.
\eeq
To see why the untwisted vacuum gets annihilated, let $\zeta$ be a local coordinate around the $j$-th zero of $\omega(z)$. Then, expand the stress tensor in even powers of $\zeta$, the field $\Phi$ in odd positive powers of $\zeta$, and $\omega(z)^{-1}$ in powers of $\zeta^{2i -2}$, $i\geq 0$. Working out the contour integral shows that only stress tensor modes $L_{n\geq -1}$  appear in the operator $\oint \xi \cdot T$. Since the untwisted vacuum $\ket{0}$ is conformally invariant, it gets annihilated by $\{L_{-1},L_0,L_1\}$. Moreover, the normal ordering of the stress tensor $T$ ensures that the $L_{n>0}$ contain only annihilation operators to the right of the creation operators. So we conclude that $L_n \ket{0} = 0$ for all $n \geq-1$. Therefore, the exponential of the interaction collapses to a single contribution from the contour $\gamma$ that does surround the branch point:
\beq
e^{-\widehat{S}_{\mathrm{int}}} \ket{\sigma} = e^{-\sum_i \oint_{\gamma_i} \xi \cdot T} \ket{\sigma} = e^{-\oint_\gamma \xi \cdot T}\ket{\sigma}~.
\eeq
This argument can easily be generalized to spectral curves with multiple branch points and twist operators. In that case, the interaction term will localize to a sum over contributions from the branch points only. 

With this localization argument, we arrive at the action of the universe field theory \eqref{KSaction}. Indeed, we can write the stress tensor on-shell as $T = \half \mathcal{J}^2$, with the point-splitting regularization
\beq
T(z) = \half \lim_{w\to z} \Big(\mathcal{J}(w)\mathcal{J}(z) - \fr{1}{(z-w)^2} \Big)~,
\eeq
and plugging this into \eqref{Sint3} we see that the interaction term is:
\beq\label{Sint}
S_{\mathrm{int}} = \half \oint_\gamma \fr{\Phi}{\omega} \mathcal{J}^2~.
\eeq
Rescaling $\omega$ by $\lambda$ gives the interaction a coupling constant. The fact that the interaction is localized to a  contour around $z = 0$ ensures that the theory is free of UV divergences which normally crop up when adding an irrelevant deformation to a CFT. When doing conformal perturbation theory and expanding the exponential of $S_{\mathrm{int}}$, the contours can be chosen to be non-intersecting so that operators are never inserted at the same point \cite{dijkgraafchiral}. In some sense, we can think of \eqref{Sint} as a `topological' interaction: the contour $\gamma$ can be deformed at will, as long as it does not cross or enclose the other zeroes of $\omega(z)$.  

Most notably, the interaction is \emph{cubic} in the fields $\Phi$ and $\mathcal{J} = \del\Phi$. We will argue that this cubic vertex represents the pair-of-pants that is used as a building block in constructing hyperbolic surfaces, which are the relevant geometries in JT gravity. The Feynman diagrams of $\mathcal{J}$-correlators are then to be viewed as the `skeletons' of the spacetime wormholes. The usual rules of summing over all possible diagrams then ensure the modular invariance of the GPI. In the next subsection, we will establish a recursion relation between the diagrams of KS theory, which will be matched to the topological recursion for JT gravity in section \ref{JTgravity}.

\subsection{Schwinger-Dyson equations}\label{sec:schwingerdyson}
The Schwinger-Dyson (SD) equations in a quantum field theory can be seen as the quantum version of the equations of motion. They are usually derived by requiring that the measure is invariant under an infinitesimal linear shift in the field variable, or equivalently, that the functional integral of a total functional derivative is zero. This gives a set of differential equations for $n$-point functions, which are sometimes taken as a definition of the theory. 

In our case, we will have a SD equation for both $\Phi$ and $\mathcal{J}$. The SD equation for $\mathcal{J}$ just imposes the quantum version of the chiral constraint. The SD equation for $\Phi \to \Phi + \delta \Phi$ is the most interesting equation: we will show that it is directly equivalent to the topological recursion. The starting point will be the full interacting partition function including sources in \eqref{KSwithsources}. Imposing that the path integral is invariant gives the SD equation:
\beq
\fr{1}{Z_\mathsf{KS}[0]}\int [d\mathcal{J}][d\Phi] \, \delta_\Phi  \exp \left[-S_\mathsf{KS}[\Phi,\mathcal{J}] - \int_{\mathcal{S}_\mathrm{JT}}  \mu_\Phi  \Phi - \int_{\mathcal{S}_\mathrm{JT}}  \mu_\mathcal{J} \mathcal{J}\right] = 0~.
\eeq
Here, we have written the functional variation $\delta_\Phi (\dots) = \fr{\delta}{\delta \Phi} (\dots) \delta \Phi$. This variation brings down two terms from the exponential:
\beq\label{SD2}
 \left \langle \delta_\Phi S_{\mathsf{KS}} + \int_{\mathcal{S}_\mathrm{JT}} \mu_\Phi \, \delta \Phi\right\rangle_{\mu_\Phi, \,\mu_\mathcal{J}} = 0~.
\eeq
To make the variational problem well-defined, we need to specify boundary conditions for the field variation $\delta \Phi$. Since $\Phi$ is odd, we will also impose that $\delta\Phi$ is odd. We further impose the regularity condition at the boundary that $\delta \Phi(z) \to 0$ as $z \to 0$. So in particular, $\delta\Phi$ cannot have a pole at the branch point. Summarizing, we demand that:
\beq
\delta \Phi \big\vert_\gamma = \textrm{odd and analytic}~.
\eeq
Let us now compute the variation of the action, carefully treating the surface and boundary contributions:
\begin{align}
\delta_\Phi S_{\mathsf{KS}}  &= \int_{\mathcal{S}_{\mathrm{JT}}} \big(- \del \overline{\del} \Phi  +\overline{\del}\mathcal{J}\big) \delta \Phi \,+ \oint_{\gamma} \left[ \fr{\omega \,\delta \Phi}{\lambda}  - \mathcal{J} \delta \Phi + \frac{\lambda}{2} \frac{\mathcal{J}^2}{\omega} \,\delta \Phi \right]~.
\end{align}
Notice that the first term inside the boundary integral vanishes, because both $\omega$ and $\delta \Phi \vert_\gamma$ are holomorphic. The second term in the boundary integral came from an integration by parts in the bulk integral. Having separated the bulk and boundary contributions to the variation in \eqref{SD2}, both should vanish separately. The SD equation in the bulk becomes:
\beq
 \left \langle - \del \overline{\del} \Phi  +\overline{\del}\mathcal{J} + \mu_\Phi \right \rangle_{\mu_\Phi,\,\mu_\mathcal{J}} = 0~,
\eeq
since the bulk variation $\delta \Phi$ was arbitrary. This just gives the quantum version of the classical e.o.m., $\mathcal{J} = \del\Phi$. By writing $\Phi$ and $\mathcal{J}$ as functional derivatives of the partition function, we can obtain the bulk SD equation in arbitrary $n$-point functions. The more interesting condition is the SD equation for the boundary term:
\beq\label{SDeqn}
\left \langle \oint_\gamma \fr{dz}{2\pi i} \left(\frac{\lambda}{2} \frac{\mathcal{J}^2(z)}{\omega(z)} - \mathcal{J}(z)\right)\delta\Phi (z)\right\rangle_{\!\mu_\Phi,\,\mu_\mathcal{J}} = 0~.
\eeq
At the boundary, the variation $\delta\Phi(z)\vert_\gamma$ is an arbitrary odd and analytic function. This means that in the Laurent expansion of the integrand all the terms with even negative powers of $z$ should vanish.
The projection to the even negative powers of $z$ is done precisely with the free two-point function: 
\beq
\mathsf{G}(z_0,z) = \fr{1}{z_0-z} - \fr{1}{z_0+z} = \big\langle\mathcal{J}(z_0) \Phi(z)\big\rangle_0^{\mathsf{c}}~.
\eeq
Using this projection, the SD equation \eqref{SDeqn} becomes
\beq\label{SD3}
\half \oint_\gamma \fr{dz}{2\pi i}\mathsf{G}(z_0,z) \left \langle \frac{\lambda}{2} \frac{\mathcal{J}^2(z)}{\omega(z)} - \mathcal{J}(z) \right\rangle_{\!\mu_\Phi,\,\mu_\mathcal{J}} = 0~.
\eeq
The second term is just the even and singular part of $\mathcal{J}(z)$. So the requirement that $\Phi$ is odd (which followed from the parity of $\omega$) automatically allows us to treat $\mathcal{J}(z)$ as an even function, and hence $\mathcal{J} = \mathcal{J}(z)dz$ is odd in $z$. This was already manifest in our on-shell identification $\mathcal{J}(z) = \del\Phi(z)$, but now we see that also in the quantum theory the structure of the SD equation gives $\mathcal{J}(z)$ the right properties of a twisted boson on the spectral curve. 

We can now use the source field $\mu_\mathcal{J}$ to write $\mathcal{J}(z)$ as a functional derivative of the KS partition function. We turn off the source for $\Phi$, since it has disappeared from the boundary SD equation. In terms of the free energy
\beq
W_\mathsf{KS}[\mu_\mathcal{J}] = \log Z_\mathsf{KS}[ \mu_\mathcal{J}]~,
\eeq
the SD equation \eqref{SD3} becomes the following functional differential equation:
\beq\label{SD6}
\fr{\delta W_\mathsf{KS}}{\delta \mu_\mathcal{J}(z_0)} \Big\vert_{\chi<0}=\fr{\lambda}{4} \oint_\gamma \fr{dz}{2\pi i} \,\fr{\mathsf{G}(z_0,z)}{\omega(z)}\left[ \fr{\delta^2 W_\mathsf{KS}}{\delta \mu_\mathcal{J}(z)\delta\mu_\mathcal{J}(z)} + \fr{\delta W_\mathsf{KS}}{\delta \mu_\mathcal{J}(z)} \fr{\delta W_\mathsf{KS}}{\delta \mu_\mathcal{J}(z)}  \right]~.
\eeq
The fact that we should pick the negative powers of $z$ in $\mathcal{J}$ has been denoted by $\chi<0$. Furthermore, there is an explicit normal ordering prescription through the point-splitting regularization for $T = \half \mathcal{J}^2$.  

In the next section, we will show that the SD equation \eqref{SD6} is equivalent to the topological recursion relation for JT gravity. The recursion relation is supplemented with initial input, given by the free one- and two-point functions \eqref{free2points} derived in the previous section:
\beq\label{eq:onetwo}
\braket{\mathcal{J}(z)}_0^{\mathsf{c}} = \omega(z)~, \quad \braket{\mathcal{J}(z)\mathcal{J}(w)}_0^{\mathsf{c}} = \mathsf{B}(z,w)~.
\eeq 
We will show that these input data also agree with those of JT gravity, namely the disk and annulus contributions.

\section{Connection to JT gravity}\label{JTgravity}

The observables in the KS theory that are relevant for JT gravity are defined as the inverse Laplace transform of $\mathcal{J}$:
\begin{equation} \label{eq:KSJTdic}
Z(\beta) \equiv \frac{1}{2\pi i}\int_{c-i\infty}^{c+i\infty}dz \, \mathcal{J}(z)\,e^{\beta z^2}~.
\end{equation}
The integration contour is along the interval $(-i\infty,i\infty)$ which is shifted slightly to the right by a small parameter $c>0$ to avoid possible poles of $\mathcal{J}$ at the imaginary axis\footnote{In the case that $\mathcal{J}$ is regular at the origin, it need not be the Laplace transform of some function. The definition \eqref{eq:KSJTdic} should then be understood in a distributional sense.}. From the gravity perspective the observables in (\ref{eq:KSJTdic}) should be thought of as creating an asymptotic boundary in spacetime (or string world-sheet) of renormalized length $\beta$. In this section, we will argue that the spacetime wormhole contributions to the JT gravity path integral will be given by \emph{connected} correlation functions of these observables in the KS theory:
\begin{equation} \label{eq:npointfunction}
\mathcal{Z}^{\mathsf{c}}_\mathrm{JT}(\beta_1,\dots,\beta_n)=\langle Z(\beta_1)\cdots Z(\beta_n) \rangle_{\mathsf{KS}}^{\mathsf{c}}~.
\end{equation}
These correlation functions can be expanded in the coupling constant $\lambda$ of the KS theory by matching $\lambda=e^{-S_0}$. The full genus expansion of the JT partition function now follows from the perturbative expansion of the KS path integral. In analogy with string field theory, each term in this expansion corresponds to a world-sheet with a fixed topology.

\subsection{Lightning review of the JT path integral}
To be somewhat self-contained, let us quickly review the gravitational path integral for JT gravity \cite{saad2019jt}. Consider Euclidean JT gravity whose action on a surface $M$ is:
\begin{equation}\label{JTaction}
S_\mathrm{JT}[g,\phi] = -S_0\,  \chi(M) - \frac{1}{2} \int_M d^2x \sqrt{g}\,  \phi (\mathcal{R}+2) - \int_{\partial M}du \sqrt{\gamma_{uu}} \, \phi (\mathcal{K}-1)~.
\end{equation}
We choose Dirichlet-Dirichlet (DD) boundary conditions, where we fix both the metric $g$ and the dilaton $\phi$ near the boundary:
\begin{equation} \label{eq:bdycond}
	g_{\partial M}=\frac{1}{\epsilon^2}~, \qquad \phi_{\partial M}=\frac{\phi_{r}}{\epsilon}~, 
\end{equation}
where $\epsilon$ is a holographic renormalization parameter, defining a cut-off surface in the bulk which approaches the boundary in the limit $\epsilon \to 0$. The boundary conditions are such that near each of the $n$ boundary components, spacetime is asymptotically Euclidean $\mathrm{AdS}_2$. The length of the thermal circle is taken to be $\beta/\epsilon$ and the boundary conditions (\ref{eq:bdycond}) give rise to a graviton mode that parametrizes the `boundary wiggles' of the cut-off surface \cite{Maldacena2016}.  
\par
\begin{figure}

\centering
   \begin{subfigure}[b]{0.24\textwidth}
       \includegraphics[width=\textwidth]{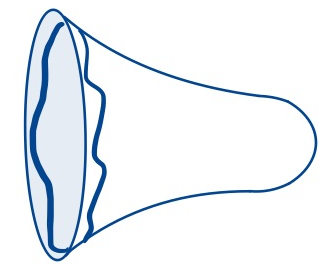}
        \caption{}
       \label{fig:table1}
   \end{subfigure} 
   \hspace{0.1\textwidth}
   \begin{subfigure}[b]{0.36\textwidth}
       \includegraphics[width=\textwidth]{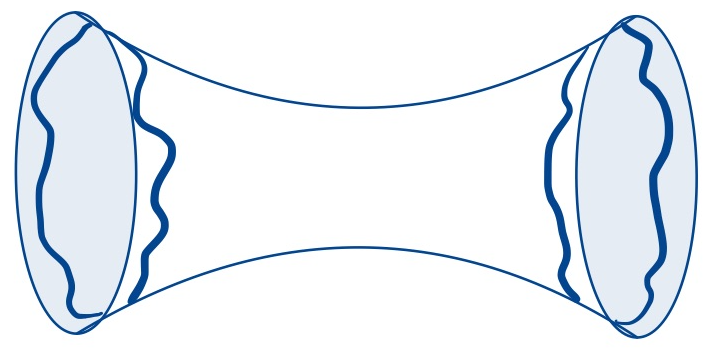}
       \caption{}
       \label{fig:table2}
    \end{subfigure}
    \caption{The disk (a) and annulus (b) geometries in JT gravity. The blue line represents the `wiggles' due to the boundary dynamics of the Schwarzian theory.}\label{fig:diskann}
\end{figure}

 The gravitational path integral is then computed as follows. The $\phi$-integral just gives a delta-functional which localizes the metric to be hyperbolic. The perturbative expansion is organized in terms of the Euler characteristic $\chi(M)=2-2g-n$, where $M$ is Riemann surface of genus $g$ with $n$ boundaries. Since the curvature is constant $\mathcal{R}=-2$, the bulk action $\int_{M} \phi (\mathcal{R}+2)$ vanishes, and the only bulk term left is the topological term $S_0\chi(M)$. Since this term is independent of the metric and dilaton, it becomes a multiplicative constant $e^{S_0\chi}$ in the path integral, which suppresses topologies with higher Euler characteristic in powers of the coupling constant $\lambda=e^{-S_0}$:
\beq\label{asymptoticexpansion}
\mathcal{Z}^{\mathsf{c}}_\mathrm{JT}(\beta_1,\dots,\beta_n) =\sum_{g=0}^\infty \lambda^{2g-2+n}Z^{\mathsf{c}}_{g,n}(\beta_1,\dots,\beta_n)~.
\eeq
Having integrated out the $\phi$-field, there is a residual path integral over the space of metrics on a surface of fixed topology and an integral for the boundary degrees of freedom. After careful analysis of the integration measure \cite{saad2019jt} the genus $g$ contribution for $\chi<0$ to the JT path integral becomes:
\beq\label{highergenus}
Z_{g,n}^{\mathsf{c}}(\beta_1,\dots,\beta_n) = \int_0^\infty \prod_{i=1}^n d\ell_i  \ell_i \, Z_{\mathsf{trumpet}}(\beta_i,\ell_i)\,V_{g,n}(\ell_1,\dots,\ell_n)~.
\eeq
Here, the residual path integral is written in terms of a bulk contribution $V_{g,n}(\bm{\ell})$, which can be viewed as a phase space volume of the bulk quantum fields, and an integration kernel
\beq \label{eq:trumpettt}
Z_{\mathsf{trumpet}}(\beta,\ell)=  \fr{1}{\sqrt{4\pi \beta}} e^{-\ell^2/(4\beta)}~,
\eeq
coming from the path integral on a trumpet geometry with one geodesic boundary and one asymptotically AdS boundary. The only exceptions to \eqref{highergenus} are the disk, $(g,n) =(0,1)$, and the annulus, $(g,n) = (0,2)$, shown in Figure \ref{fig:diskann}, which are computed separately to be:
\beq \label{eq:diskannulus}
Z_{0,1}(\beta) = \fr{1}{4\pi^{1/2}\beta^{3/2}} e^{\pi^2/\beta}~, \quad Z^{\mathsf{c}}_{0,2}(\beta_1,\beta_2) = \fr{1}{2\pi} \fr{\sqrt{\beta_1\beta_2}}{\beta_1+\beta_2}~.
\eeq
These partition functions can be derived using the one-loop exactness property of the Schwarzian theory \cite{fermioniclocalization}. We have taken the convention $\phi_r=1/2$.

\subsection{Matching KS theory with JT gravity}\label{sec:matchingtop}
First, we show that the disk and annulus partition functions are obtained using the inverse Laplace transform \eqref{eq:KSJTdic} of the free one- and two-point functions \eqref{eq:onetwo}. After that, we match the higher genus contributions for an arbitrary number of boundaries. 

\paragraph{The disk.} To obtain the one-point function of $Z(\beta)$ in the KS theory we need to compute the following inverse Laplace transform:
\begin{equation}
\langle Z(\beta) \rangle^{\mathsf{c}}_{\mathsf{KS},0} =\int_{c-i\infty}^{c+i\infty}\frac{dz}{2\pi i}\, \braket{\mathcal{J}(z)}_0\,e^{\beta z^2} = \int_{c-i\infty}^{c+i\infty}\frac{dz}{2\pi i}\, \omega(z)\,e^{\beta z^2}~.
\end{equation}
We see that the KS theory determines the leading term in the genus expansion from the holomorphic one-form, which encodes the complex structure of the spectral curve: 
\beq
\omega(z)dz = y(z)dx(z) = \fr{1}{2\pi} z\sin(2\pi z)dz~.
\eeq 
Since $\omega(z)$ is regular at $z=0$, we can set $c=0$ and substitute $z \to -iw$, giving:
\beq
\langle Z(\beta) \rangle^{\mathsf{c}}_{\mathsf{KS},0}=\frac{1}{4\pi^2}\int_{-\infty}^{\infty} dw\, w \sinh(2\pi w)\, e^{-\beta w^2} = \fr{1}{4\pi^{1/2} \beta^{3/2}} e^{\pi^2/\beta}~.
\eeq 
This expression agrees with the path integral of the disk in JT gravity.

\paragraph{The annulus.} Next, we compute the free two-point function of $Z(\beta)$ in KS theory and match it to the annulus amplitude \eqref{eq:diskannulus}. For that we need to compute the inverse Laplace transform of the free bosonic two-point function:
\beq
\langle Z(\beta_1)Z(\beta_2)\rangle^{\mathsf{c}}_{\mathsf{KS},0}= \int_{c-i\infty}^{c+i\infty}\fr{dz}{2\pi i}\fr{dw}{2\pi i} \braket{\mathcal{J}(z)\mathcal{J}(w)}^{\mathsf{c}}_0 e^{\beta_1 z^2 + \beta_2 w^2}~,
\eeq
where
\beq\label{eq:2point}
\braket{\mathcal{J}(z)\mathcal{J}(w)}^{\mathsf{c}}_0 = \mathsf{B}(z,w) = \fr{1}{(z-w)^2} + \fr{1}{(z+w)^2}~.
\eeq
Again, we may rotate the contours to the real axis, and then note that both terms in \eqref{eq:2point} give the same contribution upon sending $w\to -w$ in the second integral. Expanding $(z-w)^{-2}$ as a power series, and using gamma functions to compute the resulting Gaussian moments, it can be shown that the double integral gives:
\beq\label{doubletrumpetintegral}
\langle Z(\beta_1)Z(\beta_2)\rangle^{\mathsf{c}}_{\mathsf{KS},0}= -\fr{1}{4\pi^2}\int_{-\infty}^\infty dz dw \,\fr{e^{-\beta_1 z^2 - \beta_2 w^2}}{(z-w)^2} = \fr{1}{2\pi} \fr{\sqrt{\beta_1\beta_2}}{\beta_1+\beta_2}~.
\eeq
This matches the Euclidean wormhole contribution in JT gravity. 

\paragraph{Higher genus amplitudes.} The higher genus corrections in JT gravity are computed recursively, either using Mirzakhani's recursion for the Weil-Petersson volumes $V_{g,n}(\bm{\ell})$, or, after a Laplace transform, using Eynard and Orantin's topological recursion. In the KS theory, we can also compute higher-order corrections to connected correlation functions of $\mathcal{J}(z)$ using the SD equation \eqref{SD6}. We show that the topological recursion is retrieved as the perturbative expansion of the SD equation.   

Let us denote the connected correlation functions by
 \beq
 \mathcal{W}_n(z_1,\dots,z_n) \equiv \braket{\mathcal{J}(z_1) \cdots \mathcal{J}(z_n)}^{\mathsf{c}}_\mathsf{KS} = \fr{\delta^n W_{\mathsf{KS}}}{\delta \mu_\mathcal{J}(z_1)\cdots \delta \mu_\mathcal{J}(z_n)} \Big \vert_{\mu_\mathcal{J} =0}~.
 \eeq
We can expand the free energy $W_{\mathsf{KS}}[\mu_\mathcal{J}]$ of the KS theory in terms of  connected correlation functions as:
\beq 
W_\mathsf{KS}[\mu_\mathcal{J}] = \sum_{n=0}^\infty \int \prod_{i=1}^ndz_i \fr{\mu_\mathcal{J}(z_1)\cdots \mu_\mathcal{J}(z_1)}{n!} \mathcal{W}_n(z_1,\dots,z_n)~.
\eeq 
Plugging this expansion in to the SD equation \eqref{SD6} and comparing powers of $\mu_\mathcal{J}$, the SD equation takes the form:
 \beq\label{SD7}
 \mathcal{W}_{n+1}(z_0,z_I) = \frac{\lambda}{4} \oint_\gamma \fr{dz}{2\pi i} \,\fr{\braket{\mathcal{J}(z_0)\Phi(z)}_0}{\omega(z)} \Big[ \mathcal{W}_{n+2}(z,z,z_I) + \sum_{J_1\sqcup J_2 = I}\mathcal{W}_{1+|J_1|}(z,z_{J_1}) \mathcal{W}_{1+|J_2|}(z,z_{J_2})\Big]~.
 \eeq
 The sum in (\ref{SD7}) is over subsets $J_1\sqcup J_2 = I=\{1,\ldots,n\}$, and the multi-index notation is given by $z_{J}\equiv (z_{j})_{j\in J}$. Next, consider the perturbative expansion in powers of the KS coupling constant $\lambda$:
\beq
\mathcal{W}_n(z_1,\dots,z_n) = \sum_{g=0}^\infty \lambda^{2g-2+n} \mathcal{W}_{g,n}(z_1,\dots,z_n)~.
\eeq 
Substituting this into (\ref{SD7}) and matching the terms with the same powers of $\lambda$ we obtain a system of recursive equations:
\begin{align}
\mathcal{W}_{g,n+1}(z_0,z_I)= \underset{z\to 0}{\text{Res}}\,\,&\mathcal{K}(z_0,z)\Big[ \mathcal{W}_{g-1,n+2}(z,z,z_I) \nonumber \\ 
&+\sum_{h=0}^g\sum'_{J_1\sqcup J_2 = I}\mathcal{W}_{h,1+|J_1|}(z,z_{J_1}) \mathcal{W}_{g-h,1+|J_2|}(z,z_{J_2})\Big]~. \label{eq:toprec}
\end{align}
Here, we have defined the recursion kernel $\mathcal{K}(z_0,z)$ in terms of the twisted propagator as:
\begin{equation}
\mathcal{K}(z_0,z)\equiv\fr{\braket{\mathcal{J}(z_0)\Phi(z)}_0}{4\,\omega(z)}= \frac{1}{2} \left(\frac{1}{z_0-z}-\frac{1}{z_0+z}\right)\frac{1}{2\omega(z)}~,
\end{equation}
where $\omega(z)=\fr{1}{2\pi} z\sin(2\pi z)$. The prime indicates that terms involving $(g,n) = (0,1)$ should be excluded from the summation. We have replaced the contour integral around the branch point by a residue at $z=0$. The recursion relation is therefore determined by the pole structure of the correlation functions in the complex plane. 

Importantly, the recursion in (\ref{eq:toprec}) corresponds precisely to the topological recursion relations applied to the spectral curve $\mathcal{S}_{\mathrm{JT}}$, with input data\footnote{This can either be verified using direct computation, or the generic argument presented in \eqref{oddfirstarg}.}: 
\beq 
\mathcal{W}_{0,1}(z)=\omega(z)~,\hspace{10pt} \mathcal{W}_{0,2}(z_1,z_2)=\mathsf{B}(z_1,z_2)~.
\eeq 
The relevant background for the formalism of topological recursion is summarized in appendices \ref{toporecursion} and \ref{ch:recursion}. We can combine this result with Eynard and Orantin's observation \cite{eynard2} that the Weil-Petersson volumes $V_{g,n}$ are related to the `symplectic invariants' $\mathcal{W}_{g,n}$ by a Laplace transform:
\begin{equation} \label{eq:Wgn}
\mathcal{W}_{g,n}(z_1,\dots,z_n)= \int_0^\infty \prod_{i=1}^n d\ell_i \,\ell_i \,e^{-z_i \ell_i} V_{g,n}(\ell_1,\dots,\ell_n)~, \quad (\chi<0)~.
\end{equation}
Using our proposal (\ref{eq:KSJTdic}) for relating JT to KS, we go to the $\beta$-variable by applying the inverse Laplace transform for each $z_i$:
\begin{align}
\braket{Z(\beta_1)\cdots Z(\beta_n)}^{\mathsf{c},(g)}_\mathsf{KS} &= \int_{c-i\infty}^{c+i\infty}\prod_{i=1}^n \frac{dz_i}{2\pi i}\, e^{\beta_i z_i^2}\mathcal{W}_{g,n}(z_1,\dots,z_n) \\
&= \int_0^\infty \prod_{i=1}^n d\ell_i \,\ell_i \,V_{g,n}(\ell_1,\dots,\ell_n)\,\int_{-\infty}^{\infty}\prod_{i=1}^n \fr{dw_i}{2\pi}e^{-\beta_iw_i^2-i\ell_iw_i} \\
&= \int_0^\infty \prod_{i=1}^n d\ell_i  \ell_i \, V_{g,n}(\ell_1,\dots,\ell_n)Z_{\mathsf{trumpet}}(\beta_i,\ell_i)~.
\end{align}
This indeed agrees with the Euclidean path integral $Z_{g,n}^{\mathsf{c}}(\beta_1,\dots,\beta_n)$ in \eqref{highergenus} for the stable surfaces with $\chi<0$. Multiplying by $\lambda^{2g-2+n}$ and summing over the genus we conclude that the perturbative expansion of the universe field theory matches with the genus expansion of the gravitational path integral. Since all correlation functions can be expressed in terms of connected correlations functions, we see that the full JT gravity $n$-boundary path integral is the $n$-point function of the boundary creation operators $Z(\beta)$:
 \beq\label{KS-JT}
 \mathcal{Z}_\mathrm{JT}(\beta_1,\dots,\beta_n) =  \braket{Z(\beta_1) \cdots Z(\beta_n)}_{\mathsf{KS}}~.
 \eeq
Let us emphasize that the right-hand side is a \emph{non-gravitational} Euclidean path integral with $n$ operator insertions, whereas the left-hand side is the \emph{gravitational} path integral of JT gravity. We have thus expressed JT gravity as a Euclidean `universe field theory' on the spectral curve. This provides a non-perturbative completion (at least on a formal level) of the topological expansion of the Euclidean JT path integral. 

Given the geometric interpretation of the chiral boson $\mathcal{J}(z)$ as describing the quantum fluctuations of the target space geometry around the classical value $\omega(z)$, we can think about the `ensemble average' $\braket{\cdots}_{\mathsf{KS}}$ associated to JT gravity, roughly speaking, as describing an average over background geometries in which the JT string propagates.

\section{Non-perturbative effects} \label{boundaryconditions}
Given the KS theory description for JT gravity and its embedding in the B-model topological string theory, we can invoke intuition and tools from string theory to study non-perturbative effects of order $\mathcal{O}(e^{1/\lambda})$ in the universe field theory. We will study the insertion of certain topological D-branes \cite{vafadijkgraaf1, Dijkgraaf:2002fc, integrablehierarchies, universalcorrelators} in the target space geometry. Their effect will be doubly non-perturbative in $G_N$, as can be seen from the identification $\lambda = e^{-S_0}$. These contributions are very interesting from the point of view of gravity, as they form an indirect probe of the discreteness of the spectrum in a candidate microscopic theory. 

In section \ref{sec:branesKS}, we will study non-perturbative effects due to the D-branes. These correspond to fermionic objects in the KS theory. In section \ref{sec:branesJT}, we will give these D-branes an interpretation in JT gravity as hypersurfaces on which fixed energy boundaries can end. These boundaries are described by a boundary term in the JT gravity action that can be obtained from the standard Dirichlet-type boundary action by a Legendre transform, which on the level of the path integral becomes a Laplace transform. As an application of the D-brane formalism, we will show in section \ref{sec:application} how to obtain non-perturbative corrections to the density-density correlator, giving the `plateau' feature of the spectral form factor \cite{SFF2}.    

\subsection{Branes in KS theory}\label{sec:branesKS}

Since the basic KS field is a 2-dimensional chiral boson, we can use the familiar boson-fermion correspondence and introduce the following fermionic fields\footnote{By $\Phi(z)$ we mean the full chiral boson with both positive and negative modes. We treat it as the indefinite integral of $\mathcal{J}(z)$. For an account of the boson-fermion correspondence for twisted fields, see \cite{anguelova,mattiello}.}:  
\beq 
\psi(z) = e^{\Phi(z)}~, \quad \quad \psi^\dagger(z) = e^{-\Phi(z)}~.
\eeq 
The exponentials are normal-ordered by subtracting the OPE singularities of $\Phi(z)\Phi(w) \sim \log(z-w)$ in the expansion of the exponential. These fermionic fields have an interpretation as D-branes in the KS theory. Namely, recall that the spectral curve is embedded in the following non-compact Calabi-Yau:
\beq
\mathsf{CY}: \quad uv - y^2 + \fr{1}{(4\pi)^2}\sin^2(2\pi \sqrt{x}) = 0~.
\eeq
The base of this Calabi-Yau is the spectral curve $\mathcal{S}_{\mathrm{JT}}$. This is where the bosonic fields $\mathcal{J}$ and $\Phi$ live. The fibers over the spectral curve are defined by $u = 0$ and $v=0$: if we specify a base point on $\mathcal{S}_{\mathrm{JT}}$ these are complex one-dimensional manifolds in the geometry, which can be wrapped by topological D2-branes. In the topological string terminology, $u=0$ is wrapped by a brane, while a brane that wraps the transverse fiber $v=0$ has opposite flux and can be thought of as an `anti-brane' \cite{vafabraneantibrane}. The fibers are parametrized by a point on the spectral curve, so we can talk about a brane `inserted' at a point $\zeta \in \mathcal{S}_{\mathrm{JT}}$. 

In the topological string B-model, integrating out open strings ending on the brane deforms the geometry in which the closed strings propagate \cite{BCOV}. This can be thought of as the backreaction of a brane on the geometry, which deforms the complex structure of $\mathsf{CY}$. As before, the change in complex structure is encoded in the period integral of the holomorphic $(3,0)$-form $\Omega_\mathsf{CY}$ around a 3-cycle $\widetilde{C}$ surrounding the D-brane. The change in complex structure due to a single D-brane is found to be \cite{topologicalvertex}:
\beq\label{branedeformation}
\delta \int_{\widetilde{C}} \Omega_\mathsf{CY} = \lambda~.
\eeq
We can follow the same steps as in \eqref{periodintegral} to reduce the period integral on the complex 3-cycle surrounding the 2-dimensional brane to a period integral of $\omega$ on a contour $C(\zeta)$ surrounding the point $\zeta$. The result is simply:
\beq\label{branedeformation2}
\delta \oint_{C(\zeta)} \omega = \lambda~.
\eeq
What this equation is saying is that the insertion of a brane above the point $\zeta$ deforms the complex structure of the spectral curve $\mathcal{S}_{\mathrm{JT}}$ by a small amount $\lambda$. This is implemented in the quantum theory by a field $\psi(\zeta)$. Rescaling $\omega \to \omega / \lambda$, the property \eqref{branedeformation2} can be written as an operator product, to be read inside correlation functions:
\beq
 \oint_{C(\zeta)} dz \,\del\Phi(z) \,\psi(\zeta) = \psi(\zeta)~.
\eeq
We used the defining relation $\delta \omega = d\Phi$ for the chiral boson $\Phi$. We now recognize the operator product expansion of a complex fermion of conformal weight $h=\half$ with the holomorphic bosonic current $\mathcal{J}(z) = \del\Phi(z)$:
\beq
\del\Phi(z) \psi(\zeta) \sim \fr{1}{z-\zeta} \psi(\zeta)~.
\eeq
Conversely, we can obtain $\mathcal{J}(z)$ by taking the coincident limit of a brane and an anti-brane:
\beq\label{bosonfermioncorrespondence}
\del\Phi(z)= \lim_{z'\to z}\Big\{\psi(z') \psi^\dagger(z)\Big\}~.
\eeq
The accolades signify normal ordering, by subtracting the OPE divergence $\sim \fr{1}{z'-z}$. We conclude that the non-compact D-branes described above are indeed nothing but complex fermions on the spectral curve\footnote{This is true locally. As explained in \cite{integrablehierarchies}, for an arbitrary spectral curve $\psi(z)$ is only defined patch by patch and transforms as a wavefunction, instead of as a fermionic `half-differential' $\psi = \psi(z) \sqrt{dz}$. The wavefunction interpretation of $\psi(z)$ appears naturally from a Schr\"odinger equation satisfied by $\braket{\psi(z)}_\mathsf{KS}$, see, e.g., \cite{Gukov}.}.  

\begin{figure}
    \centering
    \includegraphics[width=7cm]{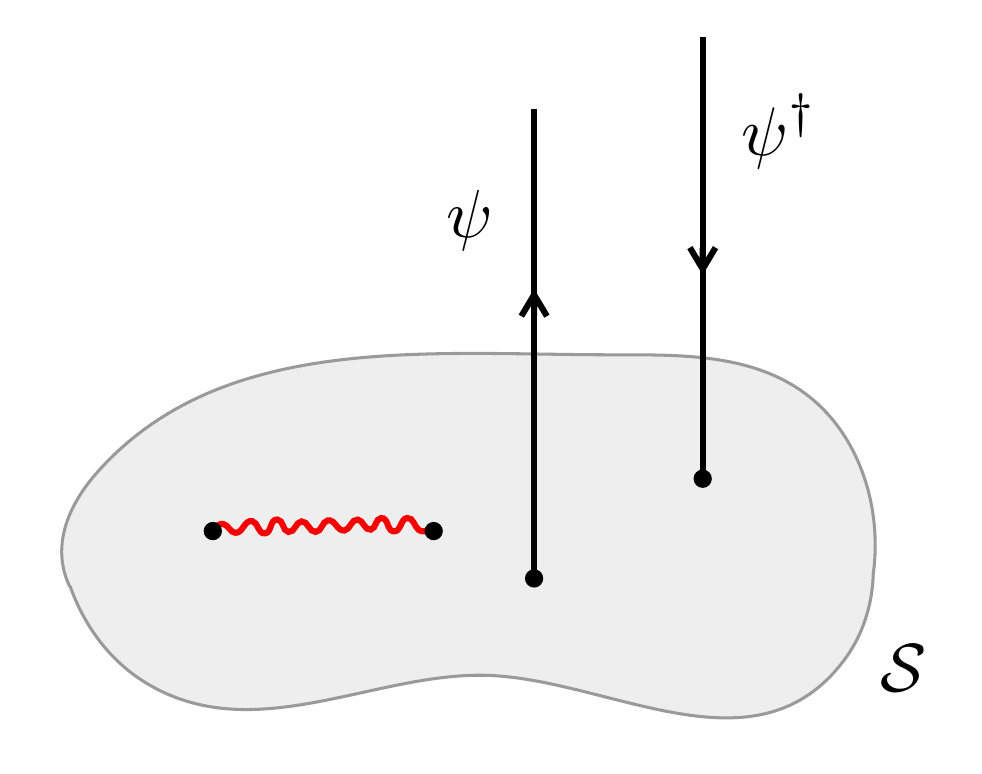}
    \caption{A pictorial representation of non-compact D-brane insertions on the spectral curve $\mathcal{S}$. The straight lines correspond to the non-compact fiber directions $u=0$ or $v=0$ which are wrapped by branes $\psi$ and anti-branes $\psi^{\dagger}$ respectively, having opposite flux as indicated by the direction of the arrow. The branch cut is denoted by a red wiggly line.}
    \label{fig:braneantibrane1}
\end{figure}

So far, we have defined the fermions on the double cover, which is also how we have presented the KS theory. However, ultimately we will be interested in extracting non-perturbative information from the fermions to correlation functions of the density of states $\rho(E)$, where $E$ is related to the base space coordinate $x = -E$. In particular, these quantities will be sensitive, at least semiclassically \cite{Maldacena:2004sn}, to the branched structure of the spectral curve. Therefore, we also want to define fermionic fields $\psi(x)$ on the spectral plane $x = z^2$. However, this requires us to choose a branch of $z = \sqrt{x}$. We therefore use the formalism developed in \cite{fukuma} to describe $\bb{Z}_2$-twisted fermions on sheeted Riemann surfaces. To connect to the $\bb{Z}_2$-twisted boson formalism outlined in appendix \ref{ch:recursion}, we will use the spectral plane variable $x$, and obtain physical quantities like $\rho(E)$ by evaluating at $x=-E$ in the end.   

\subsubsection{$\bb{Z}_2$-twisted fermions}\label{twistfermions}
We think of the spectral curve as two copies of the spectral $x$-plane, glued together along the branch cut on the negative real axis. On each sheet, labelled by indices $0,1$, we define a bosonic field, such that after a $2\pi i$ rotation the fields are rotated into each other:
\beq
\Phi_0(e^{2\pi i} x) = \Phi_1(x)~, \quad \Phi_1(e^{2\pi i} x) = \Phi_0(x)~.
\eeq
In appendix \ref{twistedfermioncomp}, we introduce explicit mode expansions for $\Phi_0$ and $\Phi_1$ and show that the KS field $\Phi(x)$ is the combination that diagonalizes the monodromy:
\beq\label{ksfield}
\Phi(x) = \fr{1}{\sqrt{2}} (\Phi_0(x) - \Phi_1(x))~, \quad x \in \bb{C} \setminus \mathbb{R}_{\leq 0}~.
\eeq
Then, $\Phi(e^{2\pi i} x) = -\Phi(x)$, so $\Phi$ is indeed an odd function of $z$. In terms of $x$, it has an expansion in only half-integer powers of $x$. So in particular it has a discontinuity across the branch cut, which we will call $\Omega(x)$. The discontinuity can be expressed alternatively in terms of the fields on opposite sheets as they approach each other on the negative real axis:
 \beq
\Omega(x) \equiv \lim_{x' \to x}(\Phi_0(x) - \Phi_1(x'))~, \quad x \in \mathbb{R}_{\leq 0}~.
\eeq   
Next, on each sheet we introduce the following bosonized fermions:
\beq
\psi_a(x)= \mathsf{c}_a e^{\Phi_a(x)}~,\quad \psi_a^\dagger(x) =  \mathsf{c}_a e^{-\Phi_a(x)} ~, \quad a = 0,1~.
\eeq
Again, the exponentials are implicitly normal ordered by subtracting the divergences. Furthermore, we have used what is known as the Jordan-Wigner trick to multiply the vertex operators by a cocycle $\mathsf{c}_a$ that ensures the correct anti-commutation between fermions on opposite sheets \cite{kostelecky}. A consistent choice of cocycles in this case is simply:
\beq
\mathsf{c}_0 = 1~, \quad \mathsf{c}_1 = (-1)^{N_f+1}~,
\eeq
where $N_f$ is the fermion number operator\footnote{It can be bosonized as $N_f = \oint \fr{dz}{2\pi i} \del\Phi_a(z) =\alpha_0$, which, being the coefficient of $\fr{1}{z}$, can be seen as the momentum of $\del\Phi_a$.}. This ensures that fermions on opposite sheets anti-commute, for example:
\begin{align}
\psi^\dagger_0(x)\psi_1(x') &= e^{-\Phi_0(x)} (-1)^{N_f+1} e^{\Phi_1(x')} = - \psi_1(x')\psi^\dagger_0(x)~.
\end{align}
For fields on the same sheet, the cocycles square to one and the anti-commutation is ensured by the OPE:
\beq
\psi_a(x)\psi_b^\dagger(x') \sim \fr{\delta_{ab}}{x-x'} + \mathrm{reg.}
\eeq
This is of course the expected OPE for fermion fields. As before, we have a boson-fermion correspondence for fermions on the same sheet:
\beq\label{samesheet}
\del\Phi_a(x) = \lim_{x'\to x} \big\{\psi_a(x') \psi_a^\dagger(x)\big\} \equiv  \lim_{x'\to x} \left(\psi_a^\dagger(x') \psi_a(x) - \fr{1}{x'-x}\right)~.
\eeq
On the other hand, for two fermions on opposite sheets, we do not have to normal order since $\Phi_0(x')\Phi_1(x)$ is regular, and we can simply add the exponentials in a single normal-ordered exponential:
\beq
\psi_0(x)\psi_1^\dagger(x) \equiv \lim_{x'\to x} \psi_0(x')\psi^{\dagger}_1(x) =  \mathsf{c}_0\mathsf{c}_1 \,e^{\Phi_0(x) - \Phi_1(x)}~.
\eeq
Usually for OPE's we implicitly demand the radial ordering $|x'|>|x|$. But here we should be careful about the ordering of the $x$-arguments when we take the coincident limit, since the points are on different sheets. We choose the convention that $\psi_0(x')$ is always left of $\psi_1(x)$ when we take the coincident limit. With this convention, the product of fields when they approach each other from opposite sheets gives the following weight one vertex operators:
\beq\label{oppositesheet}
e^{\Omega(x)} =\lim_{x' \to x} \psi_0(x') \psi_1^\dagger(x)~, \quad e^{-\Omega(x)} = \lim_{x' \to x}\psi_0^\dagger(x')\psi_1(x)~. 
\eeq
These operators will play an important role in the next section.
The fermions have the following monodromies when going around the branch point:
\beq
\bra{\sigma} \psi_0(e^{2\pi i} x) = - \bra{\sigma} \psi_1(x)~, \quad \bra{\sigma} \psi_1(e^{2\pi i}x) = -\bra{\sigma} \psi_0(x)~.
\eeq
We have multiplied from the left by the free bosonic twisted vacuum $\bra{\sigma}$, so that the cocycles $\mathsf{c}_0, \mathsf{c}_1$ become $\pm 1$, respectively. However, from now on we will leave the left-vacuum implicit. This is justified because, as we will see, to extract the non-perturbative physics we will not need the higher genus corrections from the interacting $\ket{\mathsf{KS}}$ vacuum; we will only need the free vacuum $\ket{\sigma}$ correlation functions. 

\subsection{Interpretation in JT gravity}\label{sec:branesJT}

To connect the discussion of branes on the spectral curve to JT gravity, we should introduce a type of boundary directly in the gravitational theory which can end on branes in the target space geometry. We will thus introduce a set of boundary conditions for the JT universes which do not have a fixed length, but are `hovering' in the bulk at some finite distance, and which have a fixed energy $E$. The observables in the matrix model can now be obtained from the JT path integral with this choice of modified boundary conditions, which on the level of the action amounts to a Legendre transform\footnote{This can be viewed as a particular instance of the more general result in the AdS/CFT correspondence \cite{Klebanov:1999tb}, namely that a Legendre transformation in the boundary field theory at large $N$ leads to a change in the boundary conditions for the fields on the gravity side.}. 
\subsubsection{Fixed energy boundary conditions}
Instead of fixing the dilaton and the boundary metric as in \eqref{eq:bdycond}, one can also impose Dirichlet-Neumann (DN) boundary conditions, in which one fixes both the dilaton and its normal derivative at the boundary, but leave the metric free \cite{boundaryconditions}:\beq \phi_{\partial M} = \fr{\phi_r}{\varepsilon}~, \hspace{20pt} \del_n \phi_{\del M} = \fr{\phi_r'}{\varepsilon}~.\eeq
Here, we have normalized the normal vector $n$, so that $\del_n \phi$ has the same dimensions as $\phi$. In this case, the following boundary action must be added to the bulk JT gravity action:
\beq \label{eq:DNboundaryaction}
S^\del_{DN} = -\int_{\del M} du \sqrt{\gamma_{uu}}(\del_n\phi - \phi \mcal{K})~.
\eeq
The DN boundary conditions are related to the standard DD boundary conditions by a Legendre transform. To see this, we rewrite \eqref{eq:DNboundaryaction} in the following form:
\begin{align}
S^\del_{DN} &= \fr{1}{\varepsilon} \int_{\del M} du\sqrt{\gamma_{uu}}(\phi_r - \phi_r') + \int_{\del M}du \sqrt{\gamma_{uu}}\,\phi (\mcal{K}-1)\nonumber \\
\label{legendre} &= \int_0^\beta du \sqrt{\gamma_{uu}} \, E + S^\del_{DD}[\gamma]~.
\end{align}
We have written explicitly a dependence on the boundary metric $\gamma_{uu}$ in the last term, because in the DN action $\gamma$ is kept free. We recognize the Legendre transform\footnote{Usually, the Legendre transform has a relative minus sign. The plus sign here means that on the level of the path integral, we will get an inverse Laplace transform.}, with conjugate variables $\beta$ and 
\beq\label{energy}
E \equiv \fr{ \phi_r - \phi_r'}{\varepsilon}~.
\eeq
One can show that \eqref{energy} corresponds to a fixed energy $E=\phi_r \Sch(x,u)$ in the boundary Schwarzian theory, when $\epsilon \to 0$. Therefore, the input of a particular Dirichlet type boundary in the JT path integral is some temperature $\beta$ describing a \emph{canonical} ensemble, whereas the input of a Neumann type boundary is some fixed energy $E$ describing a \emph{microcanonical} ensemble in the boundary theory.

On the level of the path integral, the Legendre transform becomes an inverse Laplace transform \cite{boundaryconditions}:
\beq \label{DN}
\mathcal{Z}_\mathrm{JT}(E) = \int_{c-i\infty}^{c+i\infty} \fr{\mcal{D}\gamma}{\mathrm{Diff} \,S^1} e^{\int_0^\beta du \sqrt{\gamma_{uu}} \, E } \mathcal{Z}_\mathrm{JT}[\gamma]~. 
\eeq
We will often omit the superscript DN, as it should be clear from using the variable $E$ that we mean the path integral with DN boundary conditions. We can go to a gauge where $\sqrt{\gamma_{uu}}$ is constant, and then we have to divide by $\beta$ to account for the time reparametrization symmetry:
\beq \label{inverselaplace}
\mathcal{Z}_\mathrm{JT}(E) =  \int_{c-i\infty}^{c+i\infty}\fr{d\beta}{\beta} e^{\beta E } \mathcal{Z}_\mathrm{JT}(\beta)~.
\eeq
For example, we can compute the DN partition function of the disk (with $\phi_r=\half$) to be
\beq\label{DNdisk}
Z_{\mathsf{disk}}(E) = \int_{c-i\infty}^{c+i\infty} \fr{d\beta}{\beta} e^{\beta E} \fr{e^{S_0}}{4\pi^{1/2}\beta^{3/2}} e^{\pi^2/\beta} = \fr{e^{S_0}}{8\pi^4}\Big(2\pi\sqrt{E}\cosh(2\pi \sqrt{E}) - \sinh(2\pi \sqrt{E})\Big)~.
\eeq
Furthermore, the trumpet partition function becomes
\beq\label{dntrumpet}
Z_{\mathsf{trumpet}}(E,\ell) = \int_{c-i\infty}^{c+i\infty} \fr{d\beta}{\beta} e^{\beta E} \fr{1}{\sqrt{4\pi\beta}} e^{-\fr{\ell^2}{4\beta}} = \fr{1}{\pi\ell}\sin(\ell\sqrt{E})~.
\eeq
Looking at the form of the higher-genus partition functions \eqref{highergenus}, we see that the $\beta$-dependence only comes in via the trumpets. So, the only modification to the perturbative formula of $\mathcal{Z}_\mathrm{JT}(E)$ will be to change the integration kernel of the trumpet to its DN counterpart:
\beq
Z_{g,n}(E_1,\dots,E_n) = \fr{1}{\pi} \int_0^\infty \prod_{i=1}^n d\ell_i  \sin(\ell_i\sqrt{E_i}) V_{g,n}(\ell_1,\dots,\ell_n)~.
\eeq
The $\ell_i$ from the gluing measure has cancelled with the $\ell_i$ in the denominator of \eqref{dntrumpet}. So $Z_{g,n}(E_1,\dots,E_n)$ is simply multiple Fourier-type transform of the Weil-Petersson volumes. We must be careful in evaluating these integrals, as the volumes $V_{g,n}$ are polynomials in $\ell_i^2$ and so the above integral in general is divergent. However, this divergence can be easily regularized, for example by introducing a small exponential regulator. 

\subsubsection{Relation to matrix integrals}

The fixed energy JT partition function $\mathcal{Z}_\mathrm{JT}(E)$ has a direct interpretation in the dual matrix model. To see this, one can write $\mathcal{Z}_\mathrm{JT}(\beta)$ as the difference of two integrals in spectral plane coordinate $x=z^2$ just above and below the negative real axis:
\beq
\mathcal{Z}_\mathrm{JT}(\beta) = - \lim_{\epsilon \to 0}\int_{-\infty}^0 \fr{dx}{2\pi i}  e^{\beta x} \Big[ \braket{\del\Phi(x+i\epsilon )}_{\mathsf{KS}} - \braket{\del\Phi(x-i\epsilon)}_{\mathsf{KS}} \Big]~.
\eeq
To obtain $\mathcal{Z}_\mathrm{JT}(E)$ from this expression, we use the following integral representation of the delta function:
\beq
\delta(E+x) = \int_{-i\infty}^{i\infty } d\beta\,  e^{\beta(E+x)}~.
\eeq
Sending $x\to -x$, the DN path integral can be expressed in terms of KS field insertions as:
\begin{align}
\mathcal{Z}_\mathrm{JT}(E) 
&=\lim_{\epsilon \to 0} \int^E dE' \int_{0}^\infty \fr{dx}{2\pi i} \delta(E'-x)\Big[ \braket{\del\Phi(-x+i\epsilon)}_{\mathsf{KS}} - \braket{\del\Phi(-x-i\epsilon)}_{\mathsf{KS}} \Big]~.
\end{align}
Since $E' \in \mathbb{R}_{\geq 0}$, the delta function sets $x = E'$. From now on we use the short-hand notation $\del\Phi(E)\equiv\del\Phi(x)|_{x=-E}$, when $\del\Phi$ is viewed as a function of $E$. This amounts to moving the branch cut to the positive real axis. On the right-hand side we then recognize the discontinuity of $\del\Phi$ across the branch cut:
\beq
\mathrm{disc}\, \del\Phi(E) \equiv \fr{1}{2\pi i} \lim_{\epsilon \to 0}\Big( \braket{\del\Phi(E + i\epsilon)}_{\mathsf{KS}} - \braket{\del\Phi(E - i\epsilon)}_{\mathsf{KS}}\Big)~.
\eeq
We find that $\del\Phi(E)$ plays the role of the \emph{resolvent} and its discontinuity across the branch cut is the \emph{density of states}. Therefore, we will match our notation with that from double-scaled matrix models and write:
\begin{align} \label{eq:dens}
\rho(E) &\equiv \mathrm{disc}\, \del\Phi(E)~, \qquad E\in \mathbb{R}_{\geq 0}~, \\ \label{eq:res}
R(E) &\equiv \del\Phi(E)~, \qquad E \in \mathbb{C} \setminus \mathbb{R}_{\geq 0}~.
\end{align}
The DN path integral can now be expressed as an insertion of the integrated density of states:
\beq \label{eq:integrateddensity}
Z(E) \equiv \int^E dE' \, \rho(E')~.
\eeq
These are the analogues of the boundary creation operators \eqref{eq:KSJTdic} in the case of DN boundary conditions. To be precise, we can obtain the JT path integral with DN boundary conditions by inserting these observables in the KS theory:
\beq 
\mathcal{Z}_\mathrm{JT}(E_1,\ldots, E_n) = \braket{Z(E_1)\cdots Z(E_n)}_{\mathsf{KS}}~.
\eeq
This completes the dictionary between JT gravity with DN boundary conditions and the KS theory. 
\begin{figure}[h]
    \centering
\vspace{10pt}
\begin{overpic}[width=9cm]{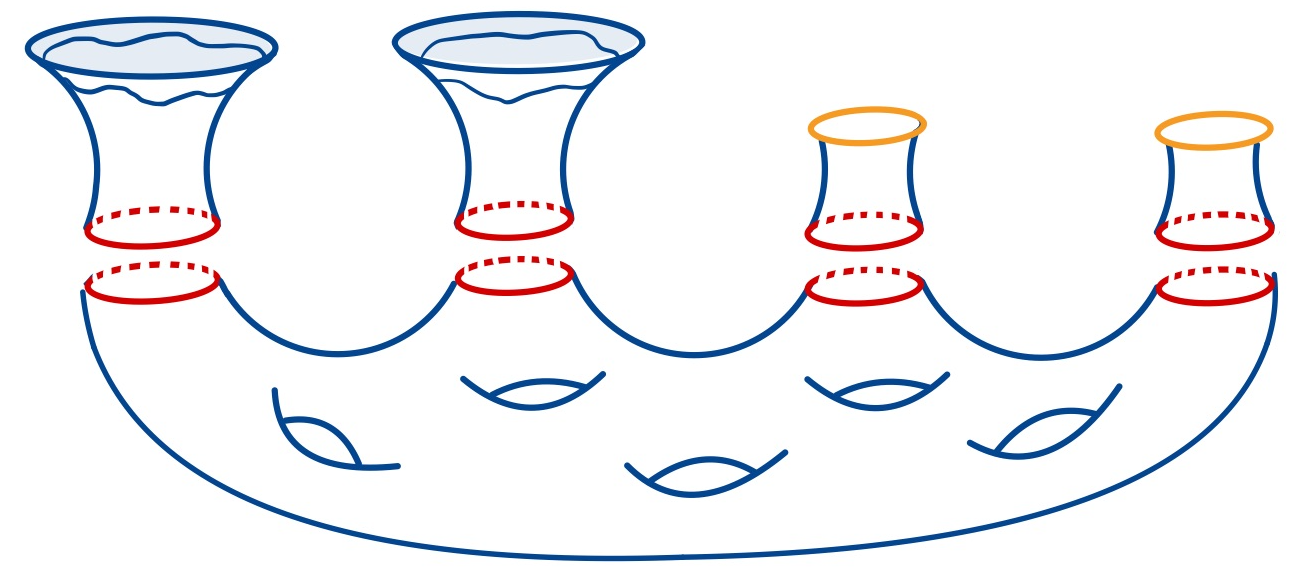}
 \put (33,50) {$Z(\beta_2)$} 
  \put (5,50) {$Z(\beta_1)$} 
   \put (58,43) {$Z(E_1)$} 
   \put (87,43) {$Z(E_2)$} 
\end{overpic}
\caption{The creation of DD and DN boundary trumpets $Z(\beta)$ and $Z(E)$ in JT gravity, indicated by blue and yellow boundaries respectively.}
    \label{fig:my_label}
\end{figure}
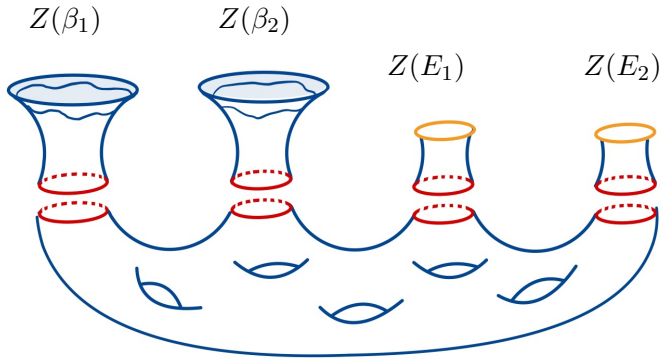

For example, the one-point function $\braket{\del\Phi(x)}_{\mathsf{KS},0} = \omega(x)$ becomes the leading order density of states:
\beq
\rho_0(E) \equiv \mathrm{disc}\,  \omega(E) = \fr{1}{4\pi^2} \sinh(2\pi \sqrt{E})~.
\eeq
Note that it has the correct universal $\sqrt{E}$-behaviour for low energy, typical of double-scaled matrix models. Integrating, we obtain the disk contribution to the DN path integral, which agrees with our previous answer \eqref{DNdisk}:
\beq 
Z_{\mathsf{disk}}(E) = \int^E dE' e^{S_0}\braket{\mathrm{disc}\,\del\Phi(E')}_{\mathsf{KS},0} =  \fr{e^{S_0}}{8\pi^4}\Big(2\pi\sqrt{E}\cosh(2\pi \sqrt{E}) - \sinh(2\pi \sqrt{E})\Big)~.
\eeq
Note that the $E$-integral came from the factor of $1/\beta$ present in the definition of the microcanonical path integral. It arose from gauge fixing the $U(1)$ symmetry of trumpet boundary. If we had assumed some marked point on the DN boundary, there would be no such factor of $1/\beta$, and $\mathcal{Z}_\mathrm{JT}(E)$ would be computed from insertions of $\rho(E)$. Such operators were considered from the matrix model point of view in \cite{eigenbranes}, where they were called `\emph{energy eigenbranes}', because they fix a particular energy eigenvalue in the matrix integral.

As another example, we can easily evaluate the contribution from two fixed energy boundaries connected by a wormhole. We do this by glueing two DN trumpets along their common geodesic boundary:
 \beq
 \int_0^\infty d\ell \,\ell \,Z_\mathsf{trumpet}(E_1,\ell)Z_\mathsf{trumpet}(E_2,\ell) = \fr{1}{2\pi^2} \log\left( \fr{\sqrt{E_1}+\sqrt{E_2}}{\sqrt{E_1}-\sqrt{E_2}}\right)~.
 \eeq
As expected, the right-hand side is the discontinuity of the free correlator $\braket{\Phi(E_1)\Phi(E_2)}_0$. Indeed, one can easily verify that taking derivatives with respect to $E_1$ and $E_2$, followed by an inverse Laplace transform, precisely returns the universal wormhole contribution \eqref{doubletrumpetintegral}. 

We can now borrow the matrix model intuition to understand why we found a twisted bosonic\footnote{The reason that we obtained a \emph{bosonic} theory, is that the collective excitations of eigenvalues in the double-scaled matrix model behave like bosons, even though single eigenvalues behave fermionically. For more on the relation between conformal field theory and double-scaled matrix models, see for example \cite{kostov3}.} field $\del \Phi$ to describe JT gravity. We have just seen that the energy $E$ that was fixed as a DN boundary condition, and which is dual to the temperature $\beta$, is related to the spectral plane coordinate as $x=-E$. The branch cut in the spectral $x$-plane is therefore mapped to the positive real axis in the $E$-plane. We can view the branch cut, and therefore the fact that $\del\Phi$ had to be $\bb{Z}_2$-twisted, as a direct consequence of a continuous eigenvalue density created by $\rho_0(E)$. This is the reason we called $x$ the `spectral plane' in the first place: a path in the $x$-plane represents the \emph{spectrum} of a double-scaled matrix model. A given DD boundary, for which the temperature is fixed, can be thought of as having an energy that is randomly drawn from a continuous statistical ensemble.

The double-scaled matrix model that gives rise to density of states $\rho_0(E)$ can be identified explicitly in the topological string theory setup. In fact, there is a precise way in which a stack of topological D2-branes wrapped around compact\footnote{Importantly these D-branes are \emph{compact} and distinct from the \emph{non-compact} D-branes introduced before.} cycles in the target space geometry \eqref{eq:calabiyau} give rise to a large $N$ matrix integral, which is dual to the closed string theory \cite{vafadijkgraaf1}. In that sense, the random matrix $H$ that leads to $\rho_0(E)$ is describing the open string degrees of freedom associated to this brane configuration. The localization of the open string field theory, which in this case is a 6-dimensional holomorphic Chern-Simons theory associated to the space-filling D6-brane, to a matrix integral can be done explicitly \cite{Dijkgraaf:2002fc} (see also \cite{Marino:2004eq}). The notation that was used heuristically in \eqref{eq:dens} and \eqref{eq:res} can then be understood more formally as a statement of the open/closed duality in topological string theory. Hence, from the perspective of the `JT string' it follows that the double-scaled matrix integral is actually dual to the KS field theory, rather than to JT gravity itself. This gives a new perspective on the role of the matrix ensemble associated to the gravitational path integral.

\subsection{Application: spectral correlation functions} \label{sec:application}
\begin{figure}
    \centering
    \includegraphics[width=7cm]{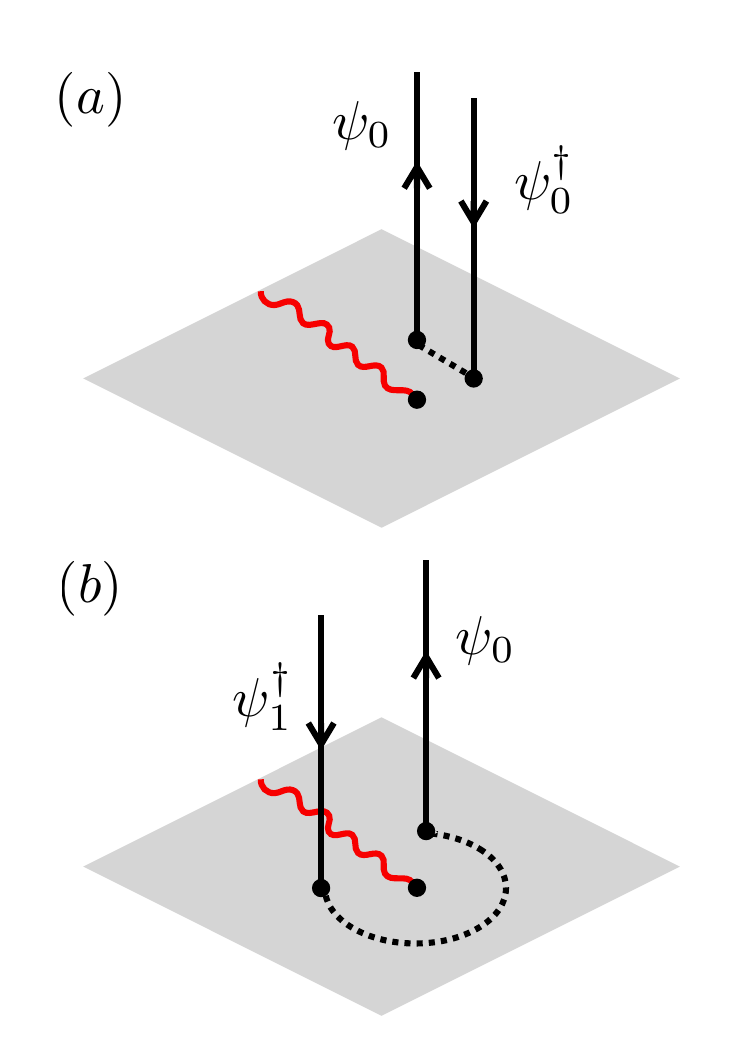}
    \caption{A pictorial representation of (a) the collision of two fermions $\psi_0$ and $\psi_0^{\dagger}$ on the same sheet, (b) the collision of two fermions $\psi_0$ and $\psi_1^{\dagger}$ on opposite sheets. Effectively, the latter is obtained from moving the anti-brane around the branch point along the striped line, and then bringing the fermions together on the branch cut (indicated by a red wiggly line).}
    \label{fig:braneantibrane2}
\end{figure}

Now we apply the formalism introduced in the previous section to extract non-perturbative corrections to the leading-order result for density and density-density correlation functions. First, recall that the density of states $\rho(E)$ is the discontinuity of the resolvent operator:
\beq\label{discont}
\rho(E) = \fr{1}{2\pi i}(\del\Phi_0(E) - \del\Phi_1(E))~.
\eeq
The (first order correction to the) collision of two branes on the same sheet $\psi_0(E) \psi^\dagger_0(E)$ leads to the insertion of a closed string state $\partial \Phi$ as in \eqref{bosonfermioncorrespondence}. The contribution coming from the interaction of branes on opposite sheets is given by 
\beq
\braket{\psi_0(E) \psi^\dagger_1(E)}_\mathsf{KS}=\braket{e^{\Omega(E)}}_\mathsf{KS}\approx e^{\frac{2\pi i}{\lambda} \int^E dE'\rho_{0}(E')}~.
\eeq
Here, we have used that the operator $\Omega(E)$ can be written as 
\beq
\Omega(E) = 2\pi i\int^E dE' \rho(E')~,
\eeq
and kept only the genus zero contribution to the expectation value. The result is entirely localized at the branch cut, and non-perturbative in the coupling constant. Intuitively, one may think about this quantity as a `geometric phase' that a brane picks up when it is transported around the branch point, see Figure \ref{fig:braneantibrane2}.

It turns out that this result captures non-perturbative contributions to the density of states if we add the following corrections to the perturbative expansion:
\begin{align}
\partial \Phi_0(E)_{\textrm{np}} &\sim \partial \Phi_0(E) +ie^{\Omega(E)}~, \label{eq:nonperturbative1} \\
\partial \Phi_1(E)_{\textrm{np}} &\sim \partial \Phi_1(E) -ie^{-\Omega(E)}~. \label{eq:nonperturbative2}
\end{align}
Crucially, the symbol $\sim$ indicates that the above operator identifications should be read inside \emph{perturbative} expectation values $\braket{\cdots}_{\mathsf{KS}}$ of the KS theory. This is the whole point of the construction: we are trying to extract non-perturbative physics using perturbative computations. The precise mechanism that underlies the above identifications is still rather mysterious, for now one should view it as an observation. We expect that the answer can be found in the open string theory dual to KS, and we hope to address this in future work. 

In terms of the matrix model $\partial \Phi(E)=R(E)$ is the resolvent, so the fermions $\psi(E)=\det(H-E)$ and  $\psi^{\dagger}(E)=1/\det(H-E)$ correspond to (inverse) determinant operators. In particular, $\psi^{\dagger}$ has singularities at the real axis and should be regularized. This leads to two fermions $\psi^{\dagger}_0$ or $\psi^{\dagger}_1$ depending on the sign in the $\pm i\epsilon$ prescription. Taking a single eigenvalue $E$ using the probe brane $\psi^{\dagger}_0$ and have it circle the branch point once results in the operator $\psi^{\dagger}_1$. In the process, the eigenvalue `feels' the force of the other eigenvalues, which is proportional to the number of eigenvalues given by $\int^E \rho_0(E')dE'$, and this effect is captured by the non-perturbative phase in  $\braket{\psi_0(E) \psi^\dagger_1(E)}_\mathsf{KS}$.

\newpage
We now make the following proposal for an observable in the universe field theory that captures non-perturbative corrections to the density of states, by analogy with \eqref{discont}:
\beq
\rho_{\textrm{np}}(E) \sim \fr{1}{2\pi i}(\partial \Phi_0(E)_{\textrm{np}} - \partial \Phi_1(E)_{\textrm{np}})~.
\eeq
The superscript `$\textrm{np}$' indicates that we have defined an observable which takes the non-perturbative corrections into account. With this proposal for a non-perturbative density of states, we can compute perturbative correlation functions in the KS theory.
\paragraph{Density correlator.}
The one-point function of $\rho_{\textrm{np}}(E)$ is computed straightforwardly:
\begin{align}
\braket{\rho_{\textrm{np}}(E)}_{\mathsf{KS}} &
\sim\braket{\rho(E)}_{\mathsf{KS}} + \frac{1}{2\pi}\left( \braket{e^{\Omega(E)}}_\mathsf{KS} + \braket{e^{-\Omega(E)}}_\mathsf{KS}\right)~.
\end{align}
In principle, we can compute this using the full interacting theory, including all genus corrections, but it turns out that the leading-order result already has the features that we are interested in, so we keep only the `disk' and `annulus' contributions. Clearly, we have 
\beq 
\braket{\rho(E)}_{\mathsf{KS}} \approx e^{S_0}\rho_0(E)~. 
\eeq 
For the other contribution we use the following identity:
\beq 
\braket{e^{\pm \Omega(E)}}_\mathsf{KS} \approx \exp \left[\pm\braket{\Omega(E)}^\mathsf{c}_0 + \fr{1}{2} \braket{\Omega(E)^2}^\mathsf{c}_0\right]~.
\eeq
The annulus contribution, which should be appropriately normal ordered, can be computed using the twisted two-point functions $\braket{\Phi_a(x)\Phi_b(x')}_\mathsf{KS}$:
\begin{align}
\fr{1}{2} \braket{\Omega(x)^2}^\mathsf{c}_0 &\equiv\fr{1}{2} \lim_{x' \to x}  \Big\langle \{\Omega(x)\Omega(x')\}\Big\rangle^\mathsf{c}_0  \nonumber \\
&= \lim_{x' \to x} \Big[\log(\sqrt{x} - \sqrt{x'}) - \log(\sqrt{x} + \sqrt{x'}) - \log(x-x')\Big] \nonumber \\
&=\log \fr{1}{4x}~.
\end{align}
Setting $x=-E$, and putting everything together, we thus find the leading-order result:
\begin{align}\label{densityofstatesapprox}
\braket{\rho_{\textrm{np}}(E)}_{\mathsf{KS}} \approx e^{S_0}\rho_0(E) - \fr{1}{4\pi E}\cos\left(2\pi \,e^{S_0}\int^E dE' \rho_0(E') \right).
\end{align}
This is the same result found by Saad, Shenker and Stanford \cite{saad2019jt}, who obtained it using matrix model techniques. As one can see, the non-perturbative effects give small oscillations on top of the perturbative leading order density of states $\rho_0(E) = \frac{1}{4\pi^2}\sinh(2\pi \sqrt{E})$, where the size of the oscillations is controlled by $\lambda = e^{-S_0}$. 

\paragraph{Density-density correlator.}
Next, we want to compute the density-density correlator:
\beq
\braket{\rho_{\textrm{np}}(E_1)\rho_{\textrm{np}}(E_2)}_{\mathsf{KS}}~.
\eeq
Of course, there will be the factorized contribution, but we will see that the interesting result comes from the connected contributions, which assemble into the so-called \emph{sine-kernel}, well-known in the matrix model literature. Expanding the expression, there are many products that we need to compute:
\begin{align}
\rho_{\textrm{np}}(E_1)\rho_{\textrm{np}}(E_2) &= \rho(E_1)\rho(E_2)  + \frac{1}{2\pi} \rho(E_1) (e^{\Omega(E_2)}+ e^{-\Omega(E_2)}) + \frac{1}{2\pi}(e^{\Omega(E_1)}+ e^{-\Omega(E_1)}) \rho(E_2) \nonumber\\[0.5em]
&\qquad+ \frac{1}{4\pi^2} (e^{\Omega(E_1)} + e^{-\Omega(E_1)})(e^{\Omega(E_2)} + e^{-\Omega(E_2)})~.
\end{align}
The singularities in the cross-terms of $\rho(E)$ with $e^{\Omega(E)} +e^{-\Omega(E)} $ cancel, while the OPE's of $e^{\Omega(E_1)}e^{\Omega(E_2)}$ and $e^{-\Omega(E_1)}e^{-\Omega(E_2)}$ are also regular as $E'\to E$. The only singular contributions come from the products $\rho(E_1)\rho(E_2)$ and $e^{\pm \Omega(E_1)}e^{\mp \Omega(E_2)}$. The first gives the perturbative contribution, keeping only the genus zero terms:
\begin{align}
\braket{\rho(E_1)\rho(E_2)}_\mathsf{KS} &\approx -\fr{1}{2\pi^2(E_1-E_2)^2} + \mathrm{reg.}
\end{align}
where we have neglected terms which are regular as $E' \to E$. This perturbative contribution to the density-density correlator is called the `ramp', because after a double Fourier transform it gives rise to the linear growth of the spectral form factor. To obtain the second term, we compute the product:
\beq
e^{\Omega(E_1)}e^{-\Omega(E_2)} =\fr{1}{(E_1-E_2)^2}\big\{e^{\Omega(E_1) - \Omega(E_2)}\big\}~.
\eeq
Here, we combined the product of normal-ordered exponentials into a single normal-ordered exponential, with the normal ordering $\{\dots \}$ given by subtracting $\log(E_1-E_2)$ from the singular products $\Phi_0\Phi_0$ and $\Phi_1\Phi_1$, leading to the multiplicative factor. Now we can take the expectation value and keep only the genus zero contributions:
\begin{align}\label{nonpertpiece}
\braket{e^{\Omega(E_1)}e^{-\Omega(E_2)}}_\mathsf{KS} \approx \fr{1}{(E_1-E_2)^2} \exp \left[\braket{\Omega(E_1)-\Omega(E_2)}_0 +\half \Big\langle\big(\Omega(E_1) - \Omega(E_2)\big)^2\Big\rangle^\mathsf{c}_0 \right]~.
\end{align}
The square in the last term of the exponent should be appropriately normal-ordered by subtracting the singular pieces, as was the case for the density correlator. It can be easily evaluated using the free two-point functions to be:
\begin{align}
\half \braket{(\Omega(x)-\Omega(y))^2}^\mathsf{c}_0 &= \fr{1}{2} \Big[\braket{\Omega(x)^2}_0^\mathsf{c} - 2\braket{\{\Omega(x)\Omega(y)\}}_0^\mathsf{c}  +\braket{\Omega(y)^2}_0^\mathsf{c} \Big] \nonumber \\
&= \log \fr{1}{4x} + \log \fr{1}{4y} - 2\left[\log\left( \fr{\sqrt{x}-\sqrt{y}}{\sqrt{x}+\sqrt{y}}\right) - \log(x-y)\right] \nonumber \\
&= \log \fr{(\sqrt{x} + \sqrt{y})^4}{16xy}~.
\end{align}
Sending $x = -E_1$ and $y=-E_2$ and plugging this into \eqref{nonpertpiece}, we find:
\begin{align}
\braket{e^{\Omega(E_1)}e^{-\Omega(E_2)}}_\mathsf{KS} &\approx  \fr{1}{(E_1-E_2)^2}  \fr{(\sqrt{E_1} + \sqrt{E_2})^4}{16E_2E_1}\,e^{\braket{\Omega(E_1)-\Omega(E_2)}_0 } \nonumber \\[0.6em]
&= \left[ \fr{1}{(E_1-E_2)^2} + \mathrm{reg.}\right]e^{\braket{\Omega(E_1) -\Omega(E_2)}_0 }~.
\end{align}
In the last line, we again expanded $E_1$ around $E_2$ and kept only the singular piece. Repeating the calculation above we find the other OPE with signs flipped:
\beq
\braket{e^{-\Omega(E_1)}e^{\Omega(E_2)}}_\mathsf{KS} \approx  \left[ \fr{1}{(E_1-E_2)^2} + \mathrm{reg.}\right]e^{-\braket{\Omega(E_1)-\Omega(E_2)}_0}~.
\eeq
Putting everything together gives the connected contribution to the non-perturbative density-density correlator:
\begin{align}
\braket{\rho_{\textrm{np}}(E_1)\rho_{\textrm{np}}(E_2)}^{\mathsf{c}}_\mathsf{KS} &\sim \braket{\rho(E_1)\rho(E_2)}_\mathsf{KS} +\frac{1}{4\pi^2}\left( \braket{e^{\Omega(E_1)}e^{-\Omega(E_2)}}_\mathsf{KS} +\braket{e^{-\Omega(E_1)}e^{\Omega(E_2)}}_\mathsf{KS}\right) \nonumber \\[0.8em]
&\approx - \fr{1}{2\pi^2(E_1-E_2)^2} \Big[1 -\cosh \left(\braket{\Omega(E_1) -\Omega(E_2)}_0\right) \Big]~.
\end{align}
Using that $\braket{\Omega(E)}_0 = \frac{2\pi i}{\lambda} \int^E \rho_0(E')dE'$ we conclude:
\beq
\braket{\rho_{\textrm{np}}(E_1)\rho_{\textrm{np}}(E_2)}^{\mathsf{c}}_\mathsf{KS} \approx -\fr{1}{\pi^2(E_1-E_2)^2} \sin^2\left(\pi e^{S_0}\int^{E_1}_{E_2} \rho_0(E') dE'\right)~.
\eeq
For $E_2\to E_1$, the integral can be approximated by $(E_1-E_2)\rho(E_2)$. Adding the disconnected piece, we arrive at the main result of this section:
\beq \label{eq:spectralform}
\braket{\rho_{\textrm{np}}(E_1)\rho_{\textrm{np}}(E_2)}_\mathsf{KS} \approx  \braket{\rho_{\textrm{np}}(E_1)}_0\braket{\rho_{\textrm{np}}(E_2)}_{0} -\fr{1}{\pi^2(E_1-E_2)^2} \sin^2\!\left(\pi e^{S_0}(E_1-E_2)\rho_0(E_2)\right)~.
\eeq
This is only an approximate answer, in the sense that we only considered genus zero contributions and were interested in the singular part of the products. The above computation was merely meant to show the universal behaviour of $\braket{\rho(E_1)\rho(E_2)}_\mathsf{KS}$ for $|E_1-E_2|\ll 1$. Here, we have shown that the non-perturbative contributions in the universal form of the sine-kernel can be understood as arising from branes in the KS theory. From the universe field theory side, these are described by (bilinears of) fermion fields, while on the JT gravity side, they describe D-branes where fixed-energy boundaries can end on.

\newpage
\section{Discussion} \label{discussion}

Let us now discuss some subtleties, open questions and directions for future research.

\paragraph{Open/closed duality.} 
We have proposed that the modified holographic dictionary which relates JT gravity to a matrix integral is in fact a consequence of a more standard open/closed duality in topological string theory. Formulating JT gravity in terms of the KS theory allows for a direct interpretation in string theory and the `ensemble average' should then correspond to the path integral in the open string field theory dual of the KS theory, as outlined in Figure \ref{fig:1}. Although we have identified the relevant quantities on both sides, a detailed account of the duality is still an open question. This requires a more careful study of both the compact and non-compact D-branes that we have introduced. More speculatively, we expect that the open string field, which can be represented by a Hermitian matrix $H$, should be viewed as the Hamiltonian of some quantum mechanical system associated to the branes in the theory. It would be interesting to see if the underlying fermionic theory of the open string degrees of freedom can be in some way related to the SYK model \cite{SYK1993}. For recent work connecting the SYK model to string theory, see \cite{Goel:2021wim}.  

\paragraph{Baby universes and $\alpha$-states.} 
Moreover, we expect that the KS theory gives a well-defined construction of the baby universe Hilbert space (as defined in \cite{marolfmaxfield}) for JT gravity. One can formally represent the path integral of KS theory in terms of an operator formalism. This would naturally lead to a notion of boundary operators $\widehat{Z}(\beta)$. In fact, we expect a slight modification of the construction by Marolf and Maxfield \cite{marolfmaxfield} in the sense that we need to consider a larger algebra of observables by adding the `canonical momentum' of $\widehat{Z}(\beta)$.
 We would then have a non-commutative algebra of observables with not only boundary creation operators but also boundary annihilation operators. This is close to the original approach taken in \cite{coleman1}. In particular, this construction would lead to a precise definition of the Hartle-Hawking state $\ket{\HH}$ with non-trivial topology, in terms of the interacting vacuum of the KS theory. We can represent this state geometrically by an integration over half the spectral curve. That is, we cut open the path integral on the slice $\mathrm{Re}(z) =0$. The resulting baby universe Hilbert space is infinite-dimensional and distinct from the baby universe Hilbert space that will be discussed in appendix \ref{ch:recursion}, since the in- and out-states are treated \emph{symmetrically}. In principle, this construction would give us a definition of the microscopic $\alpha$-states $|\alpha\rangle$ in JT gravity, and an understanding of their role in the factorization problem \cite{factorizationpuzzle}.

\paragraph{Non-perturbative effects.} 
It would be interesting to study further the non-perturbative effects that were touched upon in section \ref{boundaryconditions}. Although it seems that the brane/anti-brane perspective leads to the correct results \eqref{densityofstatesapprox} and \eqref{eq:spectralform}, the geometrical interpretation of the precise mechanism is still rather mysterious. For example, the non-perturbative correction in \eqref{eq:nonperturbative1} and \eqref{eq:nonperturbative2} decouples the the two fields $\partial \Phi_0$, and $\partial \Phi_1$ in the sense that they are not anymore related to each other by a $2\pi$ rotation. This seems to agree with the perspective that non-perturbative effects have a dramatic effect on the target space decoupling the two sheets of the branched geometry \cite{Maldacena:2004sn}. In particular, it leads to a branch cut extending over the whole real axis $[-\infty,\infty]$. It would also be interesting to understand the connection to \cite{Altland:2020ccq}, where an effective field theory for the late-time behaviour of quantum chaotic systems is presented (see also \cite{Altland:2021rqn}). For example, one could try to find an interpretation of the Altshuler-Andreev saddle and notion of causal symmetry breaking, that are important in the computation of the `plateau' feature of the spectral form factor, in the topological string theory setup. We expect these effects to become visible in the open string field theory description dual to KS theory. This is work in progress. 

\paragraph{Super JT gravity.} 
There are some generalizations of the construction which are worth studying. It would be interesting to carry out a similar analysis in the case of JT supergravity \cite{stanfordwittensuperjt,volumesandrandommatrices}. It is defined on super Riemann surfaces for which a recursion relation similar to Mirzakhani's is derived. The topological recursion for the matrix model associated to super JT is related to the Brezin-Gross-Witten and the Bessel model \cite{okuyamasuperjt}. Introducing both fermions and bosons, the super-Virasoro algebra generated by the combined stress tensor leads to super-Virasoro constraints \cite{superquantumairy}. A natural question is if these are equivalent to the `super-Mirzakhani recursion' and if we can extend the KS theory to a supersymmetric model, whose SD equations impose the super-Virasoro constraints.

\paragraph{$\overline{J}T$-deformation.} 
Another interesting perspective on the KS interaction \eqref{Sint} is as a hybrid between a marginal deformation with the stress tensor and a $T\bar{T}$-deformation. Namely, we have deformed the free theory by an irrelevant operator proportional to $ \overline{\del} \Phi\, T$. Denoting $\overline{\del}\Phi$ as the anti-holomorphic current $\overline{\mathcal{J}}$, we could call the KS interaction a `$\overline{\mathcal{J}}T$-deformation', by analogy with $T\bar{T}$ \cite{TTbar, TTbar0}. There has been much recent attention for irrelevant deformations similar to $T\bar{T}$ (see, e.g., \cite{Guica:2017lia,Bzowski:2018pcy,Guica:2019vnb,Anous:2019osb}) and their relation to integrable systems \cite{JTbar1, JTbar2, JTbar3, JTbar4}, as well as their relation to JT gravity \cite{TTbar3, TTbar2}. It would be nice to understand how this discussion fits in.

\paragraph{Pure 3d gravity as an ensemble.} 
The precise form of the KS action and its relation to JT gravity relied heavily on the interpretation of the spacetime in terms of a string world-sheet. In that sense, the construction seems to be very specific to models of 2-dimensional gravity. There is some evidence that wormholes in pure 3-dimensional gravity can also be understood in terms of some averaging prescription (research in this direction includes \cite{maxfield, Cotler:2020ugk, conicaldefects, belin2020random, wittenmaloney, Belin:2021ryy,Belin:2021ibv,Anous:2021caj}). Our derivation was fundamentally based on the universal recursive structure expressed in terms of the SD equation. If one could unearth a similar recursive structure in higher-dimensional theories of quantum gravity, this would open up a way for finding a similar field theory description. 
\paragraph{}

We hope to address some of these questions in future work.

\section*{Acknowledgements}
We would like to thank Alexander Altland, Jan de Boer, Ricardo Esp\'indola, Bahman Najian, Sergey Shadrin, Julian Sonner and Herman Verlinde for discussions. We are also grateful to Tarek Anous for providing useful comments and suggestions on an earlier version of the draft. BP and JvdH are supported by the European Research Council under the European Unions Seventh Framework Programme (FP7/2007-2013), ERC Grant agreement ADG 834878. This work is supported by the Delta ITP consortium, a program of the Netherlands Organisation for Scientific Research (NWO) that is funded by the Dutch Ministry of Education, Culture and Science (OCW). 

\appendix 
\addtocontents{toc}{\protect\setcounter{tocdepth}{1}}

\section{Free two-point functions of the universe field theory}\label{freetwopoint}
We compute the two-point functions of the free theory including sources \eqref{eq:freesources} with action \eqref{eq:freeaction}. The partition function is just a Gaussian integral in $\Phi$ and $\mathcal{J}$, so we can solve it by functional determinants. We will start with the $\Phi$-integral. Integrating by parts, the terms involving $\Phi$ are:
 \beq
\half \Phi\, \del \overline{\del}\, \Phi - \Phi \,\overline{\del}\mathcal{J} - \mu_\Phi \Phi~.
 \eeq
The Laplacian $\Delta = \del\overline{\del}$ on the spectral curve has a Green's function $\Delta^{-1}$, which can be found by first projecting to the spectral $x$-plane, and then transforming back to the double cover via $x=z^2$. There is a branch cut on the negative real axis and therefore the Green's function on the $x$-plane can be found using the method of images \cite{methodofimages}. Transforming to the double cover, $x=z^2, y=w^2$, we find the result
\beq \label{eq:delta-1}
\Delta^{-1}(z,w) = \half \ln\left\vert\fr{z-w}{z+w}\right\vert~.
\eeq
We can now solve the $\Phi$-integral by completing the square:
 \begin{align}
\int [d\Phi]\exp\left[ \half \Phi\, \del \overline{\del}\, \Phi - \Phi \,\overline{\del}\mathcal{J} - \mu_\Phi \Phi\right] = N \exp\left[-\half(\overline{\del}\mathcal{J} + \mu_\Phi)\Delta^{-1}(\overline{\del}\mathcal{J} + \mu_\Phi)\right]~. \end{align}
Here, we have used the condensed notation
\beq
A \,\Delta^{-1} B \equiv \int d^2z \int d^2w \,A(z) \Delta^{-1}(z,w) B(w)~.
\eeq
We have also denoted the functional determinant by $N$, which is just a (possibly infinite) constant that drops out, because we have normalized the partition function. Explicitly, the functional determinant is
\begin{align}
N &= \int [d\Phi] \exp\left[\half \Big(\Phi - \Delta^{-1}(\mu_\Phi + \overline{\del} \mathcal{J})\Big) \Delta \Big(\Phi - \Delta^{-1}(\mu_\Phi + \overline{\del} \mathcal{J}) \Big)\right] \\
&= \int [d\Phi'] \exp\left[-\half \Phi' (-\Delta) \Phi'\right] = \det(-\Delta)^{-1}~.
\end{align} 
In general, the determinant should be regularized, but we do not have to worry about this since we have normalized the partition function with sources. Having done the $\Phi$-integral, we are left with the integration over $\mathcal{J}$:
\beq\label{Jintegral}
 Z^{(0)}_\mathsf{KS}[\mu_\Phi, \mu_{\mathcal{J}}] = \fr{N}{ Z^{(0)}_\mathsf{KS}[0]} \int [d\mathcal{J}] \exp\left[-\half(\overline{\del}\mathcal{J} + \mu_\Phi)\Delta^{-1}(\overline{\del}\mathcal{J} + \mu_\Phi) - \mu_{\mathcal{J}} \mathcal{J}\right]~.
\eeq
Now we introduce the Green's function for the $\overline{\del}$-operator on the spectral curve. It is simply the derivative of the Green's function in \eqref{eq:delta-1}:
\beq\label{derivgreen}
\overline{\del}^{-1} = \del \Delta^{-1} = \fr{dz}{z-w} - \fr{dz}{z+w}~.
\eeq
Writing $\,\overline{\del}^{-1}\equiv \overline{\del}^{-1}(z,w) \,dz$, we perform the shift
\beq
\mathcal{J} \to \mathcal{J} - \overline{\del}^{-1} \!\cdot \mu_\Phi~,
\eeq
where the $\cdot$ means that we integrate with respect to the second argument of $\overline{\del}^{-1}$:
\beq
\overline{\del}^{-1} \! \cdot \mu_\Phi \equiv \int d^2w\, \left(\overline{\del}^{-1}(z,w) \mu_\Phi(w)\right)\,dz~.
\eeq
The exponent of \eqref{Jintegral} now becomes
\beq
-\half \overline{\del} \mathcal{J}\, \Delta^{-1} \, \overline{\del} \mathcal{J} - \mu_\mathcal{J}\big(\mathcal{J} - \overline{\del}^{-1} \! \cdot \mu_\Phi\big)~.
\eeq
Integrating the first term by parts, and using that $\overline{\del}\Delta^{-1} = \del^{-1}$, we see that the free partition function is
\beq
Z^{(0)}_\mathsf{KS}[\mu_\Phi, \mu_{\mathcal{J}}] = \fr{N}{ Z^{(0)}_\mathsf{KS}[0]} \int [d\mathcal{J}] \exp\left[\half \mathcal{J}\, \del^{-1} \overline{\del} \mathcal{J} - \mu_\mathcal{J} \mathcal{J}\right] \exp\left[\mu_\mathcal{J}\overline{\del}^{-1} \mu_\Phi\right]~.
\eeq
We can complete the square in the first exponent and do the Gaussian integral, which gives us
\beq\label{arrive}
Z^{(0)}_\mathsf{KS}[\mu_\Phi, \mu_{\mathcal{J}}] = \fr{N \,N'}{ Z^{(0)}_\mathsf{KS}[0]} \exp\left[ \half \mu_\mathcal{J} \del \overline{\del}^{-1} \mu_\mathcal{J}+\mu_\mathcal{J}\overline{\del}^{-1} \mu_\Phi\right]~.
\eeq
We have written the functional determinant $N'$ as
\begin{align}
N' &= \int [d\mathcal{J}] \exp\left[\half\Big(\mathcal{J} - (\del^{-1} \overline{\del})^{-1} \mu_\mathcal{J}\Big)\, \del^{-1}\overline{\del}\,\Big(\mathcal{J} - (\del^{-1} \overline{\del})^{-1} \mu_\mathcal{J}\Big) \right] \\
&= \int [d\mathcal{J}'] \exp \left[-\half \mathcal{J}' (-\del^{-1}\overline{\del}) \mathcal{J}' \right] = \det(-\del^{-1}\overline{\del}\,)^{-1}~.
\end{align} 
In arriving at \eqref{arrive}, we have also used that $(\del^{-1} \overline{\del})^{-1} = - \del \overline{\del}^{-1}$, as can be verified by acting from the left with $\del^{-1} \overline{\del}$ and integrating by parts. The factors $N$ and $N'$ cancel against the normalization, as can be seen by turning off the sources. Therefore, we have shown that: 
\begin{align}
Z^{(0)}_\mathsf{KS}[\mu_\Phi, \mu_{\mathcal{J}}] = \exp\left[ \half \mu_\mathcal{J} \del \overline{\del}^{-1} \mu_\mathcal{J}+\mu_\mathcal{J}\overline{\del}^{-1} \mu_\Phi\right]~.
\end{align}
In particular, this implies that there are only contractions between $\mathcal{J}$ and $\mathcal{J}$, and between $\mathcal{J}$ and $\Phi$. There is no contraction of $\Phi$ with itself. Here, $\overline{\del}^{-1}$ is given by \eqref{derivgreen} and
\beq
\del \overline{\del}^{-1} = \fr{dz\, dw}{(z-w)^2} + \fr{dz \,dw}{(z+w)^2}~.
\eeq
So, if we define the following functions: 
\beq 
\mathsf{B}(z,w) = \fr{1}{(z-w)^2} + \fr{1}{(z+w)^2}, \quad \mathsf{G}(z,w) = \fr{1}{z-w} - \fr{1}{z+w}~,
\eeq 
we arrive at the result claimed in the main text:
\beq
 \log Z^{(0)}_\mathsf{KS}[\mu_\Phi, \mu_{\mathcal{J}}]  = \int d^2z \int d^2w \left[\half \mu_\mathcal{J}(z) \mathsf{B}(z,w)\mu_\mathcal{J}(w) + \mu_\mathcal{J}(z)\mathsf{G}(z,w) \mu_\Phi(w) \right]~.
\eeq

\section{Topological recursion}\label{toporecursion}
In this appendix we give some background on the topological recursion formalism of Eynard and Orantin \cite{eynard1,eynardrandommatrices}.
The starting point of the topological recursion is the data of a spectral curve, which consists of a tuple $(\mathcal{C}, x, y, \mathcal{B})$, where
\begin{itemize}
\item $\mathcal{C}$ is a compact Riemann surface,
\item $x$ and $y$ are two analytical functions on an open domain of $\mathcal{C}$, 
\item $\mathcal{B}$ is the \emph{Bergmann kernel} of $\mathcal{C}$, which in local coordinates $z,w$, is the unique bilinear differential with a double pole at $z=w$ and no other poles:
\beq
\mathcal{B}(z,w) = \fr{dz \, dw}{(z-w)^2} + \mathrm{reg.} \quad \mathrm{as}\quad z\to w~.
\eeq
\end{itemize}
We will consider the case in which $\mathcal{C}$ is (topologically) the Riemann sphere\footnote{If $\mathcal{C}$ is a higher genus Riemann surface, one additionally needs to specify a basis of non-contractible $A$- and $B$-cycles on $\mathcal{C}$, and require that the periods of the Bergmann kernel $\mathcal{B}$ vanish on the $A$-cycles. This requirement ensures that $\mathcal{B}$ is unique: if there were two Bergmann kernels, their difference would be holomorphic and therefore constant; vanishing of the $A$-periods then implies that the constant is zero. In fact, the Bergmann kernel $\mathcal{B}$ has the important characterization as the second derivative of the Green's function for the heat equation on $\mathcal{C}$ \cite{eynard5}.} $\bb{P}^1 = \bb{C}\cup \{\infty\}$. The functions $x$ and $y$ satisfy a relation of the form $H(x,y)=0$, which defines an (algebraic) curve $\mathcal{S}$, which is also referred to as the spectral curve.
The spectral curve $\mathcal{S}$ is a branched cover of the spectral plane given by the $x$-coordinate, with branch points $a_i$ defined by $dx(a_i)=0$. The covering map corresponds to the projection to the $x$-axis, which is two-to-one in the case of JT gravity. The two `sheets' of the double cover are exchanged in the neighbourhood of a branch point by a local involution $z \to \tilde{z}$.
\subsection{The spectral curve for JT gravity} \label{app:spectralcurveJT}
We will now describe the spectral curve that is relevant for JT gravity. It is given in terms of the one-point function $\omega(x) = \fr{1}{4\pi} \sin(2\pi \sqrt{x})$ by
\beq
\mathcal{S}_{\mathrm{JT}}: \quad y^2 = \omega(x)^2~.
\eeq
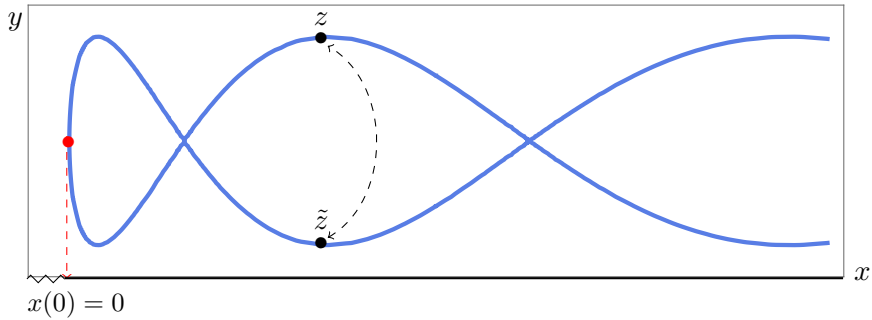
\begin{figure}[h]
\centering
\begin{tikzpicture}
    \node[anchor=south west,inner sep=0] at (0,0) {\begin{overpic}[width=0.7\textwidth]{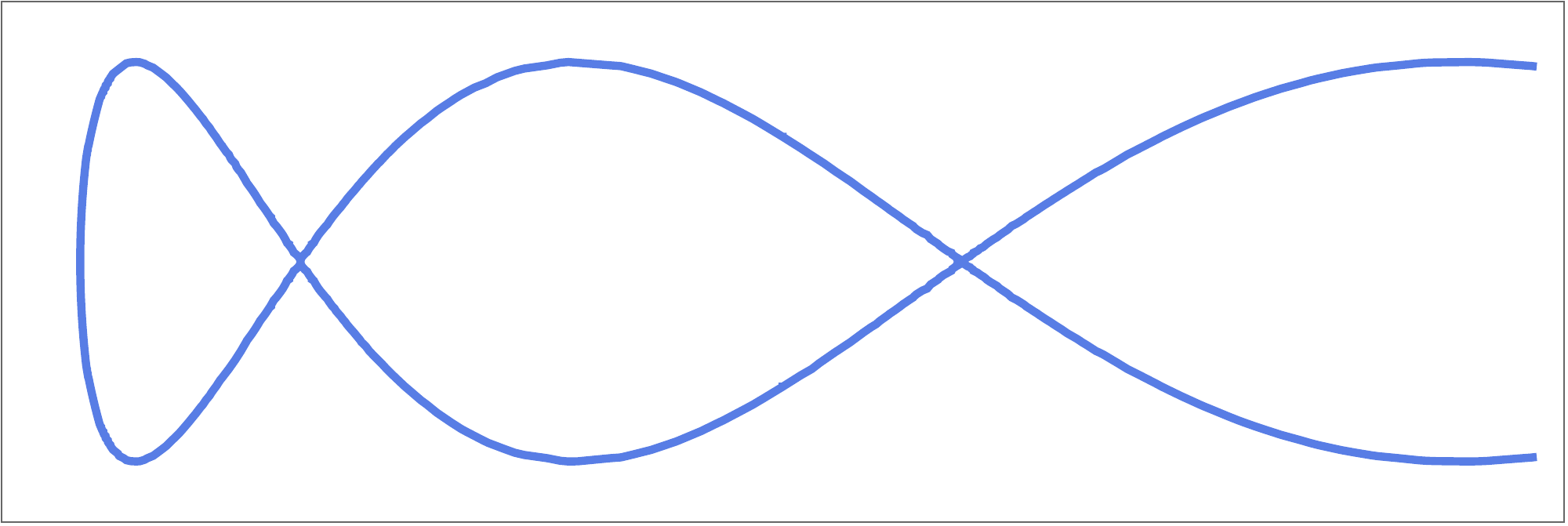}\put (35,3.42){$\bullet$} \put (4.1,15.8){\textcolor{red}{$\bullet$}} \put (35,6){$\tilde{z}$}  \put (35,28.5){$\bullet$} \put (35,31){$z$} \put (101,-0.2){$x$}  \put (-2.4,31){$y$}  \put (0,-4){{\small $x(0)=0$}} \end{overpic}};
    \draw [dashed, <->] (4,0.55) to [out=30,in=-30] (4,3.1);
    \draw [dashed, ->, color=red] (0.53, 1.9) to (0.53,0);
    \draw[decoration = {zigzag,segment length = 2mm, amplitude = 0.5mm},decorate] (0,0)--(0.5,0);
    \draw[thick, -] (0.5,0.01)--(10.85,0.01);
\end{tikzpicture}
\vspace{2em}
\caption{Spectral curve for JT gravity. There is one branch point at $z=0$, corresponding to $x=0$. The involution exchanges $z$ and $\tilde{z} = -z$.}
\label{JTspectralcurve}
\end{figure}
The spectral curve can be parametrized by a single uniformizing coordinate $z \in \bb{P}^1$:
\beq\label{jtspectraalkromme}
\mathcal{S}_{\mathrm{JT}}: \quad x(z) = z^2~,\quad y(z) = \fr{1}{4\pi} \sin(2\pi z)~.
\eeq
A real slice of this curve is plotted in Figure \ref{JTspectralcurve}. The function $x(z) = z^2$ gives $\mathcal{S}_{\mathrm{JT}}$ the structure of a branched double cover of the spectral $x$-plane. There is a single\footnote{Since the branch cut extends along the whole half-line, there is also a branch point at $\infty$. However, we will only need to know the local behavior of $y(z)$ near the branch point at $z=0$, because the topological invariants are defined as residues at the origin.} branch point at $z=0$, since $dx=2zdz$. The branch point $z=0$ gets mapped to $x=0$ on the spectral plane. The involution that exchanges the sheets of the double cover is simply $z \to -z$. 

It should be noted that the curve $H(x,y) = 0$ is not algebraic, since $\omega(x)^2$ is not a finite polynomial\footnote{Note that $\omega(x)$ has an essential singularity at $\infty$.}, so $\mathcal{S}_{\mathrm{JT}}$ is a non-compact Riemann surface. However, since it can be parametrized by a single variable $z \in \bb{P}^1$ one could possibly add a point at $\infty$, and argue that $\mathcal{S}_{\mathrm{JT}}$ effectively has genus zero. However, the notion of `genus' for such non-compact Riemann surfaces is somewhat vague, and another interesting interpretation of $\mathcal{S}_{\mathrm{JT}}$ is as a Riemann surface of \emph{infinite} genus of which infinitely many $A$-cycles have been pinched. This interpretation can be justified when we compare to the $(2,p)$-minimal string theory with $p$ an odd integer, studied for example in \cite{Seiberg}. In that case, the spectral curve was found to be an odd power $y(z) \sim z^p$, so that $H(x,y) = 0$ describes a compact Riemann surface of genus $p$. We can thus quite possibly see the JT spectral curve $y(z)\sim \sin(2\pi z)$, being an odd power series in $z$, as a `infinite linear combination' of $(2,p)$-minimal models, as was recently advocated in \cite{cliffordjohnson1, cliffordjohnson2}. 

The reason that we effectively see a genus zero spectral curve is that all the $A$-cycles have been pinched to points, at the zeros of the $\sin(2\pi z)$ where the two sheets of the double cover meet. Non-perturbative effects may cause the zeroes of $y(z)$ to `open up', adding small corrections to the right-hand side of \eqref{jtspectraalkromme} and thereby un-pinching the $A$-cycles. Since $\sin(2\pi z)$ has infinitely many zeroes, this un-pinching renders $\mathcal{S}_{\mathrm{JT}}$ a genus $p \to \infty$ Riemann surface. The $p\to \infty$ limit of minimal string theory was recently studied more thoroughly in \cite{mertensturiaci}, where many quantities were matched to quantities in JT gravity. 

\subsection{The symplectic invariants of TR}
The topological recursion associates to the data $(\mathcal{C}, x,y, \mathcal{B})$, a set of so-called \emph{symplectic invariants} $\omega_{g,n}$ on the spectral curve. They are defined recursively, starting from the initial data
\begin{align}
&\omega_{0,1}(z) \equiv y(z)dx(z)~, \hspace{15pt} \omega_{0,2}(z_1,z_2) \equiv \mathcal{B}(z_1,z_2)~.
\end{align}
The recursion kernel is defined by
\beq\label{recursionkernel}
\mathcal{K}(z_0,z) \equiv \fr{\half \int_{\tilde{z}}^{z} \mathcal{B}(z_0, \,\cdot\,)}{\big(y(z)-y(\tilde{z})\big)dx(z)}~. \vspace{1em}
\eeq
Here, the coordinates $z,w$ and the involution $z\to \widetilde{z}$ are defined locally near a branch point $a_i$. Moreover, $\fr{1}{dx(z)}$ denotes the contraction with the vector field $\left(\fr{dx}{dz}\right)^{-1} \del_z$. The notation $\int \mathcal{B}(z_0, \,\cdot\,)$ means that we integrate only with respect to the second argument. This makes $\mathcal{K}(z_0, z)$ a tensor of the type $dz_0 \otimes \del_{z}$. In other words, when acting on a multilinear differential $dz_1 \otimes\dots \otimes dz_n$, it removes a factor of $dz$ and tensors with $dz_0$. 

The topological recursion then produces multi-differentials of the form
\beq
\omega_{g,n}(z_1,\dots, z_n) = \mathcal{W}_{g,n}(z_1,\dots, z_n) \,dz_1\otimes \cdots \otimes dz_n~,
\eeq
which are defined recursively by taking residues at the branch points,
 \begin{align}\label{topologicalrecursion}
\omega_{g,n+1}(z_0 ,z_I) &= \sum_{i} \underset{z \to a_i}{\mathrm{Res}} \,\mathcal{K}(z_0,z) \Big[\underbrace{\omega_{g-1,n+2}(z,\widetilde{z},z_I)}_{\textcolor{blue}{b)}} \nonumber \\& \hspace{2cm}+ \sideset{}{'}\sum_{\substack{g_1+g_2=g \\[0.3em] J_1 \sqcup J_2 = I}} \underbrace{\omega_{g_1,1+|J_1|}(z,z_{J_1}) \,\omega_{g_2,1+|J_2|}(\widetilde{z},z_{J_2})}_{\textcolor{blue}{a)+c)}}\Big]~. 
\end{align}
We denote $z_{J}=(z_{j})_{j\in J}$ and the sum in the last term goes over all ways to partition the multi-index $I = (i_1,\dots,i_n)$ into subsets $J_1$ and $J_2$, and over all ways to distribute $g$ into $g_1+g_2$. The prime indicates that terms involving $(g,n) = (0,1)$ should be excluded from the summation. The labels $\textcolor{blue}{a)}$, $\textcolor{blue}{b)}$ and $\textcolor{blue}{c)}$ will be explained momentarily. 

\subsection{The symplectic invariants for JT gravity}
\begin{figure}[h]
\centering
\begin{overpic}[width=0.4\textwidth]{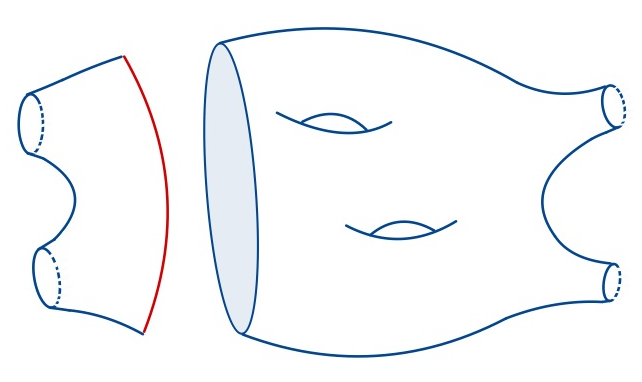}\put (1,3){$a)$}\end{overpic}\hspace{0.5cm}\begin{overpic}[width=0.36\textwidth]{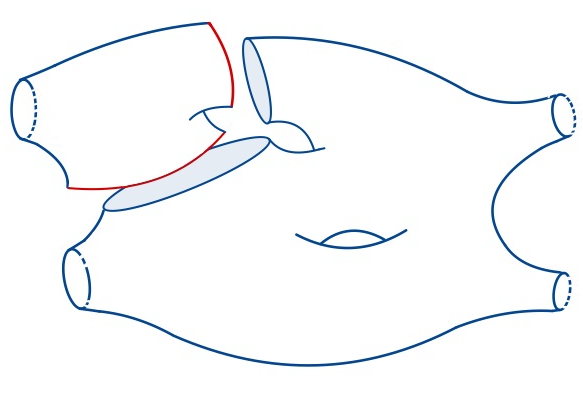}\put (1,3){$b)$}\end{overpic}\hspace{0.5cm} \begin{overpic}[width=0.36\textwidth]{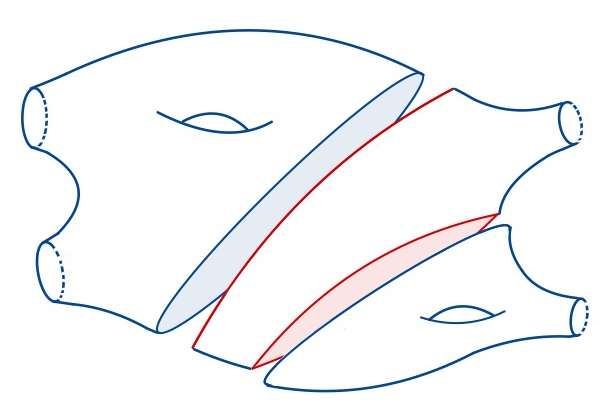}\put (1,3){$c)$}\end{overpic}
\caption{Graphical representation of the terms in Mirzakhani's recursion. Stripping off a pair-of-pants from a surface can lead to three scenarios: $a)$ the stripped surface has one boundary less than the original, but the genus stays the same, $b)$ it has one boundary more than the original, but the genus goes down by one, $c)$ the surface splits into two disconnected pieces, $M_1 \cup M_2$, such that the genera and external boundaries get distributed among $M_1$ and $M_2$.}
\label{strip}
\end{figure}

There is a structural similarity between the topological recursion \eqref{topologicalrecursion} and Mirzakhani's recursion relations \cite{Mirzakhani1, Mirzakhani2}. A detailed account of Mirzakhani's recursion goes beyond the scope of this appendix. For now we are content with a graphical representation of the recursion which expresses the Weil-Petersson volume of a hyperbolic surface in terms of smaller surfaces where a pair-of-pants has been stripped off. The terms appearing on the right-hand side of the topological recursion are packaged in a way similar to Mirzakhani's recursion, if we identify $g$ and $n$ with the genus and number of boundaries respectively. To make the comparison more transparent, we have labelled the terms by the three scenarios $\textcolor{blue}{a)}$, $\textcolor{blue}{b)}$ and $\textcolor{blue}{c)}$ which are depicted in figure \ref{strip}. The contact term $\textcolor{blue}{a)}$, which corresponds to the joining of two `baby universes', is incorporated in the topological recursion \eqref{topologicalrecursion} as the $(g_1,g_2) = (g,0)$ term of the primed sum. At each recursion step, this term is the only one that contains the Bergmann kernel $\mathcal{B}=\omega_{0,2}$.  

There is a special case in which the topological recursion is indeed equivalent to Mirzakhani's recursion. The initial data is precisely the JT spectral curve defined by equation \eqref{jtspectraalkromme} with Bergmann kernel given by
\beq
\mathcal{B}(z_1,z_2) = \fr{dz_1 \,dz_2}{(z_1-z_2)^2}~.
\eeq
Furthermore, $y(\tilde{z}) = y(-z) = -y(z)$ so the recursion kernel becomes
\begin{align}\label{eq:recursionkernel}
\mathcal{K}(z_0,z) &=\half  \int_{-z}^z \fr{1}{(z_0-z')^2}dz' \fr{dz_0 }{4  y(z) zdz}  =  \fr{1}{z_0^2-z^2} \fr{\pi}{\sin(2\pi z)} dz_0\otimes \del_z~.
\end{align}
We can expand the recursion kernel and take residues at the branch point $z=0$. Since the right-hand side of \eqref{topologicalrecursion} only contains poles of finite order, we only need to keep a finite number of terms in the expansion of $\mathcal{K}(z_0,z)$. 

It was proven that a Laplace transform of Mirzakhani's recursion gives the topological recursion in the case that the spectral curve is given by \eqref{jtspectraalkromme} (the proof can be found in appendix A of \cite{eynard1}). The relation between the Weil-Petersson volumes and the symplectic invariants for general $(g,n)$ is the following multi-Laplace transform:\footnote{The factor of $\half$ is related to the extra $\bb{Z}_2$-symmetry of the one-holed torus $(g,n) = (1,1)$ \cite{Mirzakhani1}. }
\beq
\mathcal{W}_{g,n}(z_1,\ldots,z_n)=2^{-\delta_{g,1}\delta_{n,1}}\int_0^\infty \prod_{i=1}^n d\ell_i \,\ell_i \,e^{-z_i \ell_i} V_{g,n}(\ell_1,\ldots,\ell_n)~.
\eeq
Given this relation, one can determine how the symplectic invariants are related to the JT partition functions $Z^\mathsf{c}_{g,n}(\beta_1,\ldots,\beta_n)$, as was used in the main text. 

\section{Baby universe Hilbert space and Virasoro constraints}\label{ch:recursion}

In this appendix, we give another perspective on the duality between KS theory and JT gravity by exploiting the relation to topological gravity, making the underlying Virasoro symmetry manifest. The approach is similar to the SD equation, but makes use of the operator formalism instead of path integrals and should be viewed as complementary. We show that one can reformulate the KS recursion relation in terms of a Virasoro constraint on the partition function, which is also satisfied by the Weil-Petersson volumes (when viewed as intersection numbers of the moduli space). This naturally leads to a definition of a baby universe Hilbert space. 

The Weil-Petersson volumes can be expressed in terms of the intersection theory of the moduli space via:
\beq
V_{g,n}(\bm{\ell}) =  \int_{\overline{\mathcal{M}}_{g,n}}\exp\Big(\Omega_{\mathrm{WP}} + \half \sum_{i=1}^n \psi_i \ell_i^2\Big)~,
\eeq
where $\bm{\ell}=(\ell_1,\ldots, \ell_n)$, $\Omega_{\mathrm{WP}}$ is the Weil-Petersson symplectic form on the moduli space and $\psi_i$ are $\psi$-classes on the moduli space. See, for example, \cite{DijkgraafWitten} for more details. Attaching the trumpets (we set $\phi_r = 1$ in \eqref{eq:bdycond} for convenience, which is different from the convention used in \eqref{eq:trumpettt}), and using the integral identity $\int_0^\infty x \exp(-\half ax^2)dx = a^{-1}$, we can write the genus $g$ contribution to the JT path integral with $n$ boundaries of lengths $\bm{\beta}=(\beta_1,\ldots, \beta_n)$ in the following form\footnote{All integral manipulations with the $\psi$-classes should be understood formally via the Taylor series of each function in the integrand. This is well-defined because the integral over $\overline{\mathcal{M}}_{g,n}$ only picks out the $d$-form from the Taylor expansion, where $d=6g-6+2n$ is the dimension of $\overline{\mathcal{M}}_{g,n}$.}:
\begin{flalign}
&&Z^\mathsf{c}_{g,n}(\bm{\beta}) &=  \int_0^\infty \prod_{i=1}^n d\ell_i \fr{\ell_i}{\sqrt{2\pi \beta_i}}e^{-\half \ell_i^2/\beta_i}  \int_{\overline{\mathcal{M}}_{g,n}}e^{\Omega_{\mathrm{WP}} + \half  \psi_i \ell_i^2} \nonumber &\\
&&\,&= \int_{\overline{\mathcal{M}}_{g,n}} e^{\Omega_{\mathrm{WP}}}  \int_0^\infty \prod_{i=1}^n d\ell_i \fr{\ell_i}{\sqrt{2\pi \beta_i}}\exp\left(-\half \left( \beta_i^{-1} -\psi_i\right)\ell_i^2\right) \nonumber &\\ \label{macroloop}
&&\,&=  \int_{\overline{\mathcal{M}}_{g,n}}  e^{\Omega_{\mathrm{WP}}}  \prod_{i=1}^n \sqrt{\fr{\beta_i}{2\pi}} (1-\beta_i\psi_i)^{-1}~.&
\end{flalign}
We can interpret this result in the following way. When we define JT gravity on hyperbolic surfaces without boundaries (allowing only marked points), we do not need to include any boundary terms in the action. So correlation functions in this theory are simply defined by integrating over the moduli space: 
\beq
\langle \cdots \rangle_g = \int_{\overline{\mathcal{M}}_{g,n}} e^{\Omega_{\mathrm{WP}}}( \cdots)~.
\eeq 
From \eqref{macroloop} we now see that one can create an asymptotically AdS$_2$ boundary of renormalized length $\beta$ by inserting the `observable'  
\beq
Z(\beta) = \lambda \sqrt{\fr{\beta}{2\pi}} (1-\beta \psi)^{-1}~.
\eeq
This point of view is familiar in the context of 2d topological gravity \cite{wittentopogravity1, wittentopogravity2,dijkgraafverlinde2}. Note that the observables have an extra factor of $\lambda$ compared to the main text. Summing over the genus counting parameter $\lambda = e^{-S_0}$, we can thus think of the JT gravity path integral as the correlation function of `boundary creation operators' $Z(\beta_i)$ in topological gravity:
\beq\label{topograv}
\mathcal{Z}^\mathsf{c}_{\chi<0}(\bm{\beta}) = \langle Z(\beta_1) \cdots Z(\beta_n) \rangle \sim \sum_{\chi<0} \lambda^{2g-2}  \langle Z(\beta_1) \cdots Z(\beta_n) \rangle_g~.
\eeq
The operators $Z(\beta_i)$ are called \emph{macroscopic loop operators} in the context of matrix models, see, for instance, \cite{seibergmicro} (and more recently \cite{okuyama}). We will now make the notation of \eqref{topograv} precise, by rewriting $Z(\beta)$ as a creation operator $Z_+(\beta)$ in a bosonic Fock space.

\subsection{Virasoro constraints}\label{virasoro}
To turn $Z(\beta)$ into a differential operator, we introduce the generating function $F$ for the intersection numbers of $\psi$-classes and the Weil-Petersson symplectic form\footnote{Sometimes, the convention is to replace $\Omega_{\mathrm{WP}}$ by $\kappa_1$. The $\kappa$-classes are related to the $\psi$-classes in the following way. Let $\pi :\overline{\mathcal{M}}_{g,n+1} \to \overline{\mathcal{M}}_{g,n}$ be the map that `forgets' the $(n+1)$-th marked point on a surface. Then we define $\kappa_d = \pi_*(\psi_{n+1}^{d+1})$ as the pushforward of $\psi_{n+1}^{d+1}$ to $\overline{\mathcal{M}}_{g,n}$. Wolpert \cite{wolpert} showed the remarkably simple relation that $ \Omega_{\mathrm{WP}} = 2\pi^2 \kappa_1$, so the conventions differ only by a numerical factor. Furthermore, one can add a parameter $s$ in front of $\Omega_{\mathrm{WP}}$ to generate $\kappa$-class intersections, but for JT gravity we only need $s=1$.}:
\beq
F(\bm{t}) = \sum_{g=0}^\infty \lambda^{2g-2} \int_{\overline{\mathcal{M}}_{g,n}} \exp(\Omega_{\mathrm{WP}} + \sum_i t_i \sigma_i )~,
\eeq
where the $t_i$ are `sources' for $\sigma_{d_i} \equiv \psi_i^{d_i}$. Expanding $Z(\beta)$ in a geometric series, we have:
\beq
Z(\beta) = \lambda \sqrt{\fr{\beta}{2\pi}} \sum_{k=0}^\infty (\psi \beta)^k  = \fr{\lambda}{\sqrt{2\pi}} \sum_{k=0}^\infty \sigma_{k} \beta^{k+\half}~.
\eeq
Therefore, we can write the connected JT path integral over the stable surfaces (with $\chi<0$) as differential operators acting on the generating function:
\beq
\mathcal{Z}^{\mathsf{c}}_{\chi<0}(\bm{\beta})= Z_+(\beta_1)\cdots  Z_+(\beta_n) \, F(\bm{t}) \, \Big \vert_{\substack{\bm{t}=0}}~,
\eeq
where
\beq
Z_+(\beta_i) = \fr{\lambda}{\sqrt{2\pi}} \sum_{k_i =0}^\infty \beta_i^{k_i+\half} \fr{\del}{\del t_{k_i}}~.
\eeq
We see that $F(\bm{t})$ is the generating function for \emph{connected} contributions to the JT path integral, and has the interpretation of a `free energy'. Therefore, its exponent $e^F$ has the interpretation of the full partition function of no-boundary JT gravity\footnote{With the exception of the disk and annulus.}. Acting with suitable trumpet creation operators $Z_+(\beta_i)$ on $e^F$ thus produces both connected and disconnected contributions to the path integral. The free energy $F$ generates just the spacetime wormholes, whereas $e^F$ contains both wormholes and \emph{factorized} contributions. For example, acting with $Z_+(\beta_1) Z_+(\beta_2)$ on $e^F$ produces, graphically, the two contributions in Figure \ref{disconnectedcorrelation}.
\begin{center}
\begin{figure}[h]
\hspace{1cm}
\begin{overpic}[width=0.3\textwidth]{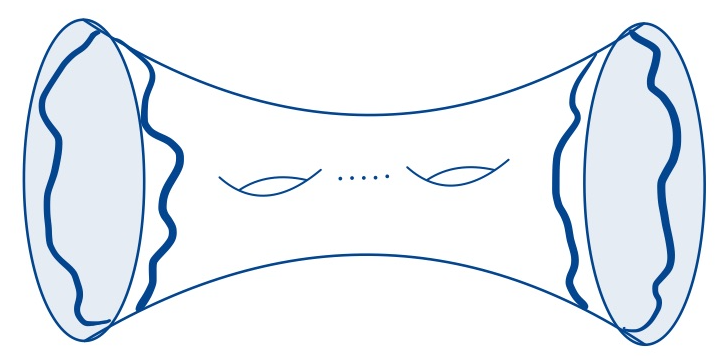} \put (-15,24) {$\sum_{g}$}\end{overpic}
\hspace{1.3cm}
\begin{overpic}[width=0.19\textwidth]{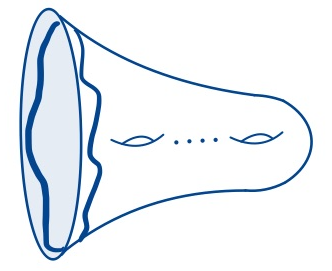} \put (-31,34) {{$\Huge \Big($}}  \put (-43,39) {+} \put (-20,39)  {$\sum_{h}$} \put (95,34) {{$\Huge \Big)$}} \end{overpic}
\hspace{1cm}
\begin{overpic}[width=0.19\textwidth]{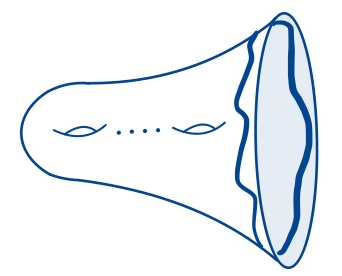} \put (-31,34) {{$\Huge \Big($}}  \put (-20,39) {$\sum_{h'}$} \put (95,34) {{$\Huge \Big)$}} \end{overpic}
\caption{Both wormholes and factorized contributions appear in $Z_+(\beta_1)Z_+(\beta_2)\exp F$.}
\label{disconnectedcorrelation}
\end{figure}
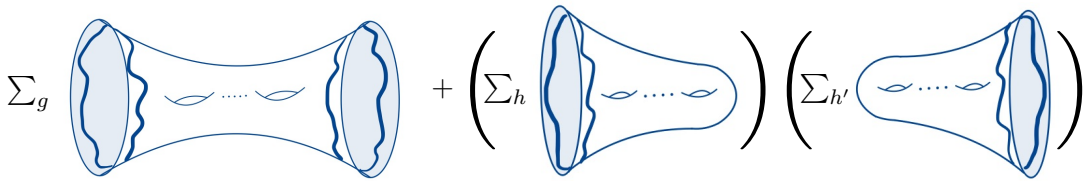
\end{center}

A natural question is how Mirzakhani's recursion can be rephrased in the operator language that we have just described. Since $F(\bm{t})$ can be used to generate Weil-Petersson volumes, a natural guess would be that the integral recursion relation of Mirzakhani can be written in a differential version as some combination of trumpet creation operators $Z_+(\beta_i)$ acting on the partition function:
\beq
\widehat{\mathcal{O}}\big[\{t_i, \del_{t_i}\}_i\big] \,e^{F(\bm{t})} = 0~.
\eeq
The precise form of the operator $\widehat{\mathcal{O}}$ can be found in \cite{safnuk}. An intermediate result reads:
\begin{align}
\quad (2k+3)!! \fr{\del F}{\del t_{k+1}}  &= \sum_{i,j=0}^\infty \underbrace{\mathcal{F}_{ijk} \,t_j \, \fr{\del F}{\del t_{i+j+k}}}_{\textcolor{blue}{a)}} \nonumber  \\ \label{diffrecursion}
&+\fr{\lambda^2}{4} \sum_{i=0}^\infty \sum_{\substack{j_1+j_2 = \\ i+k-1}} \widetilde{\mathcal{F}}_{i \,j_1j_2} \Big(\underbrace{\fr{\del^2 F}{\del t_{j_1}\del t_{j_2}}}_{\textcolor{blue}{b)}}  + \underbrace{\fr{\del F}{\del t_{j_1}} \fr{\del F}{\del t_{j_2}}}_{\textcolor{blue}{c)}} \Big) \qquad \forall\, k>0~.
\end{align}
The coefficients are given by:
\beq
\mathcal{F}_{ijk} = \fr{(2(i+j+k)+1)!!}{(2j-1)!!} \widetilde{u}_i~,
\eeq
\beq
\widetilde{\mathcal{F}}_{i\,j_1j_2} = (2j_1+1)!! (2j_2 +1)!! \, \widetilde{u}_i~,
\eeq
where the `moduli' $\widetilde{u}_i$ are defined as the Taylor coefficients of the function\footnote{A closed form can be given in terms of the Bernoulli number $B_{2i}$ by $\widetilde{u}_i = (-1)^{i-1}(2\pi)^{2i}(2^{2i+1}-4)\fr{B_{2i}}{(2i)!}$.}
\beq
 \fr{4\pi \sqrt{x}}{\sin (2\pi \sqrt{x})} \equiv \sum_{i=0}^\infty \widetilde{u}_i x^i~.
\eeq  
One proceeds by rewriting \eqref{diffrecursion} in terms of the partition function $e^F$, and rearranging terms into a single operator acting on $e^F$. To get rid of the double factorials, we rescale the source parameters $t_i$ as:
\beq
\mathfrak{t}_{2i+1} \equiv \fr{t_i}{(2i+1)!!}~.
\eeq
Including terms for $k=-1, 0$ (corresponding to the base cases $V_{0,3}$ and $V_{1,1}$ in Mirzhankani's recursion), equation \eqref{diffrecursion} is rewritten as:
\beq\label{virasoroconstraintz}
\mathcal{L}_k e^F = 0~,  \qquad \forall k\geq -1~,
\eeq
where the differential operators $\mathcal{L}_k$ are given by:
\begin{align}
\mathcal{L}_k &= -\underbrace{\half \fr{\del}{\del \mathfrak{t}_{2k+3}}}_{\textcolor{blue}{\mathrm{LHS}}} + \underbrace{\left(\fr{\mathfrak{t}_1^2}{\lambda^2} + \fr{\pi^2}{12}\right)}_{\textcolor{blue}{(1,1)}}\delta_{k,-1}  \, + \, \underbrace{\fr{\delta_{k,0}}{8}}_{\textcolor{blue}{(0,3)}}  \nonumber \\ \label{diffops}&+ \half \sum_{i,j=0}^{\infty}\!\!{\vphantom{\sum}}' (2j+1)\widetilde{u}_i \, \mathfrak{t}_{2j+1} \fr{\del}{\del  \mathfrak{t}_{2(i+j+k)+1}} + \fr{\lambda^2}{8}\sum_{i=0}^\infty \sum_{\substack{j_1+j_2 = \\ i+k-1}} \widetilde{u}_i \fr{\del^2}{\del \mathfrak{t}_{2j_1+1} \del \mathfrak{t}_{2j_2+1}}~.
\end{align}
The first term, labelled \textcolor{blue}{LHS}, comes from the left-hand side of \eqref{diffrecursion}. The next two terms, denoted by $\textcolor{blue}{(1,1)}$ and $\textcolor{blue}{(0,3)}$, arise from the torus with one hole and the pair-of-pants, respectively. The prime indicates that the term with $i=j=0$, $k=-1$ is excluded from that sum. The rest is just a rewriting of the right-hand side of \eqref{diffrecursion}. 

We would like to study the algebra associated to the infinite tower of differential equations imposed by $\{\mathcal{L}_k\}_{k\geq-1}$. The commutation relations are given by 
\beq
[\mathcal{L}_m, \mathcal{L}_n] = (m-n)\sum_{i=0}^\infty \widetilde{u}_i \mathcal{L}_{m+n+i}~.
\eeq
We now apply the simple transformation:
\beq \label{eq:Ltilde}
\widetilde{L}_k \equiv \sum_{i=0}^\infty u_i \mathcal{L}_{k+i}~,
\eeq
where $u_i$ are the reciprocal coefficients of $\widetilde{u}_i$, defined by:
\beq
\fr{\sin(2\pi \sqrt{x})}{4\pi \sqrt{x}} \equiv \sum_{i=0}^\infty u_i x^i~.
\eeq
One can show that the algebra spanned by the operators $\{\widetilde{L}_k\}$ with $k\geq -1$ is the Virasoro algebra
\beq\label{virasoroalgebra}
\big[\widetilde{L}_m, \widetilde{L}_n\big] = (m-n)\widetilde{L}_{m+n}~.
\eeq
The condition in \eqref{virasoroconstraintz} is therefore referred to as a \emph{Virasoro constraint}. The structure underlying Mirzakhani's recursion relation is a Virasoro symmetry, which expresses an underlying integrable structure, closely related to the Korteweg-de-Vries (KdV) hierarchy. For more background on its relation to intersection theory on the moduli space of Riemann surfaces, see \cite{wittentopogravity2, safnuk, liuxu, dijkgraafverlinde2}. 

\subsection{Chiral boson with a $\bb{Z}_2$ twist}\label{chiralbosontwist}
A crucial role in deriving a bosonic theory that describes JT gravity is played by the Laplace transform. Consider the boundary creation operator $Z_+(\beta)$, written in terms of the rescaled parameter $\mathfrak{t}_{2k+1}$:
\beq
Z_+(\beta) = \fr{\lambda}{\sqrt{2\pi}}\sum_{k=0}^\infty \fr{\beta^{k+\half}}{(2k+1)!!} \fr{\del}{\del \mathfrak{t}_{2k+1}}~.
\eeq
Using the identity $(2k+1)!! = \fr{2^{k+1}}{\sqrt{\pi}}\Gamma(k+3/2)$, (and setting back $\phi_r=1/2$, as was used in the main text) we find that the Laplace transform of $Z_+(\beta)$ can be written as:
\beq \label{lapl}
\int_0^\infty d\beta\, Z_+(\beta) \, e^{-\beta x}= \fr{\lambda}{2} \sum_{k=0}^\infty  x^{-k-3/2} \fr{\del}{\del \mathfrak{t}_{2k+1}}~.
\eeq
Since $\beta$ has the interpretation of a boundary length it is naturally defined on the positive real axis. The dual variable $x$ can then be taken as a complex `frequency'. The half-integer powers of $x$ appearing in \eqref{lapl}, show that the coordinate $x$ is only defined on the complex plane with a branch cut, the spectral plane. We will choose the convention that the branch cut lies on the negative real axis. When traversing a rotation of $2\pi$ around the branch point, $\sqrt{x}$ picks up a minus sign.

We want to interpret the Laplace transformed operator \eqref{lapl} as some complex scalar field $\Phi(x)$. By the above argument it should have anti-periodic boundary conditions across the branch cut:
\beq
\Phi(e^{2\pi i}x) = -\Phi(x)~.
\eeq
Therefore, it should be a \emph{$\mathbb{Z}_2$-twisted boson}. To make the correspondence precise, we introduce the following creation and annihilation operators for $k\geq 0$:
\beq\label{corresp}
\alpha_{k+\half} = \fr{\lambda}{2} \fr{\del}{\del \mathfrak{t}_{2k+1}}~, \qquad \alpha_{-k-\half} =\fr{2}{\lambda}\left(k+\half\right)\mathfrak{t}_{2k+1}~.
\eeq
These oscillators generate a representation of a twisted Heisenberg algebra. Namely, evaluating their commutator gives:
\beq\label{commu}
\big[\alpha_{n}, \alpha_{m}\big] =n \delta_{n+m,0}~, \qquad n,m \in \bb{Z}+\half~.
\eeq 
The twisted vacuum state $\ket{\sigma}$ is defined by requiring that:
\beq
\alpha_{k+\half} \ket{\sigma} = 0~, \quad \bra{\sigma}\alpha_{-k-\half} =0~, \qquad \forall k\geq 0~.
\eeq
It can be related to the vacuum of the untwisted free boson by the insertion of a twist operator $\sigma(x)$ \cite{ginsparg, dixonfriedan} at the origin and infinity:
\beq \ket{\sigma} = \sigma(0) \ket{0}~, \hspace{10pt} \bra{\sigma} = \bra{0}\sigma(\infty)~.\eeq 

The derivative of the field can be expanded in half-integer powers of $x$ as:
\beq 
\partial \Phi(x)= \sum_{k\in \bb{Z}} \alpha_{-k-\half} x^{k-\half}~.
\eeq
We split $\del \Phi(x)$ into positive and negative modes:
\beq
\del \Phi(x) = \del \Phi_-(x) + \del\Phi_+(x) =  \sum_{k=0}^\infty \alpha_{-k-\half} x^{k-\half} +  \sum_{k=0}^\infty \alpha_{k+\half} x^{-k-\fr{3}{2}}~.
\eeq
Then, we recognize the Laplace transform of the trumpet operator $Z_+(\beta)$ as the positive frequency part of $\del\Phi(x)$:
\beq
\int_0^\infty d\beta\, Z_+(\beta) \, e^{-\beta x}= \del \Phi_+(x)~.
\eeq
We see that adding a trumpet boundary of length $\beta_i$ in JT gravity corresponds to inserting $\del \Phi_+(x_i)$ at a point $x_i$ on the spectral plane, where $x_i$ and $\beta_i$ are related by the Laplace transform. We now want to relate the Virasoro constraints \eqref{virasoroconstraintz}, which we found to be equivalent to Mirzakhani's recursion, to the stress tensor of the twisted boson. 

The stress tensor $T(x)$ of the twisted theory is constructed via a normal ordering prescription:
\beq\label{normalordering}
T(x) = \half \big\{\del \Phi \del\Phi \big\}(x) \equiv \half \lim_{y\to x} \left(\del\Phi(x)\del\Phi(y) - \fr{1}{(x-y)^2}\right).
\eeq
Note that there are two notions of normal ordering. Firstly, there is the normal ordering at the level of modes, which puts all $\alpha_{-n-\half}$ to the left of the $\alpha_{n+\half}$, where $n\geq 0$. This respects the twisted vacuum $\ket{\sigma}$, and will be denoted by colons $:\!\cdots\!:$. Secondly, there is the normal ordering which subtracts the singular piece from the operator product expansion. This will be denoted by brackets $\big\{\cdots \big\}$.

The two-point function in the twisted vacuum is easily computed to be
\begin{align}
\braket{\del\Phi(x)\del\Phi(y)}_\sigma &=  \sum_{k=0}^\infty \sum_{n=0}^\infty \braket{\sigma | \big[\alpha_{k+\half}, \alpha_{-n-\half}\big] |\sigma} x^{-k-3/2} y^{n-1/2} \nonumber \\
&= \sum_{k=0}^\infty \left(k+\half\right) x^{-k-3/2} y^{k-1/2} = \half \fr{\sqrt{\fr{x}{y}} + \sqrt{\fr{y}{x}}}{(x-y)^2}~. \label{propagator}
\end{align}
This has the correct antiperiodicity in both $x$ and $y$, and it also exhibits the OPE singularity $\braket{\del\Phi(x)\del\Phi(y)} \sim (x-y)^{-2}$, as $x\to y$. From the definition of the stress tensor we find that
\beq
\braket{T(x)}_\sigma = \half \lim_{y\to x} \left\langle \del\Phi(x)\del\Phi(y) - \fr{1}{(x-y)^2}\right\rangle_\sigma = \fr{1}{16 x^2}~.
\eeq
Comparing to the other normal ordering prescription, whose expectation value is zero in the twisted vacuum by construction, we have
\beq
T(x) = \half :\! \del\Phi(x) \del\Phi(x)\!: + \fr{1}{16x^2}~.
\eeq
The stress tensor has the following mode expansion:
\beq\label{stressmodes}
T(x) = \sum_{n\in \bb{Z}} L_n x^{-n-2}~,
\eeq
where, for $n\geq -1$, we have 
\begin{align}
L_{-1} &=\sum_{k=1}^\infty \alpha_{-k-\half}\alpha_{k-\half} + \half (\alpha_{-\half})^2~,\\
L_0 &= \sum_{k=0}^\infty \alpha_{-k-\half}\alpha_{k+\half} + \fr{1}{16}~, \\
L_{n>0} &=  \sum_{k=0}^\infty \alpha_{-k-\half}\alpha_{n+k+\half} + \half \sum_{k=0}^{n-1}\alpha_{k+\half}\alpha_{n-k-\half}~.
\end{align}
These operators are related to the operators $\widetilde{L}_n$ in \eqref{eq:Ltilde} in the following way:
\beq\label{relate}
\widetilde{L}_n = L_n - \fr{1}{\lambda}\sum_{k=0}^\infty u_k \alpha_{k+n+\fr{3}{2}}~.
\eeq
To prove this, we rewrite the differential operators $\mathcal{L}_k$ of equation \eqref{diffops} in terms of the twisted creation and annihilation operators:
\begin{align}
\mathcal{L}_k &= - \underbrace{\fr{1}{\lambda}\alpha_{k+\fr{3}{2}}}_{\textcolor{blue}{\mathrm{LHS}}} + \underbrace{\left[ (\alpha_{-\half})^2 + \fr{\pi^2}{12}\right]}_{\textcolor{blue}{(1,1)}}\delta_{k,-1}  \, + \, \underbrace{\fr{\delta_{k,0}}{8}}_{\textcolor{blue}{(0,3)}} \nonumber\\
&\qquad +\, \sum_{i=0}^\infty \widetilde{u}_i \Big[\sum_{j=0}^\infty\!{\vphantom{\sum}}'  \alpha_{-j-\half} \alpha_{k+i+j + \half} + \fr{1}{2} \sum_{\substack{j_1+j_2 \\ =k+i-1}}  \alpha_{j_1+\half}\alpha_{j_2+\half}\Big]~.
\end{align}
We can incoorporate the terms multiplying $\delta_{k,0}$ and $\delta_{k,-1}$ into the sum over $i$, by realizing that $\delta_{k+i,-1}$ is only nonzero for $k=-1$ and $i=0$, while $\delta_{k+i,0}$ is nonzero for both $k=i=0$ and $k=-1, i=1$. Using the values for $\widetilde{u}_0 = 2$ and $\widetilde{u}_1 = \fr{4\pi^2}{3}$, we then find for $k\geq -1$:
\beq
 \sum_{i=0}^\infty \widetilde{u}_i \left[\half (\alpha_{-\half})^2\delta_{k+i,-1} + \fr{\delta_{k+i,0}}{16} \right] = \underbrace{\left[ (\alpha_{-\half})^2 + \fr{\pi^2}{12} \right]}_{\textcolor{blue}{(1,1)}}\delta_{k,-1}  \, + \, \underbrace{\fr{\delta_{k,0}}{8}}_{\textcolor{blue}{(0,3)}}~.
\eeq
Therefore, we recognize the operators $\mathcal{L}_k$ to be a simple transformation of the stress tensor modes:
\beq\label{stresstensormodes}
\mathcal{L}_k = - \fr{1}{\lambda}\alpha_{k+\fr{3}{2}} + \sum_{i=0}^\infty \widetilde{u}_i L_{k+i}~.
\eeq
Recalling the definition of the Virasoro operators $\widetilde{L}_n$, we thus prove equation \eqref{relate}:
\begin{align}\label{checkit}
\widetilde{L}_n &= \sum_{k=0}^\infty u_k \mathcal{L}_{k+n} = -\fr{1}{\lambda}\sum_{k=0}^\infty u_k \alpha_{k+n+\fr{3}{2}} + \sum_{k,i=0}^\infty u_k  \widetilde{u}_i L_{k+n+i} = L_n -\fr{1}{\lambda}\sum_{k=0}^\infty u_k \alpha_{k+n+\fr{3}{2}} ~,
\end{align}
where we used the fact that $u_i$ and $\widetilde{u}_i$ are reciprocal coefficients, 
$ \sum_{i=0}^k u_i \widetilde{u}_{k-i} = \delta_{k,0}$. 

We use the moduli $u_i$ to construct the following function:
\beq
\omega(x) =\frac{1}{\lambda}\sum_{k=0}^\infty u_k x^{k+\half} = \fr{1}{4\pi \lambda} \sin(2\pi \sqrt{x})~.
\eeq
We then define the shifted stress tensor $T_\omega(x)$ by translating $\del\Phi(x) \to \widetilde{\del \Phi}(x) = \del\Phi(x) - \omega(x)$:
\beq \label{eq:shiftedstresstensor}
T_\omega(x) \equiv \half \big\{\widetilde{\del\Phi} \,\widetilde{\del\Phi}\big\}(x) = T(x) - \omega(x) \del\Phi(x) + \half \omega(x)^2~.
\eeq
It has a mode expansion
\beq\label{modeexpansion}
T_\omega(x) = \sum_{n\in \bb{Z}} \widetilde{L}_n x^{-n-2}~,
\eeq
where the modes with $n\geq -1$ are precisely the Virasoro operators \eqref{relate}:
\begin{align}
\widetilde{L}_n &= \oint_0 \fr{dx}{2\pi i} x^{n+1} T_\omega (x) = \oint_0 \fr{dx}{2\pi i} x^{n+1} T(x) - \oint_0 \fr{dx}{2\pi i} x^{n+1} \omega(x) \del\Phi(x) \nonumber \\
&= L_n - \fr{1}{\lambda}\sum_{k,j=0}^\infty u_k\, \alpha_{-j-\half} \oint_0 \fr{dx}{2\pi i} x^{n+k+j+1} = L_n - \fr{1}{\lambda}\sum_{k=0}^\infty u_k \alpha_{k+n+\fr{3}{2}}~.
\end{align}
We can think of $\omega(x)$ as giving $\widetilde{\del\Phi}(x)$ a vacuum expectation value. Note that this VEV was chosen in a particular way to match Mirzakhani's recursion, but more generally we could take any set of moduli $u_i$, and treat $\omega(x)$ as a formal power series. The Virasoro constraints may then still have a geometric interpretation\footnote{In fact, much recent progress in this direction has been made in \cite{kontsevichsoibelman}, where so-called \emph{quantum Airy structures} are introduced as a generalizations of the Virasoro constraints (for an introduction to the subject, see \cite{borot1}).}, although it will not describe JT gravity. Indeed, various choices of $\omega(x)$ have been related to the generalized Kontsevich-Witten model \cite{alexandrov1, CFTtoporecursion}, topological gravity on arbitrary backgrounds \cite{dijkgraafverlinde2, DijkgraafWitten} and minimal models \cite{Seiberg, fukuma}. 

We write the generating function in the coherent state basis:
\beq
e^{F(\bm{t})} = \braket{t | \Sigma}~,
\eeq 
where the coherent state is expressed as $\bra{t} = \bra{\sigma} e^V$
with
\beq
V = \fr{2}{\lambda} \sum_{k=0}^\infty  \mathfrak{t}_{2k+1}\alpha_{k+\half}~.
\eeq
The coherent state $\bra{t}$ is a left eigenstate of the annihilation operator $\alpha_{-n-\half}$:
\beq
\bra{t} \alpha_{-n-\half} = \lambda^{-1}(2n+1)\mathfrak{t}_{2n+1} \bra{t}~.
\eeq
Therefore, when acting on $e^F$ with an operator $\mathcal{O}(\mathfrak{t}_n,\del_{\mathfrak{t}_n})$, we can bring it inside the `expectation value' $\braket{t|\dots |\Sigma}$ by converting it into oscillator language $\mathcal{O}(\alpha_{-n},\alpha_n)$. The precise relation is given by \eqref{corresp}.
 The state $\ket{\Sigma}$ is then fully determined by the Virasoro constraint:
 \beq\label{vira}
\widetilde{L}_n e^F = 0 \quad \iff \quad \widetilde{L}_n \ket{\Sigma} = 0, \qquad n\geq -1~,
 \eeq
 where on the left-hand side it is implied that $\widetilde{L}_n$ is written in terms of $\mathfrak{t}$ and $\fr{\del}{\del \mathfrak{t}}$, and on the right-hand side $\widetilde{L}_n$ is expressed in terms of twisted bosonic oscillators.
 
Looking at the mode expansion \eqref{modeexpansion} of $T_\omega(x)$, we see that the modes with $n\geq -1$ correspond to the negative powers of $x$. So instead of the infinite number of equations imposed by \eqref{vira}, we can write the Virasoro constraint as a single requirement that the expectation value of $T_\omega(x)$ is non-singular as $x\to 0$:
\beq \label{virconstraint}
 \braket{t| T_\omega(x) |\Sigma} = \mathrm{analytic~.}
\eeq
It is simply the requirement that the theory is conformally invariant at the quantum level. This is a non-trivial requirement, because in terms of the spectral variable $x$, the twisted boson $\widetilde{\del\Phi}(x)$ has a branch point at the origin. The branch point breaks the conformal invariance, and the modes of $\del\Phi$ should be `dressed' to restore conformal invariance \cite{CFTtoporecursion}. The dressing is defined through the new $SL(2,\bb{C})$-invariant state $\ket{\Sigma}$, which we can formally write as some `dressing operator' $e^{\lambda \hat{S}} \ket{\sigma}$ acting on the twisted vacuum. We would like to think of $\hat{S}$ as an interaction term in an interacting theory, with coupling constant $\lambda$, which perturbs the free theory.

In the case of the so-called \emph{topological point}, for which $\omega(x) \sim \sqrt{x}$, such an operator $\hat{S}$ was explicitly constructed in \cite{alexandrov2}. It was found to be cubic in the bosonic oscillators $\alpha_k$. For general $\omega(x)$, one can always use appropriate shift operators $V_\omega$ to obtain the solution: 
\beq
\ket{\Sigma} = e^{V_\omega} e^{\lambda \hat{S}}\ket{\sigma}~.
\eeq 

\subsection{Back to JT gravity}
We have seen that the full JT path integral on connected stable surfaces with $n$ boundaries could be obtained by acting $n$ times with a boundary creation operator $Z_+(\beta_i)$ on the free energy $F(\bm{t})$. After an $n$-fold Laplace transform this can be written as
\beq\label{afterlaplace}
\int_0^\infty \prod_{i=0}^n d\beta_i e^{-\beta_i x_i} \mathcal{Z}_{\chi<0}^\mathsf{c}(\bm{\beta}) = \del\Phi_+(x_1)\cdots \del\Phi_+(x_n) F(\mathfrak{t}) \Big\vert_{\mathfrak{t}=0}~.
\eeq 
We can use the coherent state $\bra{t}$ to bring $\del\Phi_+(x)$ inside the correlation function $\braket{t|\dots|\Sigma}$. This puts all the dependence on the sources into $\bra{t}$, and so setting $\mathfrak{t} = 0$ boils down to replacing $\bra{t}$ by the vacuum $\bra{\sigma}$. For the JT gravity path integral, including connected and disconnected spacetimes, this implies:
\beq\label{operatorrepresentation}
\int_0^\infty \prod_{i=0}^n d\beta_i e^{-\beta_i x_i} \mathcal{Z}_{\chi<0}(\bm{\beta}) = \fr{\braket{\sigma | \del\Phi_+(x_1)\cdots \del\Phi_+(x_n) |\Sigma}}{\braket{\sigma |\Sigma}}~.
\eeq 
Here, we divided by the normalization factor $\braket{\sigma | \Sigma}$, which has the effect of excluding the contributions from JT universes without boundaries. Note that the dependence on the background $\omega(x)$ is encoded in the state $\ket{\Sigma}$. 

\begin{figure}[h]
\centering
\begin{overpic}[width=0.5\textwidth]{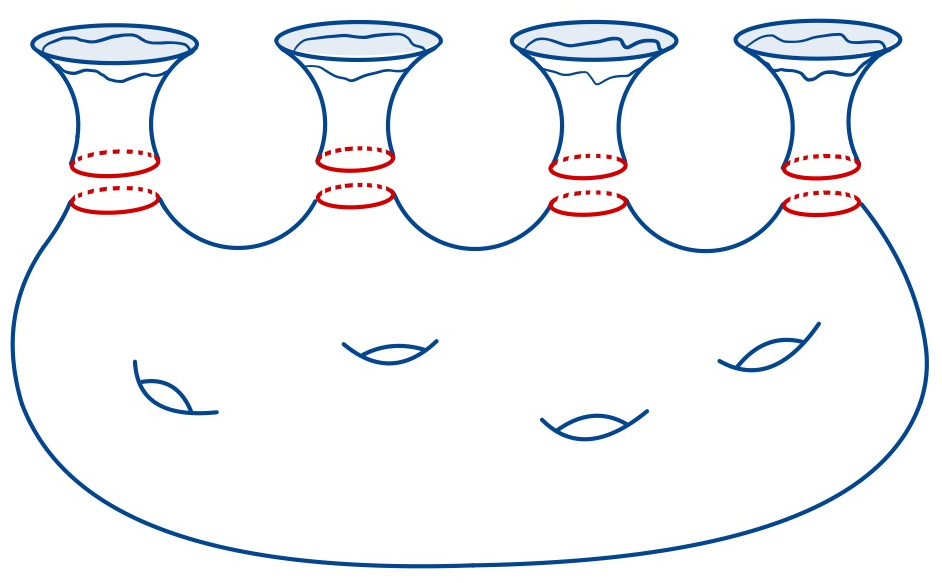} \put (-10,50) {{\Large $\bra{\bm{\beta}}$}} \put (102,25) {{\Large $\ket{\Sigma}$}}\end{overpic}
\caption{An example inner product in the one-sided baby universe Hilbert space: a genus 4 Riemann surface is prepared by $\ket{\Sigma}$, and the overlap is computed with a 4-trumpet state.}
\label{onesidedproposal}
\end{figure}

Let us interpret this formula from the point of view of the baby universe Hilbert space. The operator insertions of $\del\Phi(x)$ are radially ordered, and so one may think of equation \eqref{operatorrepresentation} as an operator representation of the JT path integral in \emph{radial quantization}. So we should read it from right to left: one in the twisted vacuum $\ket{\sigma}$, which can be thought of as a type of `Big Bang' for a (possibly disconnected) universe. In particular, the initial state has no boundaries. Then, there is a complicated splitting and joining of baby universes, determined by the interaction $\hat{S}$ in $\ket{\Sigma} = e^{\lambda \hat{S}}\ket{\sigma}$. The requirement that these processes respect modular invariance is imposed by the Virasoro constraint on $\ket{\Sigma}$. Finally, the trumpet boundaries are glued to the spacetime prepared by $\ket{\Sigma}$. The gluing is represented by the overlap 
\beq \mathcal{Z}_{\chi<0}(\bm{\beta}) = \braket{\beta_1,\dots,\beta_n|\Sigma}, \eeq 
where the normalized $n$-trumpet state $\bra{\bm{\beta}}$ is given by:
\beq 
\bra{\beta_1,\dots,\beta_n} =  \fr{\bra{\sigma} Z_+(\beta_1)\cdots Z_+(\beta_n)}{\braket{\sigma | \Sigma}}~.
\eeq
This intuitive interpretation of the formula \eqref{operatorrepresentation} is depicted in Figure \ref{onesidedproposal}. It corresponds to the particular (Euclidean) time slicing in the universe field theory in which all the asymptotically AdS$_2$ boundaries are in the infinite future. The upshot of this (Euclidean) time-slicing is that it is manifestly invariant under the large diffeomorphisms. However, the downside is that the in and the out-state are treated asymmetrically, contrary to the proposal of \cite{marolfmaxfield}: the state $\ket{\Sigma}$ is prepared from one side. Therefore, we could refer to the Hilbert space generated by the states above as the \emph{one-sided} baby universe Hilbert space. 

\subsection{The disk and the annulus}
So far, we have mainly focused on the stable $(\chi<0)$ hyperbolic surfaces. However, there are two special contributions to the Euclidean path integral coming from the disk, which represents the classical solution to the JT equations of motion, and the double-trumpet.

The double-trumpet corresponds after Laplace transform to the free two-point function that we computed in \eqref{propagator}, and the disk corresponds to $\omega(x)$, which we interpreted as the classical VEV of the twisted boson. To get the full (disconnected) JT path integral, one should consider expectation values of insertions of the full $\widetilde{\del \Phi}(x)$ on the spectral plane. To implement the shift by $\omega(x)$ we introduce a shift operator:
\beq
V_\omega = \oint_0 \fr{dx}{2\pi i} \Phi(x) \omega(x) = -\fr{1}{\lambda} \sum_{k=0}^\infty \fr{u_k}{k+\fr{3}{2}}  \alpha_{k+\fr{3}{2}}~.
\eeq
We see that translation of $\del\Phi$ by $\omega$ is implemented by conjugating with $\exp V_\omega$:
\beq
e^{-V_\omega} \del\Phi(y) \,e^{V_\omega} = \del\Phi(y) - [V_\omega, \del\Phi(y)] = \del\Phi(y) - \omega(y) = \widetilde{\del\Phi}(y)~.
\eeq 
We can use the shift operator to move around the dependence on $\omega(x)$ inside correlation functions:
\beq
\braket{\sigma | \widetilde{\del \Phi}(x_1)\cdots \widetilde{\del \Phi}(x_n) |\Sigma} = \braket{\sigma | e^{-V_\omega} \del \Phi(x_1)\cdots \del \Phi(x_n) e^{V_\omega} |\Sigma}~.
\eeq
Denoting 
\beq 
\bra{\omega} = \bra{\sigma}e^{-V_\omega} \quad \mathrm{and} \quad \ket{\Sigma_0} = e^{V_\omega} \ket{\Sigma}~,
\eeq 
the full $n$-boundary JT gravity partition function, including disk and annulus contributions, can be written as the following integral transform of an $n$-point correlation function of twisted bosons on the spectral plane:
\beq \label{fullJTpathintegral}
\mathcal{Z}(\bm{\beta}) = \int_{c-i\infty}^{c+i\infty} \prod_{i=1}^n \fr{dx_i}{2\pi i} e^{\beta_i x_i}  \fr{\braket{\omega |\del \Phi(x_1)\cdots \del \Phi(x_n) |\Sigma_0}}{\braket{\omega| \Sigma_0}}~.
\eeq
For the positive frequencies $\del\Phi_+$, the formula coincides with equation \eqref{operatorrepresentation}, because $e^{V_\omega}$ commutes with $\del\Phi_+$. For the negative frequencies, however, we get precisely the disk and annulus contributions that were missing in \eqref{operatorrepresentation}. 

\begin{figure}[h]
\centering
\vspace{1.5em}
\begin{overpic}[width=\textwidth]{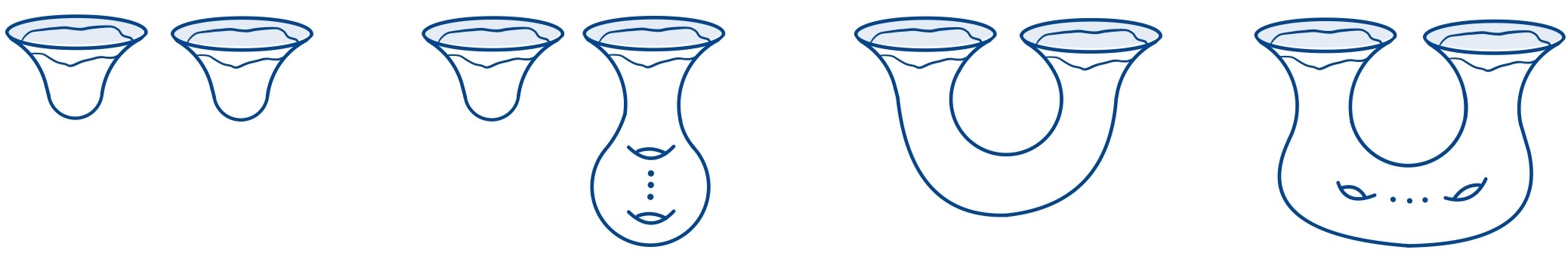}\put (23,12){$+$} \put (50,12){$+$} \put (76,12){$+$} \put (0,17) {$\overbrace{\hspace{8em}}^{\textcolor{blue}{1}}$} \put (26,17) {$\overbrace{\hspace{8em}}^{\textcolor{blue}{2}}$} \put (54,17) {$\overbrace{\hspace{8em}}^{\textcolor{blue}{3}}$} \put (80,17) {$\overbrace{\hspace{8em}}^{\textcolor{blue}{4}}$} \end{overpic}
\caption{The terms in \eqref{twopoint} correspond to distinct geometries in the JT path integral. Term $\textcolor{blue}{1}$ corresponds to two disks. Term $\textcolor{blue}{2}$ is the disconnected contribution of a disk and a sum over genus $g > 0$ Riemann surfaces with one trumpet boundary. Term $\textcolor{blue}{3}$ is the genus zero wormhole contribution, the double trumpet. And $\textcolor{blue}{4}$ is the sum over connected stable spacetime wormholes, $\mathcal{Z}_{\chi<0}(\beta_1,\beta_2)$, with two trumpet boundaries.}
\label{twopointfunction}
\end{figure}
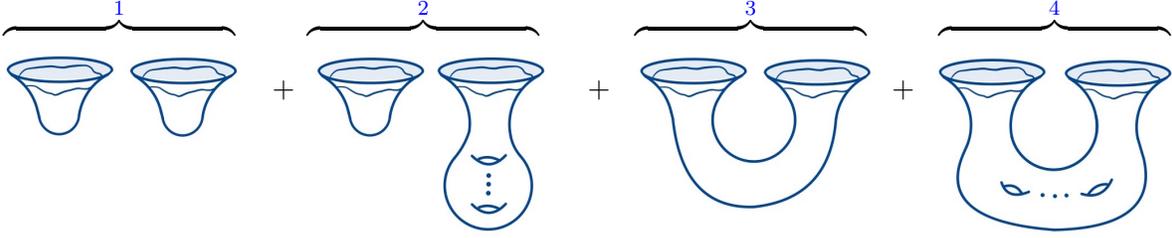

We illustrate this with an instructive example. Consider the two-point function
\beq
\braket{\del\Phi(x)\del\Phi(y)}_\Sigma \equiv \fr{\braket{\omega | \del\Phi(x) \del\Phi(y)| \Sigma_0}}{\braket{\omega | \Sigma_0}}~.
\eeq 
If we split $\del\Phi$ into positive and negative frequencies, and commute all negative modes to the left, we get four contributions. Using that $\bra{\omega}$ is a left eigenstate of $\del\Phi_-$ with eigenvalue $\omega$, and computing the commutator
\beq
\big[\del\Phi_+(x),\del\Phi_-(y)\big] =  \half \fr{\sqrt{\fr{x}{y}} + \sqrt{\fr{y}{x}}}{(x-y)^2}~,
\eeq
we find for the normalized full two-point function:
\begin{align}
\fr{\braket{\omega| \del\Phi(x) \del\Phi(y) | \Sigma_0}}{\braket{\omega | \Sigma_0}} &= \underbrace{\omega(x)\omega(y)}_{\textcolor{blue} 1} + \underbrace{\omega(x)\braket{\del\Phi_+(y)}_\Sigma + \braket{\del\Phi_+(x)}_\Sigma \omega(y)}_{\textcolor{blue} 2}\nonumber \\ \label{twopoint}& \qquad + \underbrace{\braket{\del\Phi(x)\del\Phi(y)}_\sigma}_{\textcolor{blue} 3} + \underbrace{\braket{\del\Phi_+(x) \del\Phi_+(y)}_\Sigma}_{\textcolor{blue} 4}~.
\end{align}
The terms have been labelled by the type of geometry that they represent in the JT path integral, see Figure \ref{twopointfunction}.

Let us now explicitly check that this correctly reproduces the JT gravity results. For the boundary creation operators, we have already seen that the Laplace transform of $Z_+(\beta)$ gives $\del\Phi_+(x)$, and so the integral transform 
\beq\label{integraltransform}
\del\Phi_+\mapsto  \int_{c-i\infty}^{c+i\infty} \fr{dx}{2\pi i} \,e^{\beta x} \del\Phi_+(x)
\eeq
is the standard inverse Laplace transform. However, there is no function\footnote{It can only be defined in a distributional sense, with the help of delta functions.} whose Laplace transform gives the negative modes $\del\Phi_-(x)$. Nonetheless, the integral transform in \eqref{integraltransform} may still exist. In such a case, we will still call it an `inverse Laplace transform', even though it is not actually the inverse of a convergent Laplace transform. 
\begin{figure}[h]
\centering
\begin{tikzpicture}
\path[color=navy,draw,line width=0.8pt,postaction=decorate, decoration={markings,
mark=at position 2cm with {\arrow[line width=1pt]{>}},
mark=at position 8cm with {\arrow[line width=1pt]{>}},
mark=at position 10.5cm with {\arrow[line width=1pt]{>}},
mark=at position 13.5cm with {\arrow[line width=1pt]{>}},
mark=at position 16cm with {\arrow[line width=1pt]{>}}
}
] +(2,-2.2) -- (2,2) -- (0.5,2) arc (90:180:2)  -- (1,0) arc(150:-150:0.2) -- (-1.5,-0.2) arc(-180:-90:2) -- (2,-2.2);
\node at (1.16,-0.1)[circle,fill,inner sep=1pt]{};
\node at (1.16,2.7){Im$(x)$};
\node at (3.5,-0.1){Re$(x)$};
\draw[decoration = {zigzag,segment length = 2mm, amplitude = 0.5mm},decorate] (1.16,-0.1)--(-2,-0.1);
\draw[gray,thin] (1.16,2.5) -- (1.16,-2.5);
\draw[gray,thin] (1.16,-0.1) -- (3,-0.1);

\end{tikzpicture}

\caption{The keyhole contour for evaluating the inverse Laplace transform of $\omega(x)$.}
\label{contour}
\end{figure}
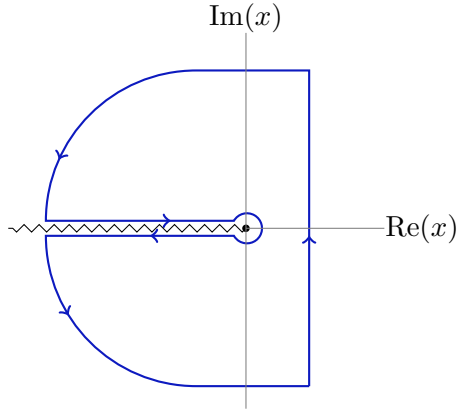

\paragraph{The disk.} Indeed, we want to compute the integral
\beq\label{integraltransform2}
 \int_{c-i\infty}^{c+i\infty} \fr{dx}{2\pi i} \,e^{\beta x} \omega(x) =  \int_{c-i\infty}^{c+i\infty} \fr{dx}{2\pi i} \,e^{\beta x} \fr{\sin(2\pi \sqrt{x})}{4\pi \lambda}~.
\eeq
Since we chose the branch cut of $\omega(x)$ to lie on the negative real axis, $e^{\beta x}$ should grow in the right-half plane\footnote{For the integral to be non-zero, the exponential $e^{\beta x}$ needs to grow in the region of the complex plane where the integrand is holomorphic. If this were not the case, we could close the integration contour in that region and conclude that the integral would be zero.}. So the condition is that $\beta >0$. The small offset $0<c\ll 1$ is included to avoid the branch point at $x=0$. To compute the integral over the offset imaginary axis, we then close the contour in the counter-clockwise direction into a so-called \emph{keyhole contour}, shown in Figure \ref{contour}.

Since the integrand is holomorphic inside the keyhole contour, the total contour integral should vanish. As we take the radius of the outer arcs to infinity, the radius of the inner arc to zero, and $c$ to zero, the only non-zero contributions to the integral come from the path along $(-i\infty, i\infty)$ and the two paths just above and below the negative real axis. Therefore, we can express the integral \eqref{integraltransform2} as
\begin{align}
 \int_{c-i\infty}^{c+i\infty} \fr{dx}{2\pi i} \,e^{\beta x} \omega(x) = \int_0^\infty \fr{dx}{2\pi i} e^{-\beta x} \omega(-x+i\epsilon) - \int_0^\infty \fr{dx}{2\pi i} e^{-\beta x} \omega(-x-i\epsilon)~,
\end{align}
where have sent $x \to -x$. Now as we take $\epsilon \to 0$, the second term gets an extra minus sign from the discontinuity of the square root across the branch cut. Using that $\sin(2\pi i \sqrt{x}) = i \sinh(2\pi \sqrt{x})$, we obtain:
\beq
 \int_{c-i\infty}^{c+i\infty} \fr{dx}{2\pi i} \,e^{\beta x} \omega(x) = 2 i \int_0^\infty \fr{dx}{2\pi i} e^{-\beta x} \fr{\sinh(2\pi \sqrt{x})}{4\pi \lambda}~.
\eeq
This integral can be evaluated using Gaussian integration. Upon setting $x = z^2$, and writing $\lambda^{-1}= e^{S_0}$, we precisely retrieve the disk partition function:
\beq
 \int_{c-i\infty}^{c+i\infty} \fr{dx}{2\pi i} \,e^{\beta x} \omega(x) = \fr{e^{S_0}}{4\pi^{1/2} \beta^{3/2}} e^{\pi^2/\beta} = Z_{\mathsf{disk}}(\beta)~.
\eeq 
\paragraph{The annulus.} We now show that \eqref{fullJTpathintegral} reproduces the correct result for the double-trumpet partition function. We want to compute
\beq
\int_{c-i\infty}^{c+i\infty} \fr{dx}{2\pi i} \fr{dy}{2\pi i} e^{\beta_1 x + \beta_2 y}\braket{\del\Phi(x)\del\Phi(y)}_\sigma = \int_{c-i\infty}^{c+i\infty} \fr{dx}{2\pi i} \fr{dy}{2\pi i} e^{\beta_1 x + \beta_2 y} \fr{\half  \left(\sqrt{\fr{x}{y}} + \sqrt{\fr{y}{x}}\right)}{(x-y)^2}~.
\eeq
Again we can close both the $x$- and $y$-contours in the left-half plane via a keyhole contour. For a fixed value of $x$, the $y$-contour may enclose a pole at $x=y$, but the residue of this pole is zero: 
\beq \mathrm{Res}_{x \to y} \langle \del\Phi(x)\del\Phi(y) \rangle_\sigma = \lim_{x\to y}\fr{1}{4}\fr{\del}{\del x} \left(\sqrt{\fr{x}{y}} + \sqrt{\fr{y}{x}} \right) = 0~.
\eeq 
So we can express the integral along the imaginary axis in terms of the discontinuity across the negative real axis. Making the substitution $x=-z^2$, $y=-w^2$, we then find:
\begin{align}\label{zwintegral}
\int_{c-i\infty}^{c+i\infty} \fr{dx}{2\pi i} \fr{dy}{2\pi i} e^{\beta_1 x + \beta_2 y}\braket{\del\Phi(x)\del\Phi(y)}_\sigma &= -\fr{1}{4\pi^2}\int_{-\infty}^\infty dz dw \, e^{-\beta_1 z^2 - \beta_2 w^2} \fr{z^2+w^2}{(z^2-w^2)^2}~.
\end{align}
We can split the integrand as a sum of two terms
\beq\label{splitintegrand}
\fr{z^2+w^2}{(z^2-w^2)^2} = \half \left[\fr{1}{(z-w)^2} + \fr{1}{(z+w)^2}\right]~,
\eeq
and then notice that both terms give the same integral, upon sending $w \to -w$ in the second term.  
Expanding $(z-w)^{-2}$ as a power series, it can be shown that the double integral in \eqref{zwintegral} precisely gives the double trumpet partition function $Z_{0,2}^\mathsf{c}(\beta_1,\beta_2)$.

This completes the proof of the dictionary \eqref{fullJTpathintegral} between the twisted boson and the full $n$-boundary JT gravity path integral. 

\subsection{Topological recursion in the twisted boson formalism}
Using our dictionary between $\mathcal{Z}^\mathsf{c}$ and operator insertions $\braket{\prod_{i=1}^n \del\Phi(x_i)}_\Sigma$, we derive that the symplectic invariants $\mathcal{\omega}_{g,n}$ can be expressed as the following correlation functions:
\begin{flalign}\label{wdictionary}
\mathcal{W}_{g,n}(\bm{z}) &= \braket{\del\Phi(z_1)\cdots \del\Phi(z_n)}^{(g)}_{\Sigma, \mathsf{c}}~,
\end{flalign}
for $(g,n)\neq (0,2)$\footnote{In defining $\mathcal{W}_{0,2}$ we subtract the singular part of the OPE
\begin{flalign}\label{normalord}
\mathcal{W}_{0,2}(z,w) =\braket{\del\Phi(z)\del\Phi(w)}^{(g=0)}_{\Sigma,\mathsf{c}} - \fr{1}{(z-w)^2}~. 
\end{flalign}
This finite renormalization is directly related to the normal ordering prescription of the twisted stress tensor.}. Here, the subscript $\mathsf{c}$ means that we take the connected correlation function, and the superscript $g$ denotes the order $\lambda^{2g-2}$ contribution. We have summarized the relations between the various quantities in Figure \ref{diagram}.
\begin{figure}[h]
\centering
\begin{tikzpicture}
\node at (0,2){$V_{g,n}(\bm{\ell})$};
\draw[thick,->] (0,1.5) -- (0,0.5) node[midway, left] {\small Trumpet};
\node at (0,0){$Z_{g,n}(\bm{\beta}\,)$};
\draw[thick,->] (1,0) -- (3,0) node[midway, above] {\small Laplace};
\node at (4.7,0){$\Big\langle\prod_{i=1}^n \del\Phi(x_i)\Big\rangle^{(g)}_\Sigma$};
\draw[thick,->] (1,2) -- (3,2) node[midway, above] {\small Laplace};
\node at (4,2){$\mcal{W}_{g,n}(\bm{z})$};
\draw[thick,->] (4,0.5) -- (4,1.5) node[midway, right] {\small Double cover};
\end{tikzpicture}
\caption{Relations between Weil-Peterson volumes $V_{g,n}$, symplectic invariants $\mcal{W}_{g,n}$, twisted boson correlators, and Dirichlet-Dirichlet JT partition functions.}
\label{diagram}
\end{figure}

The operators $\del\Phi(z)$ are related to the twisted boson on the spectral plane $\del\Phi(x)$ by substituting $x=z^2$ and performing a coordinate transformation as 1-forms:
\beq
\del\Phi(z) dz = \del\Phi(x)dx~.
\eeq
Since $\Phi(x)$ is expanded in half-integer powers of $x$, $\Phi(z)$ is expanded in odd powers of $z$. Taking into account the extra factor of $z$ from $dx = 2zdz$, we see that $\del\Phi(z)$ is even. 

To make contact with the Virasoro constraints, we will rewrite the topological recursion in a more transparent way. First, we introduce a generating function for insertions of $\del\Phi(z)$ on the double cover:
\beq\label{generatingfunc}
Z_\Sigma[\mu] = \left\langle \exp \oint_0 \fr{dz}{2\pi i} \mu(z) \del\Phi(z) \right\rangle_\Sigma~.
\eeq
We split $\del\Phi(z)$ into positive and negative modes $\del\Phi_+(z)$ and $\del\Phi_-(z)$. Using that $e^{A+B} = e^{A}e^B e^{-\half [A,B]}$ for $[A,B]$ central, we can split the exponential inside the generating function \eqref{generatingfunc} as:
\beq
Z_\Sigma[\mu] =\left\langle \omega \left\vert e^{\oint \fr{dz}{2\pi i} \mu(z)\del\Phi_-(z)} e^{\oint \fr{dz}{2\pi i} \mu(z) \del\Phi_+(z)} e^{\half \oint \fr{dz_1}{2\pi i} \fr{dz_2}{2\pi i} \mu(z_1)\mu(z_2) \big[\del\Phi_+(z_1), \del\Phi_-(z_2)\big]} \right\vert \Sigma_0\right\rangle~.
\eeq
The commutator between positive and negative modes gives 
\beq
\big[\del\Phi_+(z), \del\Phi_-(w)\big] = \fr{1}{(z-w)^2} + \fr{1}{(z+w)^2} = \braket{\del\Phi(z)\del\Phi(w)}_\sigma~,
\eeq
and $\del\Phi_-(z)$ also pick up a commutator from acting on $\bra{\omega} = \bra{\sigma}e^{V_\omega}$:
\beq
\big[V_\omega, \del\Phi_-(z)\big] = \omega(z)~.
\eeq
We can therefore write the logarithm of the generating function as:
 \begin{empheq}{align}
\log Z_\Sigma [\mu] &=\underbrace{ \oint_0 \fr{dz}{2\pi i} \mu(z)\omega(z)}_{\textcolor{blue}{\mathrm{Disk}}} + \underbrace{\half \oint_0 \fr{dz_1}{2\pi i} \fr{dz_2}{2\pi i} \mu(z_1)\mu(z_2) \braket{\del\Phi(z_1)\del\Phi(z_2)}_\sigma}_{\textcolor{blue}{\mathrm{Annulus}}}  \nonumber\\
&\hspace{5cm}+ \underbrace{\log \left\langle \exp \oint_0 \fr{dz}{2\pi i} \mu(z) \del\Phi_+(z)\right\rangle_{\!\Sigma}}_{\textcolor{blue}{\mathrm{Stable}}} ~.
\end{empheq}
The first two terms represent the connected contributions from the disk and the annulus, and the last term contains the contributions from all the stable $(\chi <0)$ surfaces. So, we see that the connected correlation functions can be expressed as functional derivatives of $W_\Sigma[\mu] \equiv \log Z_\Sigma[\mu]$ with respect to the sources:
\begin{align}
\braket{\del\Phi(z_1)\cdots \del\Phi(z_n)}_{\Sigma, \mathsf{c}}&= \fr{\delta}{\delta \mu(z_1)} \cdots \fr{\delta}{\delta \mu(z_n)} W_\Sigma [\mu] \Big\vert_{\mu=0} \nonumber  \\[1em]
&= \omega(z) \delta_{n,1} + \braket{\del\Phi(z_1)\del\Phi(z_2)}_\sigma \delta_{n,2} + \braket{\del\Phi_+(z_1) \cdots \del\Phi_+(z_n)}_{\Sigma,\mathsf{c}}~.
\end{align}
We can use the generating function $Z_\Sigma[\mu]$ to directly show the equivalence between the Virasoro constraints and the topological recursion. The topological recursion can be written as a functional differential equation for the generating functional of connected correlation functions $W_\Sigma[\mu]$:
\beq\label{TRdiffeq}
\fr{\delta W_\Sigma}{\delta \mu(z_0)}\Big\vert_{\chi<0} =\,\, \underset{z \to 0}{\mathrm{Res}} \,\fr{\braket{\del\Phi(z_0)\Phi(z)}_\sigma}{2\,\omega(z)} \fr{1}{2} \left[ \fr{\delta W_\Sigma}{\delta \mu(z)} \fr{\delta W_\Sigma}{\delta \mu(z)} + \fr{\delta^2 W_\Sigma}{\delta \mu(z)\delta\mu(z)} \right]~.
\eeq
Here, it is assumed that the second derivative $\fr{\delta^2}{\delta \mu(z)^2}$ is normal ordered according to \eqref{normalord}. It is also implicit that the $(g,n) = (0,1)$ term is excluded. To show that this equation indeed generates the topological recursion, we expand $W_\Sigma$ in powers of the coupling constant $\lambda$:
\begin{align}
W_\Sigma[\mu] =  \sum_{g=0}^\infty \lambda^{2g-2} \sum_{n=0}^\infty \fr{\mu(z_1)\cdots \mu(z_n)}{n!} \braket{\del\Phi(z_1)\cdots \del\Phi(z_n)}_{\Sigma, \mathsf{c}}^{(g)}~.
\end{align}
If we insert this expansion into \eqref{TRdiffeq} and compare powers of $\lambda^g \mu^n$, we find the following recursion for $2g-2+n\geq 0$:
\begin{align}\label{recursion3}
\mathcal{W}_{g,n+1}(z_0 ,z_I) &= \underset{z \to 0}{\mathrm{Res}} \,\fr{\braket{\del\Phi(z_0)\Phi(z)}_\sigma}{4 \,\omega(z)}   \Big[\mathcal{W}_{g-1,n+2}(z,z,z_I) \\& \hspace{2cm}+ \sideset{}{'}\sum_{\substack{g_1+g_2=g \\[0.3em] J_1 \sqcup J_2 = I}} \mathcal{W}_{g_1,1+|J_1|}(z,z_{J_1}) \,\mathcal{W}_{g_2,1+|J_2|}(z,z_{J_2}) \Big]~. \nonumber
\end{align}
The normal ordering prescription \eqref{normalord} has taken care of the term $\mathcal{W}_{0,2}(z,z)$ corresponding to the computation of the one-holed torus amplitude $\mathcal{W}_{1,1}(z)$:
\beq
\mathcal{W}_{0,2}(z,z) \equiv \lim_{w\to z} \left(\braket{\del\Phi(z)\del\Phi(w)}_\sigma - \fr{1}{(z-w)^2} \right) = \fr{1}{4z^2}~.
\eeq
At first sight, this recursion looks slightly different from the topological recursion \eqref{topologicalrecursion}. The $\mathcal{W}_{0,2}$ obtained in the twisted boson formalism after normal ordering is $(z+w)^{-2}$, whereas the Bergmann kernel was $(z-w)^{-2}$. Furthermore, we have dropped the minus signs for $\widetilde{z} =-z$. However, the recursion \eqref{recursion3} actually computes the \emph{same} invariants as the topological recursion \eqref{topologicalrecursion}. This can be checked either by direct computation, or by making the following observations:
\begin{itemize}
\item The recursion kernel \eqref{recursionkernel} is odd in its first argument, $
\mathcal{K}(-z_0,z) = -\mathcal{K}(z_0,z)$. The topological recursion \eqref{topologicalrecursion} then implies that all the symplectic invariants except $\omega_{0,2}$ are also odd in their first argument:
\beq\label{oddfirstarg}
\omega_{g,n+1}(-z_0, z_I) = -\omega_{g,n+1}(z_0,z_I)~.
\eeq
Therefore, the functions $\mathcal{W}_{g,n}(\bm{z})$ are all even in their first argument. In \cite{eynard4} it is furthermore shown that the invariants $\omega_{g,n}$ are symmetric multi-differentials, from which it follows that the $\mathcal{W}_{g,n}(\bm{z})$ are even in each argument. This, of course, agrees with our observation that the fields $\del\Phi(z)$ are even in $z$.
\item The recursion kernel \eqref{recursionkernel} is even in its second argument, $\mathcal{K}(z_0,-z) = \mathcal{K}(z_0,z)$. Writing $\mathcal{K}(z_0,z) = \kappa(z_0,z) \,dz_0\otimes \partial_z$, we conclude that $\kappa(z_0,z)$ is an odd function of $z$. Since all the other $\mathcal{W}_{g,n}(\bm{z})$ are even, the kernel $\kappa(z_0,z)$ projects to the even part in $z$ of $(z-w)^{-2}$. But the even part in $z$ of $(z-w)^{-2}$ is the same as the even part of $(z+w)^{-2}$. Therefore, we have for all $(g,n) \neq (0,2)$:
\beq
\underset{z \to 0}{\mathrm{Res}} \, \kappa(z_0,z) \fr{1}{(z-w)^2} \mathcal{W}_{g,n}(z, z_I) = \underset{z \to 0}{\mathrm{Res}} \, \kappa(z_0,z) \fr{1}{(z+w)^2} \mathcal{W}_{g,n}(z, z_I)~.
\eeq
This allows us to replace the Bergmann kernel by the regularized two-point function. 
\end{itemize}
Lastly, note that the recursion kernel $\mathcal{K}(z_0,z)$ for JT gravity matches the recursion kernel in \eqref{recursion3}.  Namely, we have:
\beq
\fr{\braket{\del\Phi(z_0) \Phi(z)}_\sigma}{4\,\omega(z)} = \half\left(\fr{1}{z_0 - z}-\fr{1}{z_0+z}\right) \fr{\pi \lambda}{2z \sin(2\pi z)} = \kappa(z_0,z)~,
\eeq
which agrees with \eqref{recursionkernel}.

We would like to show that the topological recursion, rewritten in the form \eqref{recursion3}, is equivalent to the Virasoro constraint for the twisted stress tensor. To do so, we write the functional differential equation \eqref{TRdiffeq} as an operator equation (to be read inside correlation functions):
\beq
\del\Phi_+(z_0) =  \underset{z \to 0}{\mathrm{Res}} \,\fr{\braket{\del\Phi(z_0)\Phi(z)}_\sigma}{2\,\omega(z)} \fr{1}{2} \big\{\del\Phi(z) \del\Phi(z)\big\}~.
\eeq
To see that this equation generates \eqref{TRdiffeq}, one replaces the operator $\del\Phi(z)$ by a functional derivative $\fr{\delta}{\delta \mu(z)}$ and act on $Z_\Sigma = e^{W_\Sigma}$. On the right-hand side we now recognize the stress tensor $T(z) = \half \big\{\del\Phi \del\Phi\big\}(z)$. Multiplying by $\omega(z_0)$ and integrating around zero we find:
\beq\label{eq5}
\oint_0 \fr{dz_0}{2\pi i} \fr{\omega(z_0) \del\Phi(z_0)}{w - z_0} -\oint_0 \fr{dz_0}{2\pi i}\fr{\omega(z_0)}{w - z_0} \oint_0 \fr{dz}{2\pi i}\fr{\braket{\del\Phi(z_0)\Phi(z)}_\sigma}{2\,\omega(z)} T(z) = 0~.
\eeq
We have written the residue as a contour integral, and introduced a point $w$ with $|w| > |z_0|$ to project onto the negative powers of $z_0$. Furthermore, we have replaced $\del\Phi_+$ by $\del\Phi$, which can be done because $\omega$ is holomorphic. 
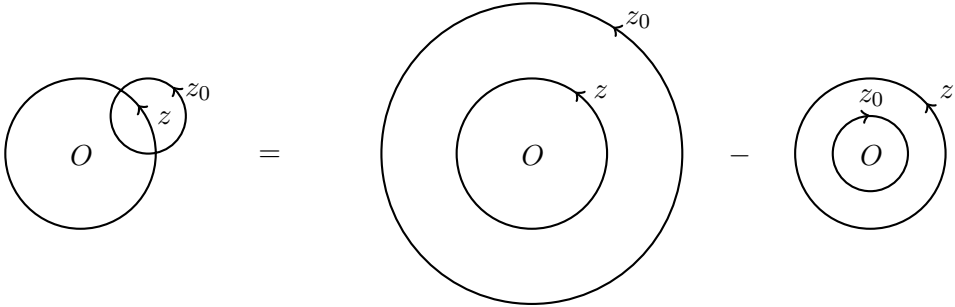
\begin{figure}[h]
\centering
\begin{tikzpicture}
\path[draw,line width=0.8pt,postaction=decorate, decoration={markings,
mark=at position 0.7cm with {\arrow[line width=1pt]{>}},
mark=at position 9.1cm with {\arrow[line width=1pt]{<}}
}
] (10:1) +(2,0) arc (0:360:1)  +(0.4,0.5) arc (0:-360:0.5);

\path[draw,line width=0.8pt,postaction=decorate, decoration={markings,
mark=at position 2cm with {\arrow[line width=1pt]{>}},
mark=at position 18cm with {\arrow[line width=1pt]{<}}
}
] (10:1) +(9,0) arc (0:360:2)  +(-1,0) arc (0:-360:1);

\path[draw,line width=0.8pt,postaction=decorate, decoration={markings,
mark=at position 0.9cm with {\arrow[line width=1pt]{<}},
mark=at position 8.8cm with {\arrow[line width=1pt]{<}}
}] (10:1) +(12,0) arc (0:360:0.5)  +(0.5,0) arc (0:-360:1);

\node at (2,0.15) {$O$};
\node at (4.5,0.15) {$=$};
\node at (8,0.15) {$O$};
\node at (12.5,0.15) {$O$};
\node at (10.75,0.15) {$-$};
\node at (8.9,1) {$z$};
\node at (9.4,1.95) {$z_0$};
\node at (13.5,1) {$z$};
\node at (12.5,0.92) {$z_0$};
\node at (3.1,0.65) {$z$};
\node at (3.55,1) {$z_0$};
\end{tikzpicture}
\caption{The contour deformation argument for $\oint_{0} dz \oint_z dz_0 = \oint_0 dz_0 \oint_0 dz - \oint_0 dz \oint_0 dz_0$.}
\label{contourdef}
\end{figure}

Next, we deform the $z$ and $z_0$ contours in the second term according to the contour deformation argument in Figure \ref{contourdef}. Using the fact that $\omega(z_0)$ is holomorphic and $\braket{\del\Phi(z_0) \Phi(z)}_\sigma$ has no poles at $z_0=0$, we conclude that the $\oint_0 dz \oint_0 dz_0$ integral vanishes. So, the second term of \eqref{eq5} can be written as
\begin{align}
\oint_0 \fr{dz_0}{2\pi i}\fr{\omega(z_0)}{w - z_0} \oint_0 \fr{dz}{2\pi i}\fr{\braket{\del\Phi(z_0)\Phi(z)}_\sigma}{\omega(z)} T(z) &= \half \oint_0 \fr{dz}{2\pi i} \oint_z \fr{dz_0}{2\pi i} \nonumber  \fr{\braket{\del\Phi(z_0)\Phi(z)}_\sigma}{w-z_0} \fr{\omega(z_0)}{\omega(z)} T(z) \\[1em]
&=\half \oint_0 \fr{dz}{2\pi i} \left(\fr{1}{w-z} - \fr{1}{w+z}\right) T(z)~.
\end{align}
The integration kernel $\half((w-z)^{-1} - (w+z)^{-1})$ is odd in $z$, so it projects to the even negative powers of $T(z)$. However, since $T(z)$ is already even in $z$, the kernel simply projects to the negative powers of $T(z)$. Plugging this result into \eqref{eq5}, we obtain
\beq
\oint_0 \fr{dz}{2\pi i} \fr{1}{w - z}  \Big[\omega(z) \del\Phi(z)- T(z)\Big] = 0~.
\eeq
Since $\omega(z)$ is holomorphic, we can freely add $\half \omega(z)^2$ inside the brackets. Doing so gives
\beq
\oint_0 \fr{dz}{2\pi i} \fr{1}{w - z} T_\omega(z) = 0~,
\eeq
where $T_\omega(z)$ is defined in \eqref{eq:shiftedstresstensor}. In correlation functions this gives the requirement \eqref{virconstraint} that the expectation value of $T_\omega$ is analytic, which is precisely the Virasoro constraint. 

\subsection{$\bb{Z}_2$-twisted fermions}\label{twistedfermioncomp}
We provide an operator formalism for the twisted fermion fields introduced in section \ref{twistfermions}. Firstly, we introduce $\Phi_0(x)$ and $\Phi_1(x)$ on the two sheets, which we expand in both integer and half-integer powers of $x$:
\begin{align}
\partial \Phi_0(x)&= \fr{1}{\sqrt{2}}\sum_{n\in \bb{Z}} \alpha_n x^{-n/2 -1}~,\\
\partial \Phi_1(x)&= \fr{1}{\sqrt{2}}\sum_{n\in \bb{Z}} (-1)^n\alpha_n x^{-n/2 -1}~.
 \end{align}
where the oscillators satisfy the commutation relation $[\alpha_n, \alpha_m] = \fr{n}{2} \delta_{n+m}$. As one can see, the fields rotate into each other due to the square root:
\beq
\Phi_0(e^{2\pi i} x) = \Phi_1(x)~, \quad \Phi_1(e^{2\pi i} x) = \Phi_0(x)~.
\eeq
Taking the difference leaves only odd $n = 2k+1$ in the sum, and so the following combination diagonalizes the monodromy:
\beq
\del\Phi(x) \equiv \fr{1}{\sqrt{2}}(\del\Phi_0(x) - \del\Phi_1(x)) = \sum_{k\in \bb{Z}} \alpha_{2k+1} x^{-k-\fr{3}{2}}~.
\eeq
If we rename $\alpha_{2k+1} \equiv \alpha_{k+\half}$, we see that $[\alpha_{k+\half}, \alpha_{l+\half}] = (k+\half) \delta_{k+l}$, which matches precisely with our definition of the twisted boson in \eqref{commu}. Now we compute the vacuum two-point functions in the usual way:
\begin{align}
\braket{\del\Phi_0(x)\del\Phi_0(y)}_\sigma &= \fr{1}{2} \sum_{n=0}^\infty \sum_{m=1}^\infty \braket{\sigma | \alpha_n \alpha_{-m} |\sigma}  x^{-n/2-1}y^{m/2-1} \nonumber\\
&= \half \sum_{n=1}^\infty  \fr{n}{2} x^{-n/2-1}y^{n/2-1}= \fr{(\sqrt{x} + \sqrt{y})^2}{4\sqrt{xy}} \fr{1}{(x-y)^2}~.
\end{align}
This has the correct behaviour of bosonic two-point functions as $x\to y$. Integrating with respect to $x$ and $y$, we get:
\beq
\braket{\Phi_0(x)\Phi_0(y)}_\sigma = \log(\sqrt{x}-\sqrt{y}) =\log(x-y) + \mathrm{reg.}
\eeq
The same answer is found for $\braket{\del\Phi_1\del\Phi_1}_\sigma$ and $\braket{\Phi_1\Phi_1}_\sigma$. Next, we compute the free two-point functions for bosons on opposite sheets:
\begin{align}
\braket{\del\Phi_0(x)\del\Phi_1(x)}_\sigma &= \half \sum_{n=1}^\infty \fr{n}{2} (-1)^n  x^{-n/2-1}y^{n/2-1} \nonumber \\
&= - \fr{1}{4\sqrt{xy} (\sqrt{x}+\sqrt{y})^2}~,
\end{align}
which is regular as $x \to y$. So in particular, we do not have to normal order combinations of $\Phi_0$ and $\Phi_1$. Integrating, we obtain the two-point function:
\beq
\braket{\Phi_0(x)\Phi_1(y)}_\sigma = \log(\sqrt{x}+\sqrt{y})~.
\eeq
The same results hold for $\braket{\del\Phi_1 \del\Phi_0}_\sigma$ and $\braket{\Phi_1 \Phi_0}_\sigma$. For consistency, we check that the KS field $\Phi(x)$ has the free two-point function that we derived in \eqref{propagator}:
\begin{align}
\braket{\Phi(x)\Phi(y)}_\sigma &= \fr{1}{2} \Big(\braket{\Phi_0(x)\Phi_0(y)}_\sigma + \braket{\Phi_1(x)\Phi_1(y)}_\sigma - \braket{\Phi_0(x)\Phi_1(y)}_\sigma - \braket{\Phi_1(x)\Phi_0(y)}_\sigma\Big) \nonumber \\
&= \log(\sqrt{x}-\sqrt{y}) - \log(\sqrt{x}+\sqrt{y})~.
\end{align}
Having computed the bosonic correlators, we can go on to study the bosonized fermions. 
First, let us compute the OPE between two fermions on the same sheet:
\beq
\psi_a(x)\psi_a(y)^\dagger = \big\{e^{\Phi_a(x)}\big\}\big\{e^{-\Phi_a(y)}\big\} \sim \fr{1}{x-y}e^{\Phi_a(x) - \Phi_a(y)}  \sim \fr{1}{x-y}~.
\eeq
The symbol $\sim$ means that we have only kept singular terms in the limit that $x\to y$. We used the OPE computed above that $\Phi_a(x)\Phi_a(y) \sim \log(x-y)$. As one can see, the cocycles have squared to one. By the boson-fermion correspondence, we have for $a=0,1$:
\beq
\del\Phi_a(x) = \lim_{y\to x} \big\{\psi_a(x) \psi_a^\dagger(y)\big\} \equiv  \lim_{y\to x} \left(\psi_a^\dagger(x) \psi_a(y) - \fr{1}{x-y}\right)~.
\eeq
For two fermions on opposite sheets, we do not have to normal order since $\Phi_0(x')\Phi_1(x)$ is regular, and we can simply add the exponentials in a single normal-ordered exponential:
\beq
\psi_0(x)\psi_1^\dagger(x) \equiv \lim_{x'\to x} \psi_0(x')\psi_1(x) =  \mathsf{c}_0\mathsf{c}_1 \big\{e^{\Phi_0(x) - \Phi_1(x)}\big\}~.
\eeq
\newpage
\bibliographystyle{JHEP}
\bibliography{mybibliography}

\end{document}